\documentclass[]{pasj01}
\usepackage{tabularx}
\usepackage{siunitx}
\usepackage{url}
\usepackage{threeparttable}
\usepackage{enumitem}

\Received{}
\Accepted{}
 
\begin{document}

\title{
    Giant molecular clouds and their Type classification in M74: Toward understanding star formation and cloud evolution}

\author{Fumika \textsc{Demachi}\altaffilmark{1}%
    \thanks{f.demachi@a.phys.nagoya-u.ac.jp}}
\altaffiltext{1}{Department of Physics, Nagoya University, Furo-cho, Chikusa-ku, Nagoya 464-8601, Japan}

\author{Yasuo \textsc{Fukui}\altaffilmark{1}}

\author{Rin I. \textsc{Yamada}\altaffilmark{1}}

\author{Kengo \textsc{Tachihara}\altaffilmark{1}}

\author{Takahiro \textsc{Hayakawa}\altaffilmark{1}}

\author{Kazuki \textsc{Tokuda}\altaffilmark{2,3}}
\altaffiltext{2}{Department of Earth and Planetary Sciences, Faculty of Sciences, Kyushu University, Nishi-ku, Fukuoka 819-0395, Japan}
\altaffiltext{3}{National Astronomical Observatory of Japan, National Institutes of Natural Science, 2-21-1 Osawa, Mitaka, Tokyo 181-8588, Japan}

\author{Shinji \textsc{Fujita}\altaffilmark{4}}
\altaffiltext{4}{The Institute of Statistical Mathematics, 10-3 Midori-cho, Tachikawa, Tokyo 190-8562, Japan}

\author{Masato I. N. \textsc{Kobayashi}\altaffilmark{5}}
\altaffiltext{5}{I. Physikalisches Institut, Universit\"{a}t zu K\"{o}ln, Z\"{u}lpicher Str. 77, D-50937 K\"{o}ln, Germany}

\author{Kazuyuki \textsc{Muraoka}\altaffilmark{6}}
\altaffiltext{6}{Department of Physics, Graduate School of Science, Osaka Metropolitan University, 1-1 Gakuen-cho, Naka-ku, Sakai, Osaka 599-8531, Japan}

\author{Ayu \textsc{Konishi}\altaffilmark{6}}

\author{Kisetsu \textsc{Tsuge}\altaffilmark{7}}
\altaffiltext{7}{Institute for Advanced Study, Gifu University, 1-1 Yanagido, Gifu 501-1193, Japan}

\author{Toshikazu \textsc{Onishi}\altaffilmark{6}}

\author{Akiko \textsc{Kawamura}\altaffilmark{3}}


\newcommand{\twelveco}{\mbox{$^{12}$CO}}
\newcommand{\thirteenco}{\mbox{$^{13}$CO}}
\newcommand{\twelvecoh}{\mbox{$^{12}$CO($J$ = 2--1)}}
\newcommand{\twelvecohh}{\mbox{$^{12}$CO($J$ = 3--2)}}
\newcommand{\thirteencoh}{\mbox{$^{13}$CO($J$ = 2--1)}}
\newcommand{\twelvecol}{\mbox{$^{12}$CO($J$ = 1--0)}}
\newcommand{\thirteencol}{\mbox{$^{13}$CO($J$ = 1--0)}}
\newcommand{\ceighteenol}{\mbox{C$^{18}$O($J$ = 1--0)}}
\newcommand{\ceighteenoh}{\mbox{C$^{18}$O($J$ = 2--1)}}
\newcommand{\csl}{\mbox{CS($J$ = 1--0)}}
\newcommand{\msun}{\mbox{$M_\odot$}}
\newcommand{\ergs}{\mbox{erg~s$^{-1}$}}
\newcommand{\kms}{\mbox{km~s$^{-1}$}}
\newcommand{\kkms}{\mbox{K~km~s$^{-1}$}}
\newcommand{\vlsr}{\mbox{$V_\mathrm{LSR}$}}
\newcommand{\nhtwo}{\mbox{$N_\mathrm{H_2}$}}
\newcommand{\nco}{\mbox{$N_\mathrm{^{13}CO}$}}
\newcommand{\tmb}{\mbox{$T_\mathrm{mb}$}}
\newcommand{\tex}{\mbox{$T_\mathrm{ex}$}}
\newcommand{\hone}{\mbox{H{\sc i}}}
\newcommand{\LHa}{\mbox{$L_{\mathrm{H\alpha}}$}}
\newcommand{\htwo}{H{\sc ii}}
\newcommand{\red}{\textcolor{red}}
\newcommand{\blue}{\textcolor{blue}}
\newcommand{\green}{\textcolor{green}}

\maketitle

\begin{abstract}
    We investigated the giant molecular clouds (GMCs) in M74 (NGC~628), using data obtained from the PHANGS project.
    We applied the GMC Types according to the activity of star formation: Type I without star formation, Type II with H$\alpha$ luminosity (\LHa) less than $10^{37.5}~\ergs$, and Type III with \LHa \ greater than $10^{37.5}~\ergs$.
    A total of 432 GMCs were identified, with 59, 201, and 172 GMCs, for Type I, II, and III, respectively.
    The size and mass of the GMCs range from 23 to 238 pc and $10^{4.9}$ to $10^{7.1}$ M$_{\odot}$, indicating that the mass and radius increase from Type I to III.
    Clusters younger than 4 Myr and \htwo\ regions are concentrated within 150 pc of a GMC, indicating a tight association between these young objects and GMCs.
    The virial ratio decreases from Type I to Type III, indicating that Type III GMCs are the most gravitationally relaxed among the three.
    We interpret that the GMCs evolve from Type I to Type III, as previously observed in the LMC.
    Based on a steady- state assumption, the estimated evolutionary timescales of Type I, II, and III are 1, 5, and 4 Myr, respectively.
    We assume that the timescale of Type III is equal to the age of the associated clusters, indicating a GMC lifetime of 10 Myr or longer.
    Although Chevance et al. (2020, MNRAS, 493, 2872) investigated GMCs using the same PHANGS dataset of M74, they did not define a GMC, reaching an evolutionary picture with a 20 Myr duration of the non-star-forming phase, which was five times longer than 4 Myr.
    We compare the present results with those of Chevance et al. (2020) and argue that defining individual GMCs is essential for understanding GMC evolution.
\end{abstract}

\section{Introduction}
The formation and evolution of giant molecular clouds (GMCs), the main sites of star formation, are the fundamental processes that drive galaxy evolution.
In early CO studies on galaxies, the low spatial resolutions of the millimeter telescopes hampered the resolution of the individual GMCs in galaxies, and GMCs were observed as an unresolved complex of CO emission above kiloparsec scales.
These early CO images did not resolve the star formation process in GMCs.
Schmidt's law was proposed to describe the large--kpc scale relationship between star formation and \hone~gas \citep{Schmidt1959} and was later extended to CO gas by \citet{Kennicutt1998}, which is known as the Kennicutt--Schmidt law (K--S law).
The K--S law states that the star formation rate (SFR) is proportional to the 1.4th power of the gas column density in galaxies and applies to all galaxies in the universe, including high redshifts.
Despite the success of the K--S  law as an empirical law, the physical mechanisms that govern star formation remain unknown.

\subsection{Type classification of resolved GMCs in the Large Magellanic Cloud and M33}
The first resolved GMCs in a single galaxy were obtained in the 2.6 mm \twelvecol \ emission at 40 pc resolution over the whole Large Magellanic Cloud (LMC hereafter) with the NANTEN 4 m telescope by \citet{Fukui1999} (F99 hereafter).
The LMC is located at the closest distance of 50 kpc from the sun among the galaxies, and a small telescope with a diameter of 4 m was able to resolve GMCs with a typical size of 100 pc.
The NANTEN survey mapped some 150 GMCs and found that the GMCs are classified into three types; Type I, Type II, and Type III by the difference in star formation activity; Type I shows no sign of star formation; that is, it is not associated with \htwo \ regions or stellar clusters. Type II is associated with small \htwo \ regions only, but with no stellar clusters, where small \htwo \ regions correspond to those of H$\alpha$ luminosity \LHa \ $\sim 10^{36\text{--}37} \ \ergs$, and Type III is associated with stellar clusters and large \htwo \ regions with $\LHa > 10^{37} \ \ergs$, suggesting active, on-going formation of massive clusters.

The classification is based on high-mass stars and clusters, including high-mass stars because low-mass stars are not available as signposts at a distance.
The association of these young objects was revealed by their tight spatial correlation with GMCs through careful inspection of optical images (F99).
It is obvious that the Type classification was intended to provide a first empirical classification scheme of GMCs, and all the datasets including the H$\alpha$ emission and the clusters younger than 10 Myr were utilized.
It is important to note that the classification was performed without considering any evolutionary aspects.
After the Type classification was completed, F99 pursued possible implications of the GMC Types and reached an interpretation that the Types likely represent an evolutionary sequence of GMCs which is expressed as [Type I $\rightarrow$ Type II $\rightarrow$ Type III] over the GMC lifetime.
Subsequently, \citet{Yamaguchi2001} renewed a comparison between GMCs and \LHa \ and found that the boundary \LHa \ between Type II and Type III is raised slightly to $10^{37.5} \ \ergs$.
\citet{Fukui2008} (F08 hereafter) conducted a follow-up NANTEN survey and cataloged GMCs with a sensitivity twice that of the initial results.
Based on F08, \citet{Kawamura2009} (K09 hereafter), confirmed the GMC Types and derived the timescales of Types I, II, and III as 6, 12, and 7 Myr, respectively, corresponding to a total GMC lifetime of 20--30 Myr.
Evidently, a Type III GMC is rapidly disrupted by cluster feedback to evolve into naked clusters without gas (F99, K09).
The Type classification of GMCs by F99 indicated a breakdown of the K--S law at the 100 pc scale because the GMC Types clearly indicate that, in contrast to the K--S law, the SFR is not determined only by the column density at the GMC scale.
The mass and size of the GMCs of each Type are not significantly different, whereas the high-mass star formation levels differ significantly.
Therefore, it is clear that the K–S law is not applicable to GMCs at the 100 pc scale if the GMC Types in the LMC are universal.

Subsequently, the GMC Types were confirmed in M33, the second-nearest neighbor, using the IRAM \twelvecoh \ survey by \citet{Gratier2012} and \citet{Corbelli2017}.
These authors showed that the Type classification of GMCs, with some minor modifications, is applicable to M33, and estimated the GMC lifetime to be 17 Myr.
The M33 results suggest that GMC Types may be common to galaxies. Prior to these studies, \citet{Onodera2010} observed GMCs in part of M33 and showed that the relationship between SFR and $\mathrm{H_2}$ column density varies depending on the resolution, and that the K--S law breaks down at 100 pc.

\subsection{Resolved GMCs in the other galaxies without Type classification}
Until the early 2010s, resolved GMCs were obtained only for several nearby galaxies, including the LMC and M33, because the high resolution required to extend the resolved study further to other galaxies was difficult to achieve \citep{FukuiKawamura2010}.
It was only in the mid-2010s that CO observations of galaxies at a resolution of 100 pc were made for those galaxies located within 10 Mpc, which include NGC 300 at a distance of 2 Mpc observed with ALMA in \twelvecoh \ \citep{Faesi2018,Kruijssen2019}, M83 at 5 Mpc observed with ALMA, \twelvecol \ \citep{Freeman2017}, NGC~4736 at 5 Mpc, NGC~6846 at 7 Mpc and NGC~4826 at 8 Mpc observed by CARMA and NRO-45m, \twelvecol \ \citep{DonovanMeyer2013}, M101 at 7 Mpc and NGC~6946 at 7 Mpc observed by CARMA, \twelvecoh \ \citep{Rebolledo2015}, M51 at 8 Mpc (PAWS, \cite{Schinnerer2013,Colombo2014}).

These galaxies were observed at the GMC scale, whereas only M51 was a grand design spiral among these galaxies.
Although the Type classification of GMCs was not carried out for these galaxies, more distant interacting galaxies, the Antennae Galaxies, at 22 Mpc were mapped with ALMA in \twelvecoh.
Further, the GMC Type classification was applied to the overlapping region by \citet{Whitmore2014}, yielding an insight into the different levels of star formation activities where star formation is strongly affected by collisional interaction.

Most recently, PHANGS-ALMA collaboration made a systematic effort to map resolved GMCs in 70 galaxies up to a distance of 10--20 Mpc and obtained physical parameters, including the molecular mass of the resolved GMCs and their evolutionary timescales.
The collaboration includes the determination of distance \citep{Anand2021} and fitting rotation curve \citep{Lang2020}, and determination of the CO(3--2)/CO(2--1) ratio and CO(2--1)/CO(1--0) ratio \citep{denBrok2021,Leroy2022}.
The physical parameters were compared by adjusting the resolution to the same values \citep{Leroy2016}.
\citet{Sun2018} showed that the virial ratio ($\alpha_\mathrm{vir}$) is in a range of $\sim$1.5--3.0 for 15 galaxies including M74.
\citet{Sun2020} suggested that interstellar medium properties are affected by environmental conditions at the cloud scale by comparing the central, bar, and arm regions.
In addition, \citet{Sun2022} showed that the molecular gas properties of a host galaxy are well represented by the average properties on the kpc scale.
Furthermore, the relationship between molecular gas and star-formation regions was investigated.
\citet{Kreckel2018} used data at a resolution of 50 pc and confirmed that the K--S law breaks down.
\citet{Schinnerer2019} and \citet{Pan2022} investigated the correlation between CO and H$\alpha$ to quantify the relationship between molecular gas and the star forming regions on a pixel-to-pixel basis at various resolution and confirmed that CO to H$\alpha$ correlation becomes strong below kiloparsec scales.
Based on the statistical methods proposed by \citet{KruijssenLongmore2014, Kruijssen2018, Chevance2020, Kim2021, Kim2022, Ward2022}, a GMC lifetime of 10--30 Myr was derived.
This lifetime is consistent with that estimated by K09 and \citet{Corbelli2017}, whereas the non-star-forming phase of \citet{Chevance2020} is significantly longer than that of K09 and \citet{Corbelli2017}.
The method employed by \citet{Chevance2020} does not consider a GMC as a discrete entity and treats the interstellar matter as consecutive medium by measuring the fraction of CO and H$\alpha$ within an aperture of varying radius centered on the peak positions of CO and H$\alpha$.
This is in contrast with the preceding works by F99, K09, and \citet{Corbelli2017}, which were based on defining each GMC using the resolved CO images at the highest resolution achieved.

\subsection{Goal of a GMC study of galaxies toward a universal scheme of GMC Types}
GMC Types were shown to be useful for GMC classification in the LMC and M33.
Despite this, negligible effort has been invested to test the GMC Type classification in most of the resolved GMCs mapped.
Our ultimate goal is to establish the GMC classification as a universal tool to learn the essential role of GMCs in galaxy evolution.
To achieve this goal, our immediate step is to develop an extendable practical scheme for GMC Type classification.
The original classification scheme of F99 utilized \htwo \ regions and clusters to characterize the level of high star-formation activity in a GMC.
However, it has a limitation owing to the availability of clusters because a uniform study of stellar clusters in a single galaxy is not yet available for many galaxies.
On the other hand, the H$\alpha$ emission is more uniformly surveyed in galaxies.
By recollecting that the original GMC Types adopted the level of active high mass star formation by referring to the \htwo \ regions and clusters, and the difference between Type II and Type III was also characterized by $\LHa \sim 10^{37.5} \ \ergs$ empirically not only by the young clusters.
Accordingly, in the present study, we use \LHa \ only to classify GMC Type II and Type III instead of using clusters.
This scheme is a working hypothesis that requires testing in galaxies for which cluster information is available; however, it allows us to apply the method to more numerous galaxies.
This classification maintains the original idea of F99 that the Types represent the degree of high-mass star formation and will significantly widen the galaxy sample with a balanced view supplemented by the associated clusters.

\subsection{M74 as the primary target galaxy in the present study}
In the present study, we focus on M74 (NGC~628), which is a nearby (d = 9.7 Mpc, \cite{Anand2021}), SAc type grand-design spiral galaxy.
The galaxy has a small inclination angle and is suitable for the detailed study of galaxy evolution, including star formation and interstellar medium (ISM) evolution.
The galaxy is known to exhibit a moderate SFR of $\sim 2 \ M_\odot \ \mathrm{yr^{-1}}$ over the whole galaxy \citep{Sanchez2011}.
CO emissions were mapped and presented in several papers prior to the ALMA era (e.g.,  \cite{WakkerAdler1995,Helfer2003,Leroy2009,Rebolledo2015}).
The ALMA observations were made at a $1” \ \sim$ 50 pc resolution in \twelvecoh \ as part of a PHANGS-ALMA survey (PI: E. Schinnerer; co-PIs: A. Hughes, A. K. Leroy, A. Schruba, and E. Rosolowsky).
The M74 CO maps with ALMA were published by \citet{Leroy2016,Sun2018,Kreckel2018,Utomo2018}.
The Multi Unit Spectroscopic Explorer (MUSE) optical Integral Field Units (IFU) instrument on the Very Large Telescope (VLT) provided H$\alpha$ images of galaxies.

Large projects, including THINGS, HERACLES, SINGS, KINGFISH, and EMPIRE \citep{Walter2008,Leroy2009,Kennicutt2003,Jimenez-Donaire2019} observed M74 at various wavelengths.
Wide-field Camera 3 (WFPC3) / Advanced Camera for Surveys (ACS) observations by the Hubble Space Telescope (HST) Legacy Extragalactic UV Survey (LEGUS) program \citep{Calzetti2015} provide the observational data on stellar clusters, and high-resolution, wide--field optical Integral Field Unit (IFU) imaging was carried out by the VIRUS-P Exploration of Nearby Galaxies (VENGA) survey and Spectro-Imageur \`a Transform\'ee de Fourier pour l'\'Etude en Long et en Large des raies d'\'Emission (SITELLE) at Canada–France–Hawaii Telescope (CFHT) \citep{Blanc2013b,Rousseau-Nepton2018}.

The goal of this study is to measure the physical conditions of molecular clouds and quantitatively describe the associated star-formation activity by applying the GMC Type classification in a clear grand-design spiral galaxy for the first time.
By comparing GMCs and young objects, we aim to construct a picture of GMCs and star formation therein.
The remainder of this paper is organized as follows.
Section \ref{data} presents the observational data used in the study.
Section \ref{propeties} and \ref{extinction} describes the properties of the molecular gas and \htwo \ regions.
In Section \ref{discussions}, we discuss the GMC Type classification and high-mass star formation.
The conclusions are summarized in Section \ref{conclusions}.
\section{Data}\label{data}
\subsection{\twelvecoh}
We used the \twelvecoh~data PHANGS--ALMA data release v1.0 \citet{Leroy2021b} observed with 12 m and 7 m dishes of ALMA in the total power array (TP).
M74 was observed in mosaic with a velocity resolution of $2.54~\mathrm{km~s^{-1}}$, angular resolution of $1\farcs12$ ($\sim 53~\mathrm{pc}$), and the $1\sigma$ noise level of $115~\mathrm{mK}$.
For details of data reductions, see \citet{Leroy2021a}.
Further, to minimize the effect of the noise in the GMC identification, ``Strict mask'' (see for details \cite{Rosolowsky2006, Leroy2021a}) was employed to pick up voxels with $>4\sigma$ and adjoining voxels above $2\sigma$.

\subsection{\htwo ~regions}
\begin{figure*}[htbp]
    \begin{center}
        \includegraphics[width=0.6\linewidth,clip]{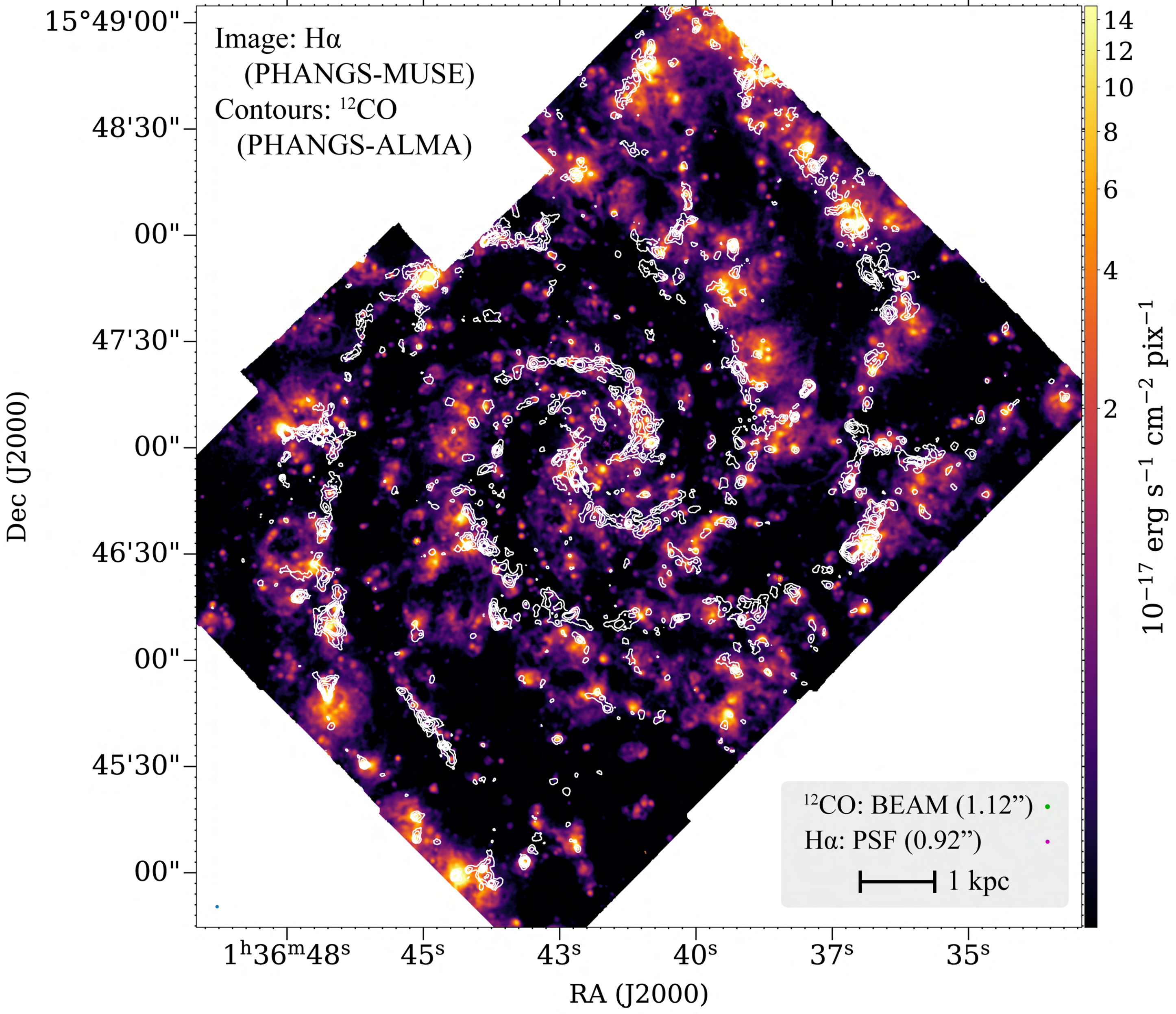}
    \end{center}
    \caption{Distribution of the \twelvecoh ~emission \citep{Leroy2021b} superposed on the H$\alpha$ image \citep{Emsellem2022}. The white contours indicate the \twelvecoh ~integrated intensity. The lowest contour is $5~\kkms$, and the other contours are drawn in intervals of $1~\kkms$. The CO emission shows spiral arms similar to those of the H$\alpha$ emission, while the two distributions show spatial shifts.}
    \label{COcontour_HaMap}
\end{figure*}

We used the H$\alpha$ data from the PHANGS--MUSE Large program \citep{Emsellem2022} taken with the ESO Science Archive Facility.
PHANGS--MUSE consists of optical observations made with a surface spectrometer Multi Unit Spectroscopic Explorer (MUSE) installed on the ESO Very Large Telescope (VLT).
One grid is $0\farcs2$ with a resolution of $0\farcs92$ (Gaussian PSF), and the typical sensitivity in H$\alpha$ flux ($3\sigma$) is $2\times 10^{-18}$ erg s$^{-1}$ cm$^{-2}$ spaxel$^{-1}$.
To obtain the physical parameters of the ionized gas by suppressing the effect of stellar radiation, we used data from which continuum emission was subtracted (for details, see \cite{Emsellem2022}).
To judge the association between a GMC and an \htwo\ region via pixel-to-pixel comparison, we used the same grid size of $0\farcs4$ for both H$\alpha$ and \twelvecoh.
These two distributions are shown in Figure \ref{COcontour_HaMap}.

\subsection{Star clusters}\label{data_cluster}
\begin{figure}[htbp]
    \begin{center}
        \includegraphics[width=0.85\linewidth,clip]{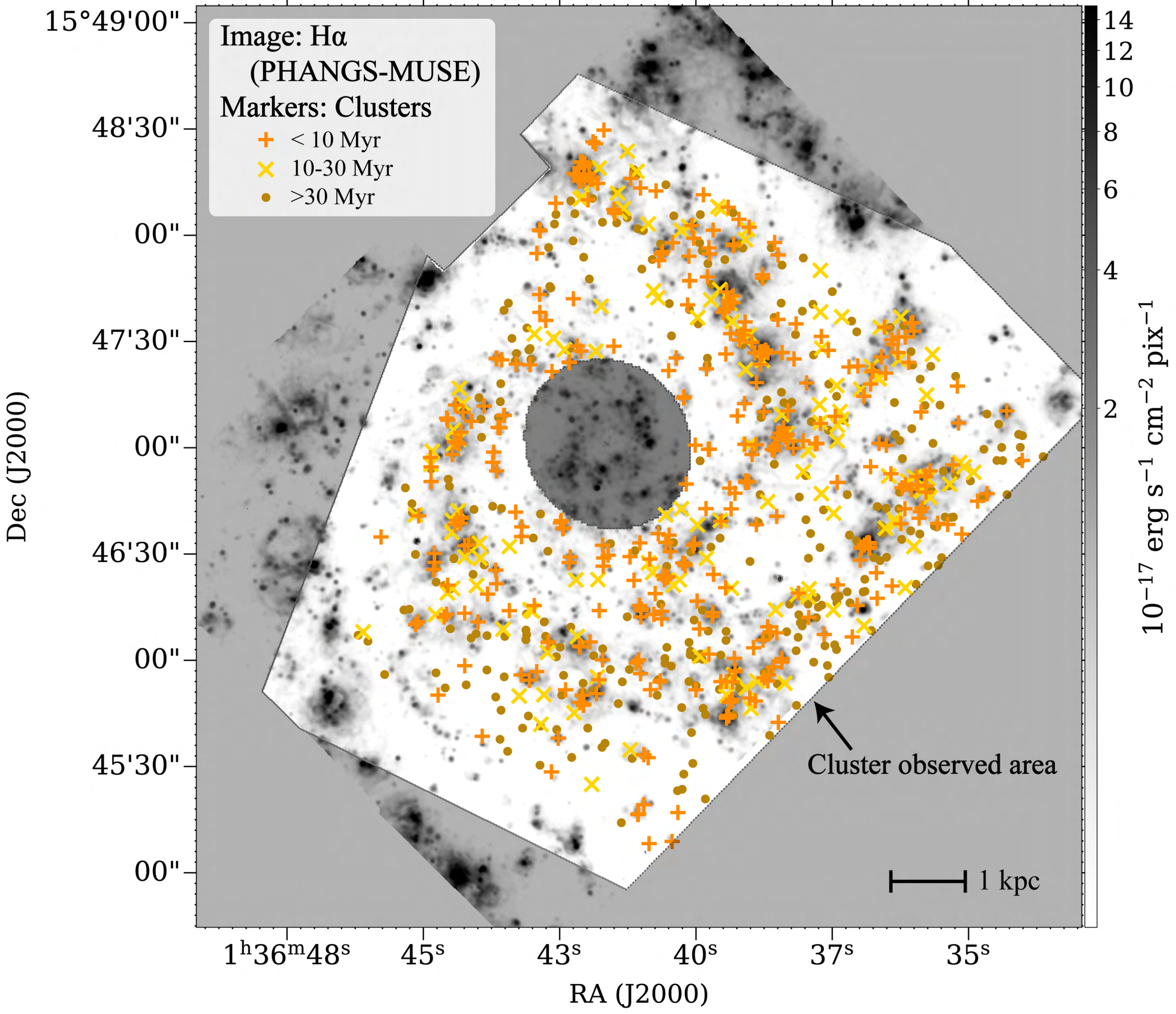}
    \end{center}
    \caption{Position of the clusters observed with HST \citep{Adamo2017} and selected in section \ref{data_cluster} is superposed on the H$\alpha$ image. The gray dotted lines show the area observed with HST, and the central gray circle shows the mask for the central area defined by \citet{Querejeta2021}.
        The orange crosses indicate 374 clusters with Age $< 10~\mathrm{Myr}$; the yellow crosses indicate 135 clusters with Age $=10$--$30~\mathrm{Myr}$, and the brown dots show 309 clusters with Age $ > 30~\mathrm{Myr}$.}
    \label{cluster_map}
\end{figure}

We used the catalog of star clusters by \citet{Adamo2017}, based on The Legacy Extra Galactic UV Survey (LEGUS) \citep{Calzetti2015}.
LEGUS made simultaneous observations covering the near UV to I band and optical observations with WFC3 and ACS using HST, and provided homogeneous datasets of local star-forming galaxies.
The PSF of WFC3 had a full width at half maximum (FWHM) of $\sim 0\farcs08$.
The WFC3 data were gridded to $0\farcs04~\mathrm{pix^{-1}}$, the same as the pixel size, and aperture photometry was performed in a circle of four pixels in radius.
For details of data reduction and cluster identification, see \citet{Calzetti2015} and \citet{Adamo2017}.
The present study used the following clusters: clusters younger than 10 Myr are 374, and those of 10--30 Myr are 135.
An overlay of the clusters on the H$\alpha$ distribution is shown in Figure \ref{cluster_map}.

\subsection{JWST 21$\mathrm{\mu}$m}
We used 21 $\mathrm{\mu}$m data from PHANGS--JWST survey \citep{Lee2023} obtained by Mid-Infrared Instrument (MIRI).
The data covers a large area of M74 with resolution of $0\farcs67$ (PSF).
See \citet{Lee2023} for more details on data reduction.
To judge the association between a GMC and a 21 $\mathrm{\mu}$m region by pixel-to-pixel comparison, we used the same grid size of $0\farcs4$ for both 21 $\mathrm{\mu}$m and \twelvecoh.
\section{Physical parameters of GMCs and \htwo~regions}\label{propeties}
\subsection{GMCs}
\subsubsection{Analysis of GMCs}\label{analysisGMC}
The GMCs were identified using \texttt{PYCPROPS}\footnote{$\langle$ \url{https://github.com/PhangsTeam/pycprops} $\rangle$} \citep{Rosolowsky2021}.
\texttt{PYCPROPS} is a Python package based on CPROPS \citep{Rosolowsky2006}.
This algorithm first determines the local maximum as a kernel of identification and then allocates the surrounding pixels to determine the distribution of the local structure of the position--position--velocity (PPV) data cube.
\citet{Rosolowsky2021} identified GMCs for 10 galaxies, including M74, but they smoothed the CO data to 90 pc and added noise artificially so that the surface brightness sensitivity became $75~\mathrm{mK}$ per $2.5~\mathrm{km~s^{-1}}$ channel for all galaxies for comparison.
In the present work we need a GMC catalog of M74 to compare GMCs with the LMC, for which the resolution is 40 pc and $1 \sigma$ noise level is $0.4~\mathrm{K~km~s^{-1}}$ (K09).
Therefore, we performed GMC identification independently and achieved a resolution and noise level similar to those of LMC.
In the present study, we tuned the parameters such that the GMC is not divided into too small, unnatural pieces (for details see Appendix.\ref{cprops}).
Because the H$\alpha$ observed area is smaller than that of CO, we exclude a GMC defined by \texttt{PYCPROPS} if it is extended beyond the H$\alpha$ observed area.
Furthermore, the GMCs in the central region of a galaxy have a larger velocity dispersion than those in the disk, indicating that they are of different populations.
Thus, we excluded them by applying the center region mask defined by \citet{Querejeta2021}.
This mask is defined based on eye inspection of the excess of the Spitzer $3.6~\mathrm{\mu m}$ and ALMA $\twelvecoh$.

\subsubsection{Physical properties of GMCs}
\begin{figure}[htbp]
    \begin{center}
        \includegraphics[width=0.85\linewidth,clip]{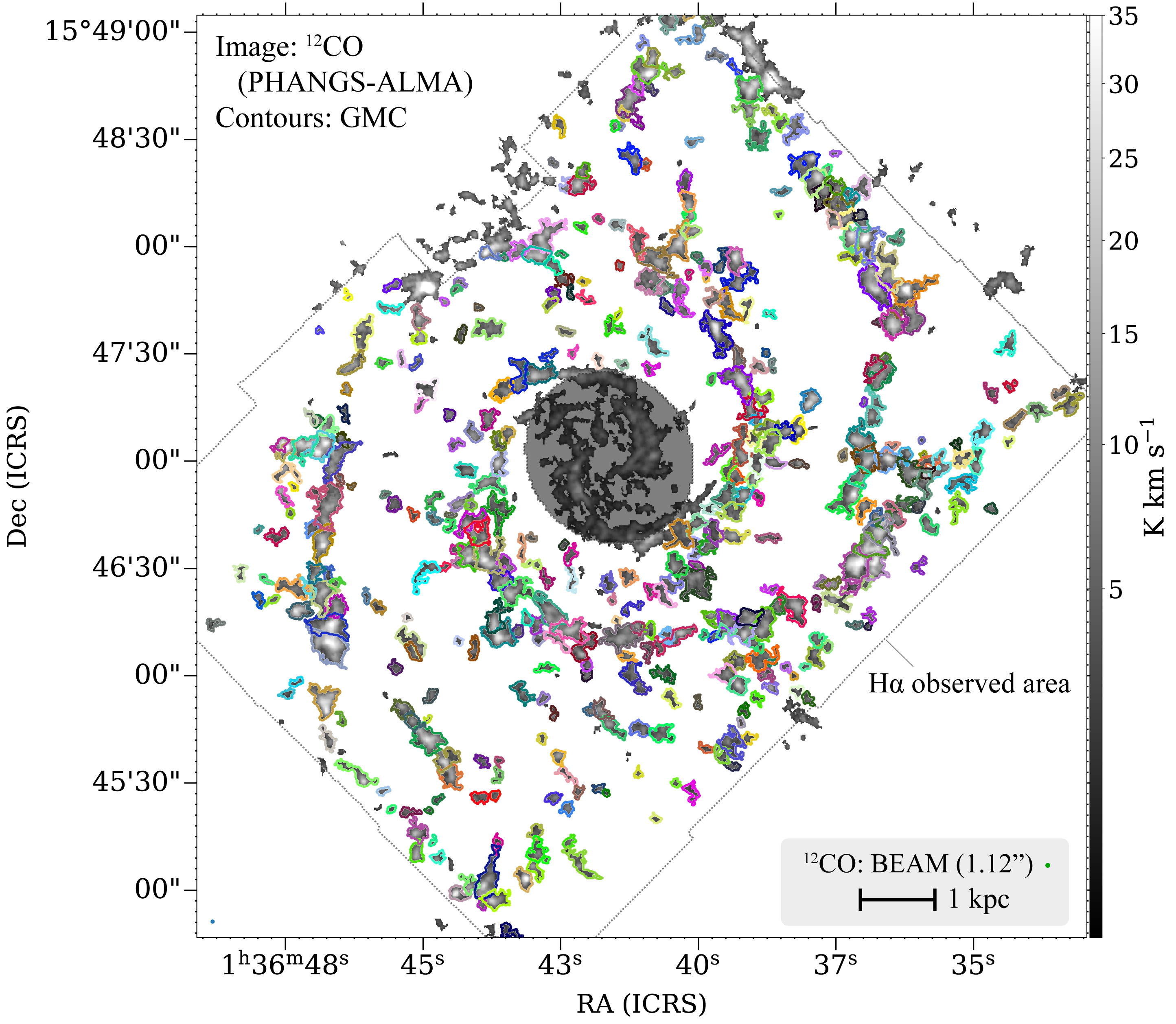}
    \end{center}
    \caption{432 GMCs identified by \texttt{PYCPROPS} are shown by color contours superposed on the \twelvecoh \ integrated intensity map (Figure \ref{COcontour_HaMap}). The gray dotted lines indicate the observed area in H$\alpha$ and the central gray circle indicates the masked region for the galactic center region. }
    \label{GMC_COmap}
\end{figure}

\begin{figure*}[htbp]
    \begin{center}
        \includegraphics[width=0.85\linewidth,clip]{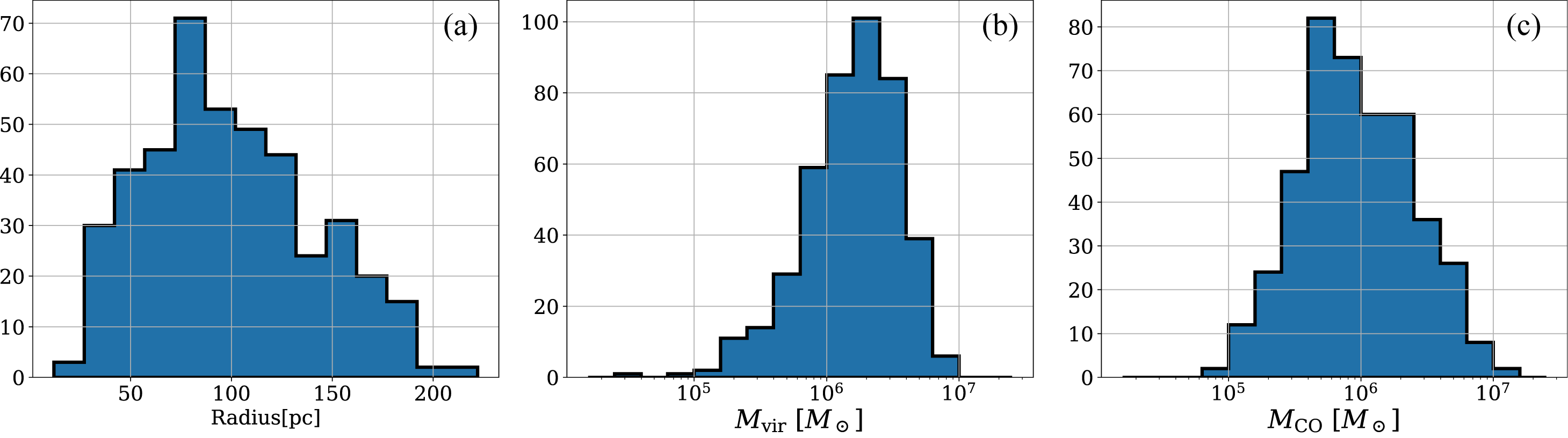}
    \end{center}
    \caption{Histogram of (a) the radius of GMCs, (b) the virial mass of GMCs ($M_\mathrm{vir}$), and (c) the CO luminous mass ($M_\mathrm{CO}$).}
    \label{R_Mvir_Mco}
\end{figure*}

Here we summarize the main properties of GMCs.
Figure \ref{GMC_COmap} shows the GMCs identified by \texttt{PYCPROPS}.
The total number of GMCs is 432, except for the masked central region.
First, we estimate the radius of the GMC.
The envelope of a GMC is extrapolated by correction, which extends the boundary surface up to $T_{\rm edge} = 0$ K (see for details \cite{Rosolowsky2006}) to suppress the effects of sensitivity.
The GMC distribution is deconvoluted using a circular Gaussian beam.
As a result, we obtain second moments $\sigma_{\rm maj}$ and $\sigma_{\rm min}$ in the $x$ and $y$ axes and the coefficient $\eta = \sqrt{2\ln 2}$ which comes from the mass distribution within the GMC under the assumption of 2D Gaussian.
Then, we calculate a radius as $R = \eta\sqrt{\sigma_{\rm maj}\sigma_{\rm min}}$ and a typical radius is found to be 90 pc in a range of 23--238 pc as shown in Figure \ref{R_Mvir_Mco}(a).

Then, we calculate the virial mass $M_\mathrm{vir} = 5\sigma_v^2R_{3D}/G$ for each GMC.
$\sigma_v$ is the velocity dispersion of \twelvecoh\ and is expressed as follows: $\sigma_v = \sqrt{\sigma_\mathrm{extrap}^2-\sigma_\mathrm{chan}^2}$ where $\sigma_\mathrm{extrap}$ is that extrapolated as $T_\mathrm{edge}=0$.
$\sigma_\mathrm{chan}$ is the Gaussian channel width and calculated as $\sigma_\mathrm{chan} = \Delta v/\sqrt{2\pi}$ ($\Delta v$ is channel width).
$G$ is the gravitational constant, and $R_{3D}$ is the three-dimensional mean radius.
If the radius $R$ in the disk orientation is smaller than half of the scale height of the galaxy $H$, we adopt $R_{3D} = R$.
However, if $R$ exceeds half of the scale height $H$, then we adopt $R_{3D} = \sqrt[3]{(R^2H)/2}$ because the depth cannot be estimated from $R$. We assume $H = 100$ pc.\
The distribution of $M_\mathrm{vir}$ is shown in Figure \ref{R_Mvir_Mco}(b).

Next, we estimate the luminous mass of a GMC as calculated by using the relationship $M_{\rm CO} = \alpha_\mathrm{CO}^{2-1} L_{\rm CO}$, where $L_{\rm CO}$ is the CO luminosity, and $\alpha_\mathrm{CO}^{2-1}$ is a conversion factor of \twelvecoh\ intensity into the luminous mass.
We adopt $\alpha_\mathrm{CO}^{2-1}= 5.6\ M_{\odot}\ \rm{pc}^{-2}\ (\rm{K\ km\ s}^{-1})^{-1}$ derived by \citet{Sandstrom2013} (see also \cite{Yasuda2023}).
The distribution of $M_\mathrm{CO}$ is shown in Figure \ref{R_Mvir_Mco}(c); a typical $M_\mathrm{CO}$ is $10^{5.93} M_\odot$.

Further details on the derivation method for the physical parameters are provided by \citet{Rosolowsky2021}.

\subsection{\htwo ~regions}\label{analysisHII}
\citet{Santoro2022} presented a catalogue of \htwo ~regions in 19 galaxies but did not present size of the \htwo ~region.\footnote{After completing our analysis and preparing this manuscript, a complete catalog including information on the area of \htwo\ regions was released by \citet{Groves2023}. We quickly compared that catalog to the \htwo\ regions we identified using \texttt{Astrodendro} and confirmed that it did not significantly affect the results of the Type classification.}
Therefore, we identified \htwo ~regions in the present study.
For this purpose, we applied \texttt{Astrodendro}\footnote{$\langle$ \url{http://www.dendrograms.org/} $\rangle$}
algorithm \citep{Rosolowsky2008} which incorporates dendrogram, an algorithm that analyzes structures using stratified cluster analysis.
\texttt{Astrodendro} defines the distribution of a minimum unit near a peak as “leaf” \citep[see, Fig.1]{Rosolowsky2008}.
The H$\alpha$ emission in galaxies accompanies weak extended components originating from the diffuse ionized gas (DIG) and localized components originating from the individual \htwo ~regions \citep{Haffner2009}.
To exclude the DIG components, we adopted a leaf that extracteds local peaks as individual \htwo ~regions.
\texttt{Astrodendro} uses three parameters \texttt{min\_value}, \texttt{min\_delta} and \texttt{min\_npix} to define leaf \citep{Rosolowsky2021}.
\texttt{min\_value} specifies the lowest value for defining the distribution, and \texttt{min\_npix} specifies the minimum number of pixels.
\texttt{min\_value} and \texttt{min\_npix} are parameters for defining one leaf, on the other hand, \texttt{min\_delta} is for dividing two leaves.
\texttt{min\_delta} is the standard value to define leaf and leaf should be divided if the difference between peaks is larger than \texttt{min\_delta}.
We adopted \texttt{min\_value}$= 2\sigma$ ($\sigma = 70 \times 10^{-20} \ \ergs \ \mathrm{cm^{-2} \ spaxel}$ is the typical noise level), \texttt{min\_delta}$= \sigma$, \texttt{min\_npix}$= 17$ (the number of pixels comparable to the resolution) in the present work.

\begin{figure}[htbp]
    \begin{center}
        \includegraphics[width=0.85\linewidth,clip]{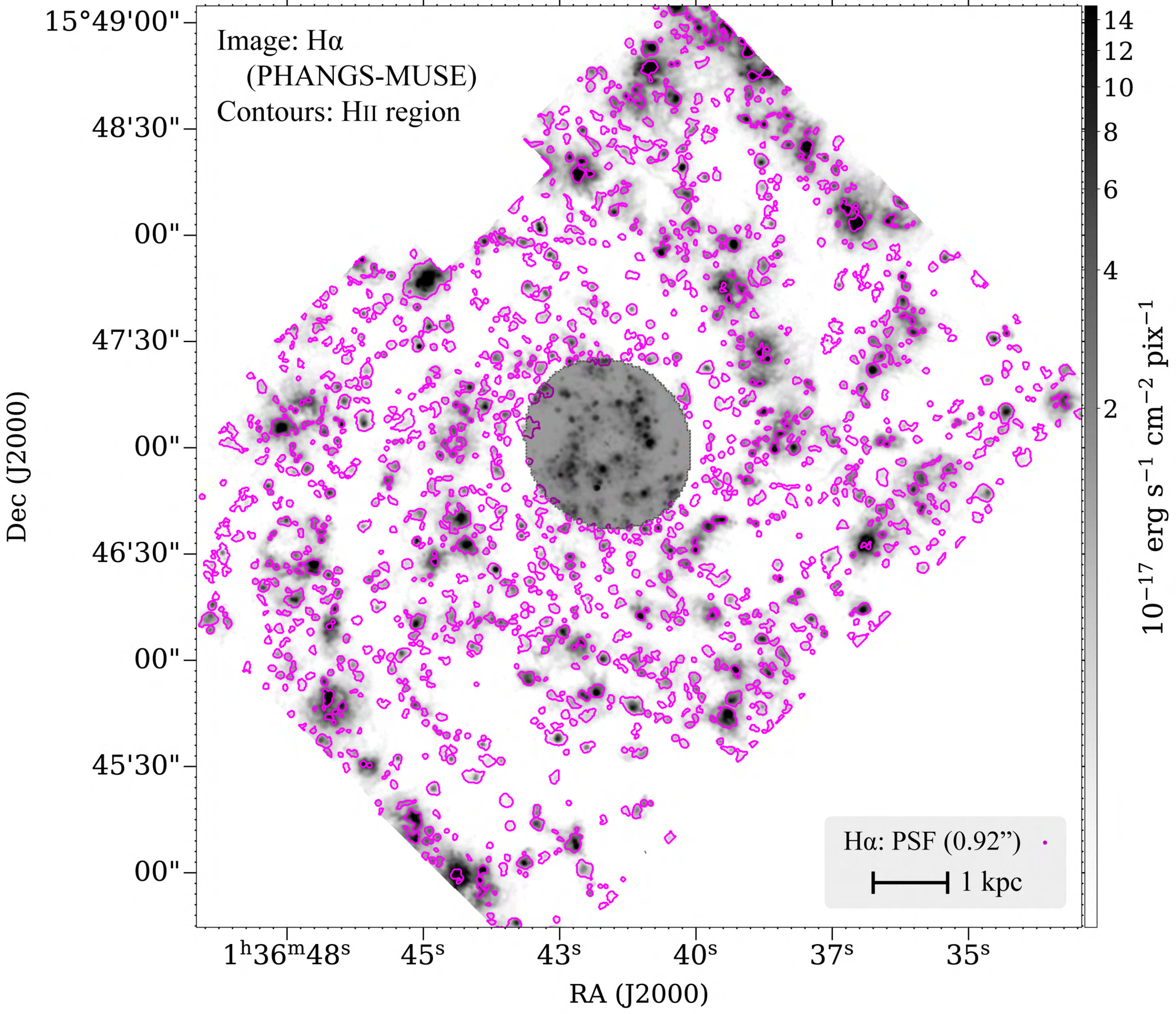}
    \end{center}
    \caption{1351 \htwo ~regions (leaf) identified by \texttt{Astrodendro} are shown by magenta contours on the H$\alpha$ image.}
    \label{HII_Hamap}
\end{figure}

\begin{figure*}[htbp]
    \begin{center}
        \includegraphics[width=0.7\linewidth,clip]{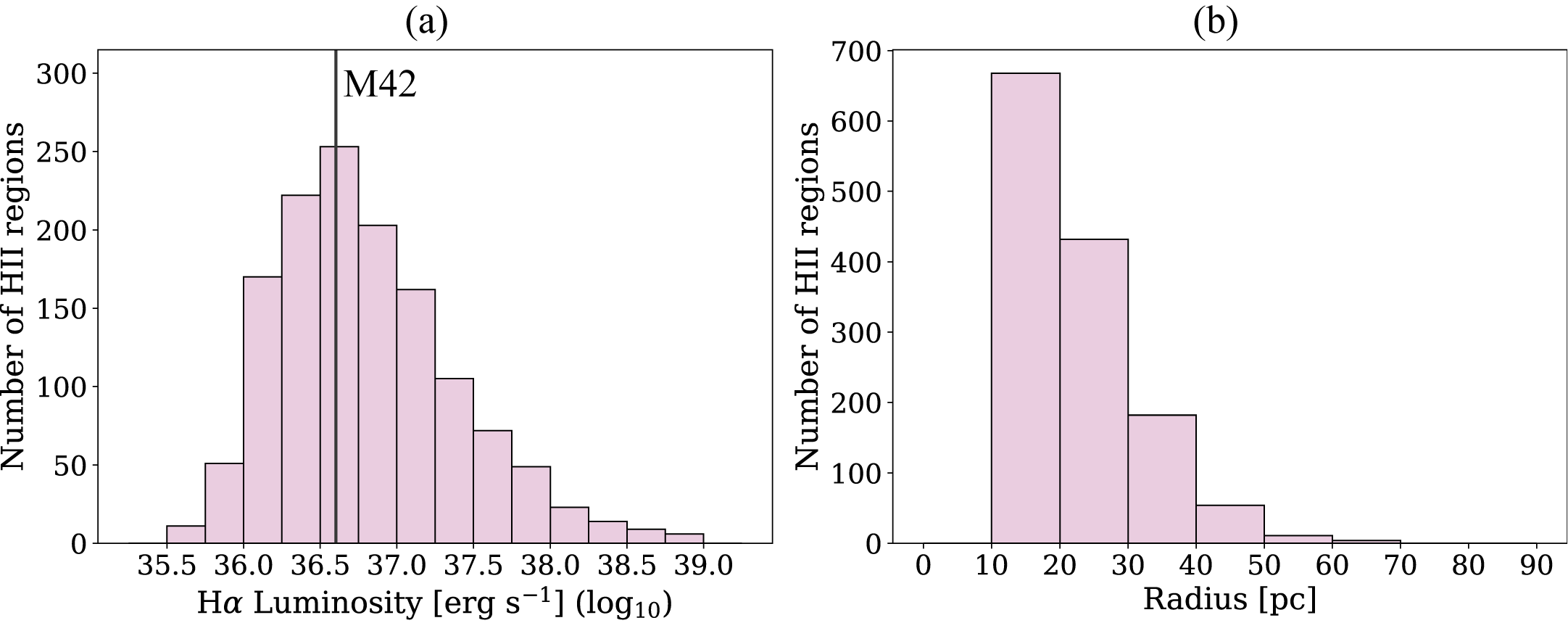}
    \end{center}
    \caption{Distributions of (a) the H$\alpha$ luminosity and (b) radius of the \htwo ~regions. The vertical line in (a) is the luminosity of M42, $4\times 10^{36}~\mathrm{erg~s^{-1}}$ \citep{Gebel1968}.
    The H$\alpha$ luminosity of the \htwo ~regions is in a range of $10^{35.65}$--$10^{39.47}~\ergs$ and their size in a range of 11--70 pc.}
    \label{HII_prop}
\end{figure*}

Figure \ref{HII_Hamap} shows the \LHa\ distribution of the leaf identified in M74 by \texttt{Astrodendro}, where the total number of leaves is 1351.
Figure \ref{HII_prop}(a) shows that \LHa\ of a leaf is in a range $10^{35.65}$ -- $10^{39.47}~\ergs$, and we assume that the \LHa\ of a leaf is equal to \LHa\ of the \htwo\ region.
So, we find that a typical high mass star forming region like M42 with $L_{\mathrm{H\alpha}} = 4 \times 10^{36}~\ergs$ \citep{Gebel1968} are detected reliably.
Figure \ref{HII_prop}(b) shows the distribution of the radii of the \htwo\ regions in the range of 10--70 pc, where the majority (70 \%) are less than 40 pc.
Based on the separation from the GMCs to the \htwo\ regions and between the GMCs (see section \ref{association} and Figures \ref{dist_GMC_from_GMC} and \ref{dist_HII_from_GMC}), we require higher resolution ($<40$ pc) CO data to determine the association between them more accurately.

\subsection{GMC Type classification based on \htwo~regions}
\subsubsection{The boundary value of \LHa\ for Type classification}
In the present study, we use \LHa\ of the \htwo \ regions for the GMC Type classification.
F99 noted that Type II GMC is associated with HII regions with H$\alpha$ luminosity (\LHa) less than $10^{37}~\ergs$  and that Type III GMCs with \LHa\ greater than $10^{37}~\ergs$.
\citet{Yamaguchi2001} confirmed this trend and found that the boundary value of \LHa\ between the two Types is $10^{37.5}~\ergs$.
In the present work, we adopt $\LHa=10^{37.5}~\ergs$ as a boundary value between Type II and Type III.

\subsubsection{Association of a GMC with \htwo~regions}\label{association}
\begin{figure}[htbp]
    \begin{center}
        \includegraphics[width=0.8\linewidth,clip]{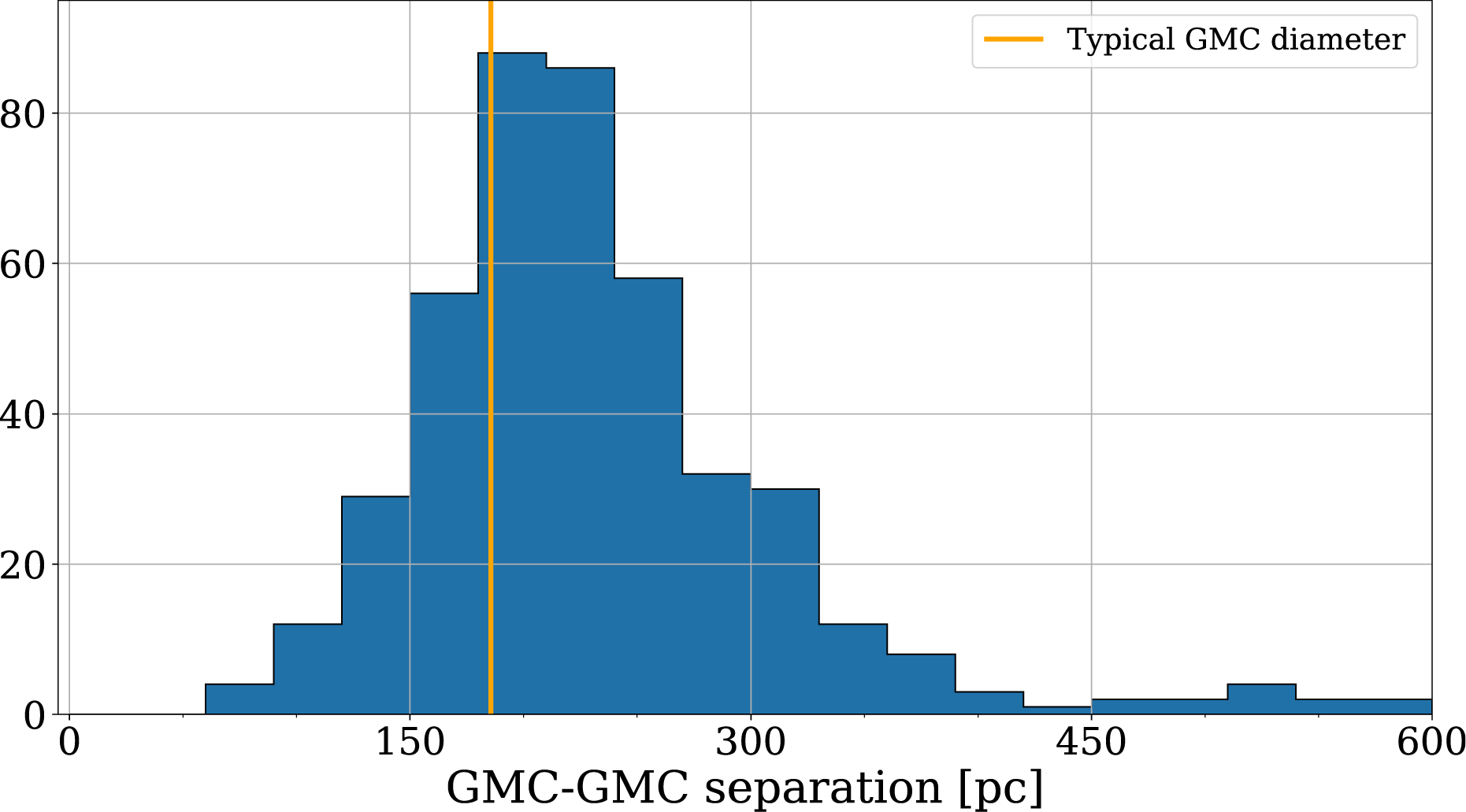}
    \end{center}
    \caption{Histogram of the GMC--GMC separation. The separations are derived from the projected distance between the center of GMCs. The orange line indicates the typical GMC diameter of 186 pc.}
    \label{dist_GMC_from_GMC}
\end{figure}

\begin{figure}[htbp]
    \begin{center}
        \includegraphics[width=0.8\linewidth,clip]{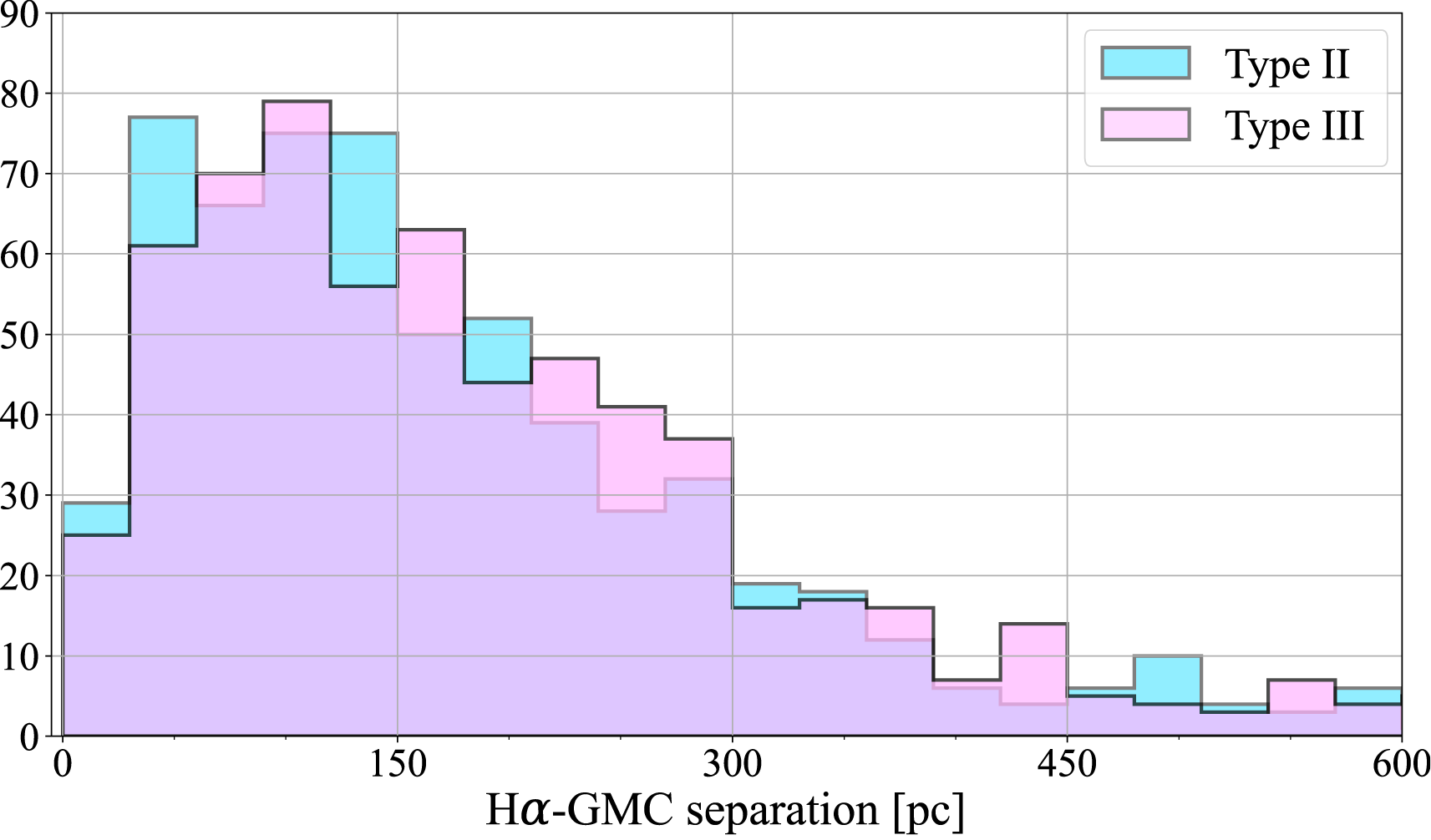}
    \end{center}
    \caption{Histogram of the separation between the nearest GMCs and \htwo\ regions. If the nearest GMC is Type II or Type III, the separation is shown in blue or red, respectively. They show 47\% of the \htwo\ region located within 150 pc from GMC.}
    \label{dist_HII_from_GMC}
\end{figure}

To determine the association of a GMC with an \htwo\ region, we investigate whether the GMC boundary defined by \texttt{PYCPROPS} overlaps with the \htwo\ region boundary (leaf) defined by \texttt{Astrodendro}.
If such an overlap is found, they are determined to be associated with each other.
It is possible that more than two GMCs overlap with the \htwo\ region.
This is because the GMC--GMC separation shown in Figure \ref{dist_GMC_from_GMC} has a peak at 200 pc (with a quartile range from 185 to 265 pc), which corresponds to approximately twice the median radius of a GMC and provides a yardstick for separation from a GMC within which association is probable.
In this case, we measured the separation of the GMCs from the \htwo\ region and identified the nearer GMC to be associated.
Using this method, most of the GMCs were classified, while 5\% of them were not determined association with \htwo \ regions exactly.
These GMCs were classified by visual inspection (see Appendix \ref{classification_apd} for details).
The separation between the GMC and associated \htwo\ regions is shown in Figure \ref{dist_HII_from_GMC} and 47\% of the \htwo\ regions are located within 150 pc of the nearest GMCs.
We identified 373 GMCs associated with 722 \htwo\ regions, with some GMCs are associated with only a few \htwo\ regions.

\subsubsection{GMC Type classification}\label{classification}
\begin{figure*}[htbp]
    \begin{center}
        \includegraphics[width=0.8\linewidth,clip]{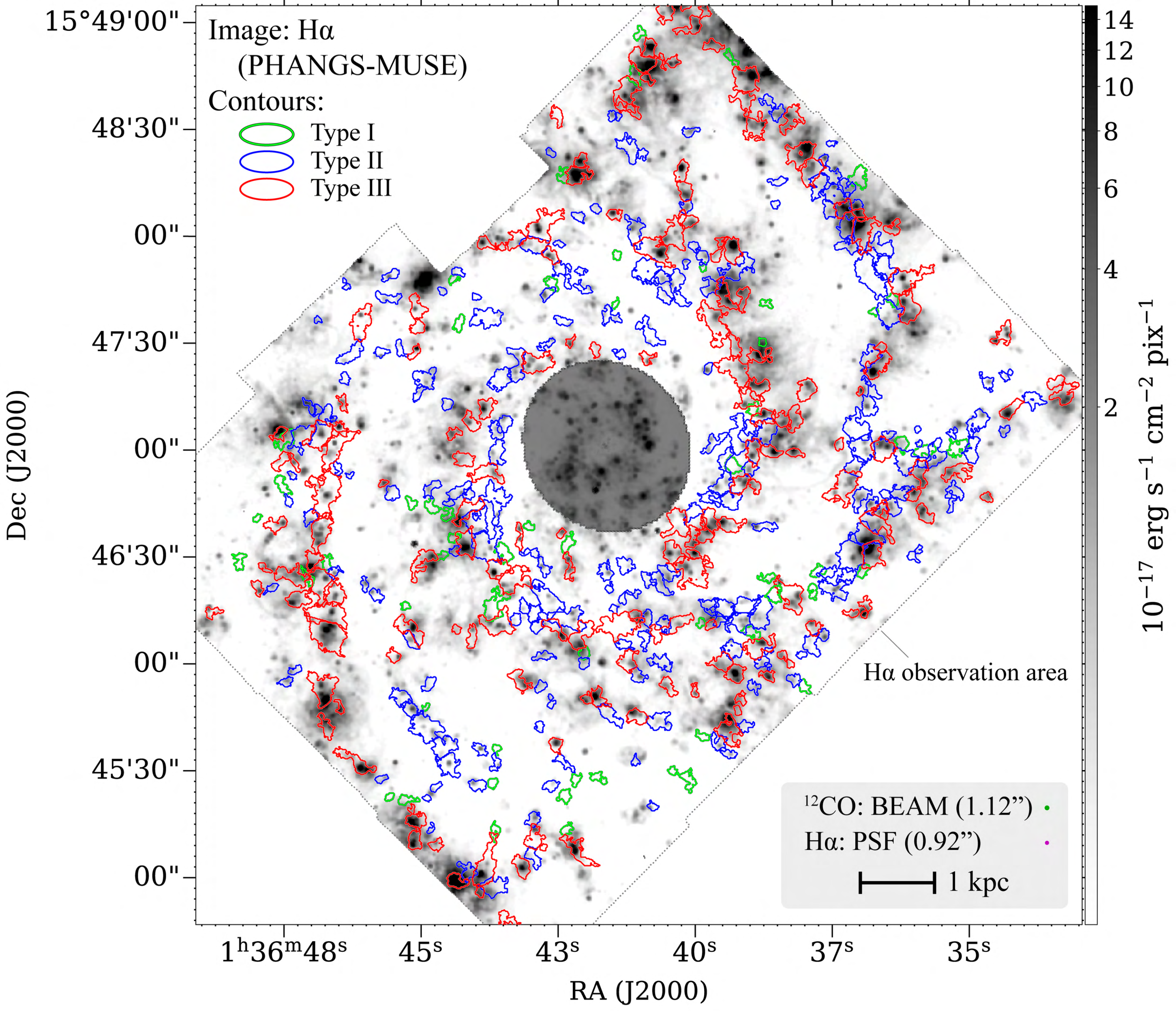}
    \end{center}
    \caption{GMCs classified into three Types I (without \htwo~regions), II (with \htwo~regions with H$\alpha$ luminosity $L_{\mathrm{H\alpha}} < 10^{37.5}~\mathrm{erg~s^{-1}}$), and III (with \htwo~regions with $L_{\mathrm{H\alpha}} > 10^{37.5}~\mathrm{erg~s^{-1}}$) are shown by green, blue and red contours, respectively, on H$\alpha$ image. The gray dotted lines indicate the observed area in H$\alpha$ and the central gray circle indicates the masked region for the galactic center region. We obtained 59 Type I, 201 Type II, and 172 Type III.}
    \label{TypeGMC_HaMap}
\end{figure*}

Following the results of the associated \htwo ~regions with a GMC, we are able to calculate $L_{\rm{H\alpha}}$ in each GMC and classify GMCs into three Types, Type I, Type II, and Type III according to the value of $L_{\rm{H\alpha}}$.
We identified 59 Type I GMCs, 201 Type II GMCs, and 172 Type III GMCs.
Figure \ref{TypeGMC_HaMap} shows the distribution of the GMCs of each Type in three colors, illustrating that their distribution, especially that of Type III, follows the spiral arms.

\begin{figure*}[htbp]
    \begin{center}
        \includegraphics[width=0.7\linewidth,clip]{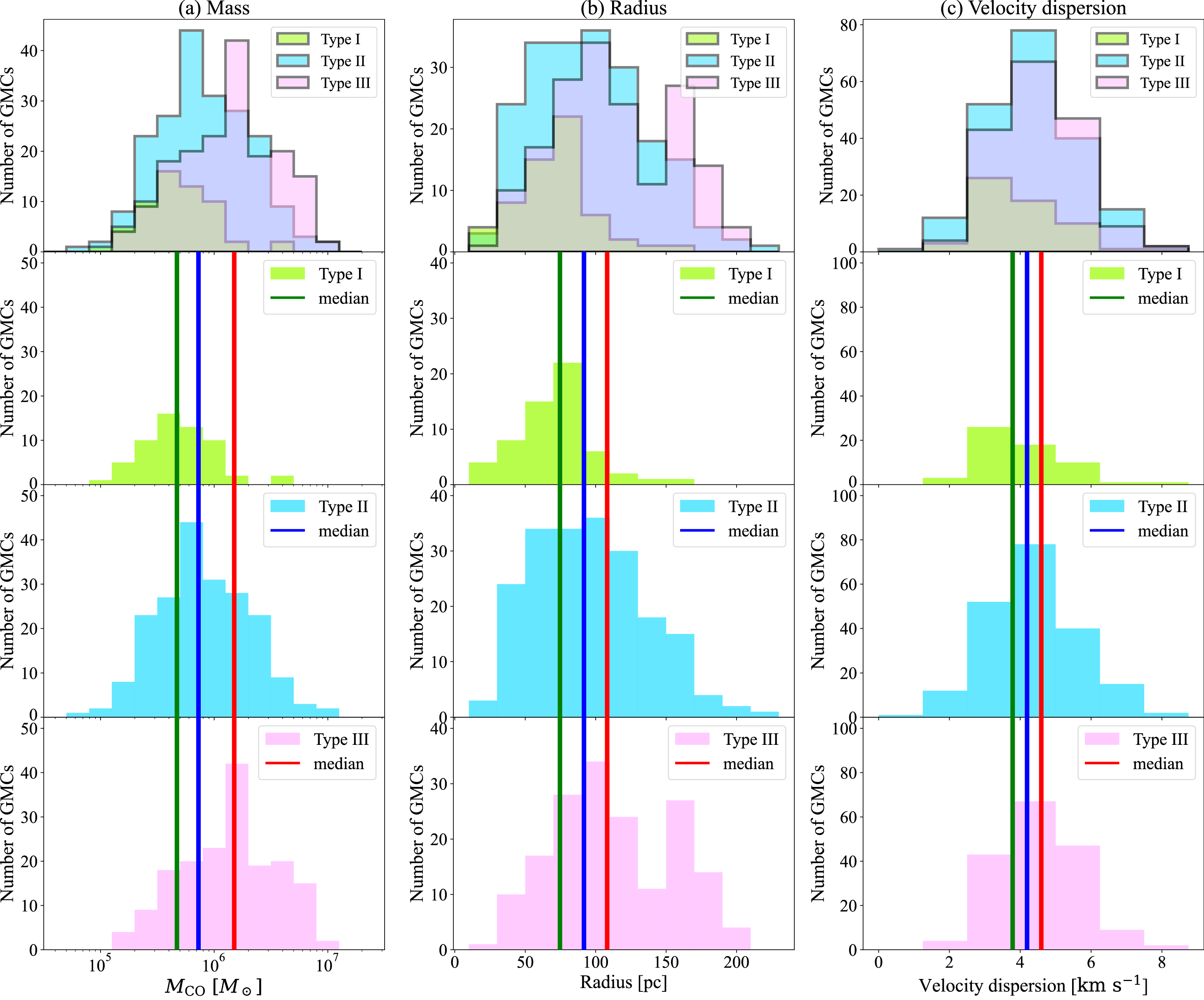}
    \end{center}
    \caption{Histogram of (a) mass, (b) radius, and (c) velocity dispersion. Results of these physical properties for all the GMC, Type I, II, and III, are shown from top to bottom. In the top panel, the histograms for the three Types of GMCs are stacked. The light green, light blue, and pink vertical lines represent the median values of Type I, II, and III, respectively.}
    \label{GMC_prop}
\end{figure*}

Figure \ref{GMC_prop} and Table \ref{GMC_prop_table} show the three physical parameters, $M_{\mathrm{CO}}$, $R$ and $\sigma_v$, of GMCs.
For each Type, the median masses are $4.8\times10^5,M_\odot$, $7.3\times10^5,M_\odot$, and $1.5\times10^6,M_\odot$, respectively, and the median radii are 74 pc, 92 pc, and 108 pc, respectively.
These two parameters increase with Type, but their increments are small compared with their dispersions.
Similar trends are observed for the LMC and M33.
The median velocity dispersions are $3.81$ km s$^{-1}$, $4.19$ km s$^{-1}$, and $4.59$ km s$^{-1}$, respectively, indicating that it shows no clear systematic trend.

\begin{table*}[htbp]{
        \centering
        \tbl{Physical properties of the GMCs.}{
            \begin{tabularx}{0.65\linewidth}{ccccccc}
                \hline
                    &                     &         &                 & Type I & Type II & Type III \\
                \hline
                    &                     & Median  & ($10^5M_\odot$) & 4.75   & 7.27    & 15.1     \\
                (1) & Mass                & Average & ($10^5M_\odot$) & 6.38   & 12.6    & 20.6     \\
                    &                     & Std.    & ($10^5M_\odot$) & 6.12   & 14.2    & 19.1     \\
                \hline
                    &                     & Median  & (pc)            & 74     & 92      & 108      \\
                (2) & Radius              & Average & (pc)            & 71     & 96      & 114      \\
                    &                     & Std.    & (pc)            & 27     & 40      & 44       \\
                \hline
                    &                     & Median  & (\kms)          & 3.81   & 4.19    & 4.59     \\
                (3) & Velocity dispersion & Average & (\kms)          & 4.03   & 4.36    & 4.57     \\
                    &                     & Std.    & (\kms)          & 1.17   & 1.29    & 1.13     \\
                \hline
            \end{tabularx}}
        \begin{tabnote}
            Lines: (1) Luminous mass ($M_{\mathrm{CO}}$). (2) Radius of GMCs (3) Velocity dispersion of GMCs.
        \end{tabnote}
        \label{GMC_prop_table}
    }
\end{table*}

\subsection{Association with clusters}
\label{sec_association}
We use the clusters cataloged by \citet{Adamo2017} to identify clusters associated with GMCs.
The ages of clusters are determined by their spectral properties and can provide a “clock” for estimating the timescales of cloud evolution.
\citet{Adamo2017} listed clusters with age every 1 Myr for 1--10 Myr and every 10 Myr for 10--100 Myr in an area shown in Figure \ref{cluster_map}.
This area contains 47 Type I GMCs, 169 Type II GMCs, 127 Type GMCs, and 343 in total as shown in Table \ref{GMC_table}.

\begin{figure*}[htbp]
    \begin{center}
        \includegraphics[width=0.7\linewidth,clip]{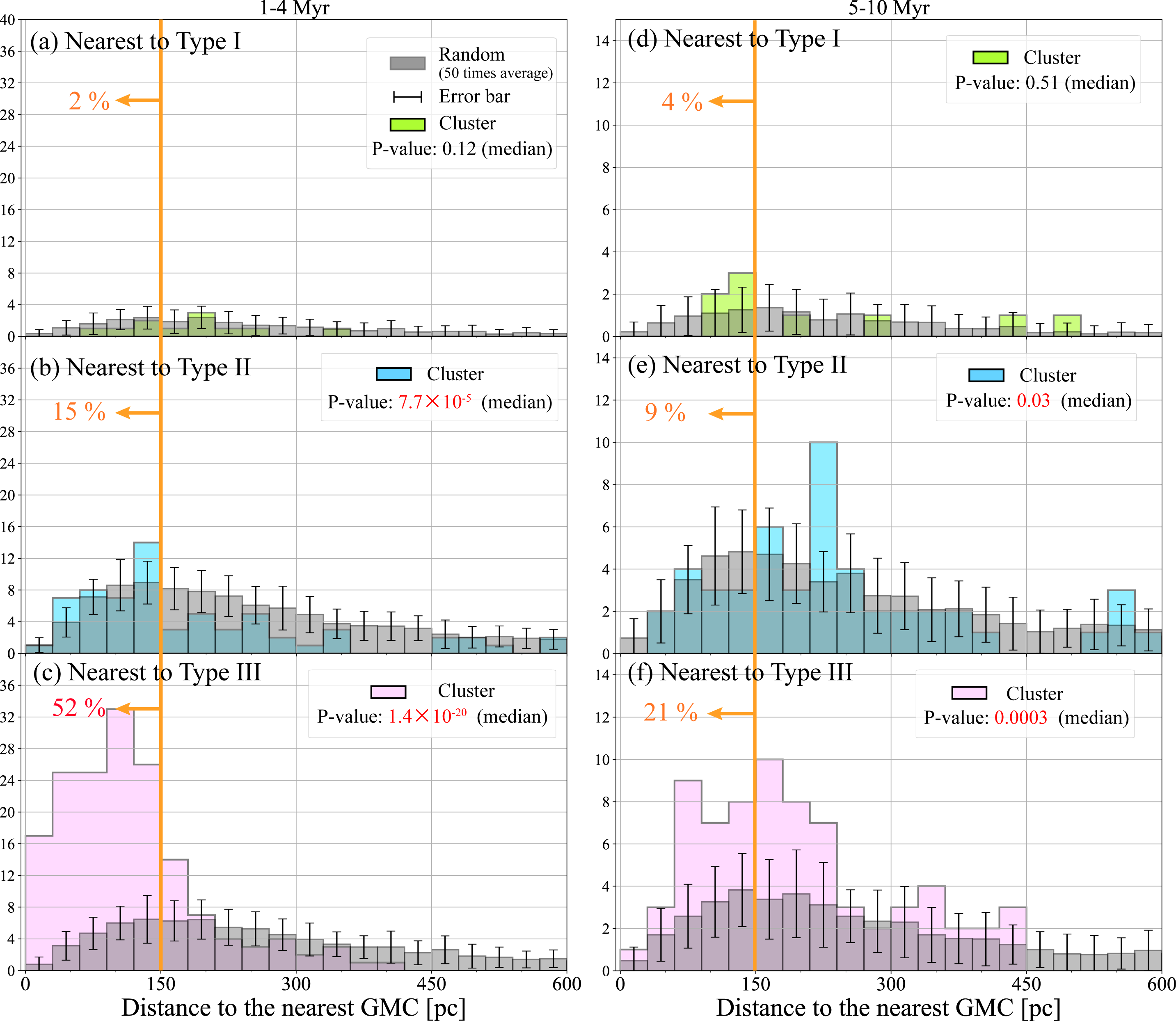}
    \end{center}
    \caption{Histograms of the separation between clusters and the nearest GMCs.
        (a)--(c) for 1--4 Myr clusters and GMCs of Type I, II, and III, respectively, and (d)--(f) for 5--10 Myr clusters and GMCs of Type I, II, and III, respectively.
        If the nearest GMC is Type I, II, and III, the separation is shown by filled areas in light green, light blue, and pink, respectively.
        The vertical lines in orange indicate a distance of 150 pc from a GMC and the digit shows the fraction of clusters located within 150 pc of a GMC.
        The histograms compare the random distribution of clusters of the same number over the analyzed region 50 times.
        The average number and standard deviation of random distributions of each bin are shown by the shaded area and error bar in each panel.
        We used the K--S test to compare them by each GMC Type; the median P-values are shown in each panel.}
    \label{dist_from_GMC}
\end{figure*}

We examined the clusters of all the age ranges every 1 Myr for association with GMCs, and, consequently, found that only the clusters with age $\leq$ 4 Myr (“1--4 Myr cluster” etc. hereafter) and Type II and Type III GMCs show significant association.
To illustrate the analysis, we present only the cases of 1--4 Myr and 5--10 Myr as summarized in Figure \ref{dist_from_GMC} and Table \ref{cluster_table}.
The total number of 1--4 Myr clusters shown in Figure \ref{dist_from_GMC} is 243 and the number of clusters nearest to each GMC Type is as follows: 11 for Type I, 67 for Type II, and 165 for Type III.
We then measured the separation of the central position of the GMC from the cluster, as shown in Figure \ref{dist_from_GMC}.
The three panels in Figure \ref{dist_from_GMC}(a)--(c) show the distribution of the separation of 1--4 Myr clusters from the nearest GMCs by filled areas.
This indicates that Type II and Type III GMCs are distributed mostly within 150--180 pc of the clusters, whereas Type I GMCs exhibit no such trend.
Especially 52\% of the 1--4 Myr clusters are located within 150 pc of Type III GMCs.
It is important to test whether the distribution is significantly different from a random distribution.
We tested this by generating random distributions of clusters of the same number over the analyzed region using the Python \texttt{random} module and constructed the distribution of the separation from the GMCs, as shown by the shaded areas in Figure \ref{dist_from_GMC}(a)--(c).
We generated a random distribution 50 times and verified the results.
We used the K--S test (Kolmogorov--Smirnov test) to compare the two distributions by calculating the P-value  (by SciPy function \texttt{ks\_2samp}, \cite{Virtanen2020}).
If the P-value is less than 0.05, we determined that the two distributions were not generated from the same probability distribution.
The median P-value is 0.12 for Type I GMC, $7.7\times10^{-5}$ for Type II GMC, and $1.4 \times 10^{-20}$ for Type III GMC as listed in Table \ref{cluster_table} and Figure \ref{dist_from_GMC}.
Therefore, we considered that the 1--4 Myr clusters showed a significant association with Type II and Type III GMCs.
For the 5--10 Myr clusters we did the same as shown in the three panels of Figure \ref{dist_from_GMC}(d)--(f), and confirmed that they showed no significant association with GMCs. 
The 5--10 Myr clusters near Type III have a P-value less than 0.05, whereas only 21\% of them show concentration within 150 pc of a GMC.
The fraction is less than a half of that of the 1--4 Myr clusters, and we cannot consider that they are not significantly associated with Type III.
Similarly way, \citet{Turner2022} showed that in 11 galaxies, 4--6 Myr clusters were associated with gas clouds.
The distribution of clusters no longer exhibited a pattern distinguishable from random distribution as they aged.

There are 141 Type II GMCs and 109 Type III GMCs in the observed area of HST and JWST.
For these GMCs, we compared the three color-synthesized images of HST (275 nm, 415 nm, and 814 nm) and the near-infrared JWST image (3.6 $\mathrm{\mu m}$) and inspected by eye if the point-like HST sources that were not identified as clusters are located toward H$\alpha$ sources.
As a result, we found that 106 Type II GMCs (75\%) and all Type III GMCs are associated with both HST point sources, which include star clusters identified by \citet{Adamo2017}, and the JWST near-infrared sources.
The remaining 13 Type II GMCs are associated with only JWST near-infrared sources, 2 Type II GMCs with only HST sources, and 1 with no associated sources.
The HST sources that were not identified as clusters may be too small clusters including high-mass stars, which were identified only as an HST point source.
It means these GMCs with \htwo \ regions may have high-mass stars.

\begin{table*}[htbp]{
        \centering
        \tbl{Number of GMCs used for the present comparison.}{
            \begin{tabularx}{0.8\linewidth}{cXcccc}
                \hline
                    &                                                           & Type I & Type II & Type III & All       \\
                \hline
                (1) & All GMC (removed central region)                          &        &         &          & 580 (535) \\
                (2) & Comparison with \htwo \ regions                           & 59     & 201     & 172      & 432       \\
                (3) & Comparison with clusters                                  & 47     & 169     & 127      & 343       \\
                (4) & Comparison with $21 \ \mathrm{\mu m} $ leaves             & 45     & 154     & 127      & 326       \\
                (5) & Comparison with $21 \ \mathrm{\mu m}$ leaves and clusters & 40     & 141     & 109      & 290       \\
                \hline
            \end{tabularx}
        }
        \begin{tabnote}
            Lines: (1) All GMCs (central region removed).
            (2) GMCs used for comparison with \htwo\ regions.
            (3) GMCs used for comparison with clusters.
            (4) GMCs used for comparison with 21 $\mathrm{\mu m}$ leaves.
            (5) GMCs used for comparison with 21 $\mathrm{\mu m}$ leaves and clusters.
        \end{tabnote}
        \label{GMC_table}
    }
\end{table*}

\begin{table*}[htbp]{
        \centering
        \tbl{Association of clusters and GMCs.}{
            \begin{tabularx}{0.85\linewidth}{cclccc}
                \hline
                    &                   &                                                                & Type I       & Type II                    & Type III                     \\
                \hline
                (1) & 1--4 Myr cluster  & $N_\mathrm{cluster}^*$                                         & 11           & 67                         & 165                          \\
                    &                   & $N_\mathrm{<150pc}^\dagger$ (fraction)                         & 4 (2\%)      & 37 (15\%)                  & 126 (52\%)                   \\
                    &                   & \begin{tabular}{l} Median of P-value $^\ddagger$ \end{tabular} & 0.12         & $7.7\times 10^{-5}$        & $1.4\times 10^{-20}$         \\
                    &                   & \begin{tabular}{l} (IQRs) \end{tabular}                        & (0.05--0.26) & ($10^{-4.7}$--$10^{-3.5}$) & ($10^{-15.7}$--$10^{-12.4}$) \\
                \hline
                (2) & 5--10 Myr cluster & $N_\mathrm{cluster}^*$                                         & 9            & 50                         & 72                           \\
                    &                   & $N_\mathrm{<150pc}^\dagger$ (fraction)                         & 5 (4\%)      & 12 (9\%)                   & 28 (21\%)                    \\
                    &                   & \begin{tabular}{l} Median of P-value $^\ddagger$ \end{tabular} & 0.51         & 0.03                       & $10^{-2.5}$                  \\
                    &                   & \begin{tabular}{l} (IQRs) \end{tabular}                        & (0.22--0.69) & (0.01--0.07)               & ($10^{-4.2}$--$10^{-2.9}$)   \\
                \hline
            \end{tabularx}}
        \begin{tabnote}
            Lines: (1) Number of 1--4 Myr clusters. (2) Number of 5--10 Myr clusters.
            $*$ The number of clusters nearest to GMC for each Type. For example, if the nearest GMC is Type I, the number is counted in column ``Type I."
            $\dagger$ The number of clusters within 150 pc from the nearest GMC.
            $\ddagger$ P-values derived by K--S test of comparison between cluster and the random distribution shown in each panel in Figure \ref{dist_from_GMC}.
            We generated random distribution 50 times and derived P-values.
            If the median P-value $<$ 0.05, the difference between the two distributions is significant in the sense that the two distributions are not drawn from the same probability distribution.
        \end{tabnote}
        \label{cluster_table}}
\end{table*}

\subsection{Virial ratio variation}
\begin{figure*}[htbp]
    \begin{center}
        \includegraphics[width=0.8\linewidth,clip]{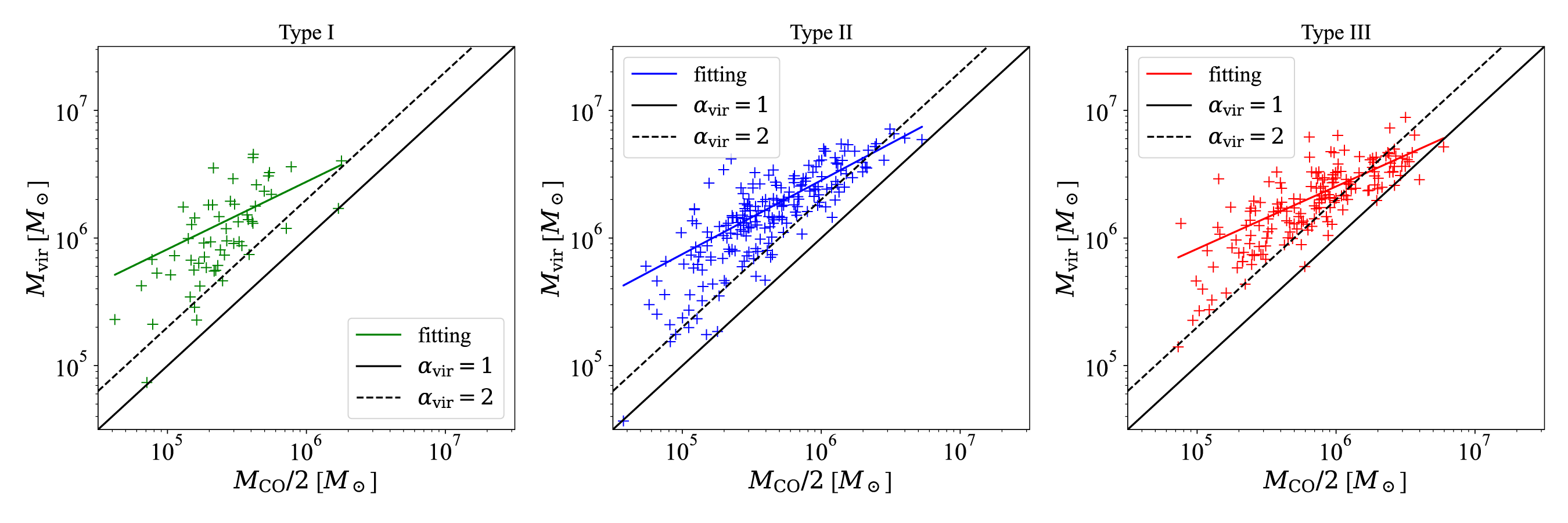}
    \end{center}
    \caption{Scatter plots between ${M_\mathrm{CO}/2}$ and $M_{\mathrm{vir}}$ of (a) Type I, (b) Type II, and (c) Type III.
        The black solid line indicates $\alpha_\mathrm{vir} = M_\mathrm{vir}/(M_\mathrm{CO}/2)=1$ line and the black dashed line indicates $\alpha_{\mathrm{vir}}=2$ line.
        The green, blue, and red lines represent the regression lines obtained by the least squares fitting as follows;
        Type I: $\log M_\mathrm{vir} = 0.53 \log(M_\mathrm{CO}/2) + 3.3$,
        Type II: $\log M_\mathrm{vir} = 0.59 \log(M_\mathrm{CO}/2) + 3.0$,
        Type III: $\log M_\mathrm{vir} = 0.49 \log(M_\mathrm{CO}/2) + 3.5$.}
    \label{Mco_Mvir}
\end{figure*}

\begin{figure}[htbp]
    \begin{center}
        \includegraphics[width=1.0\linewidth,clip]{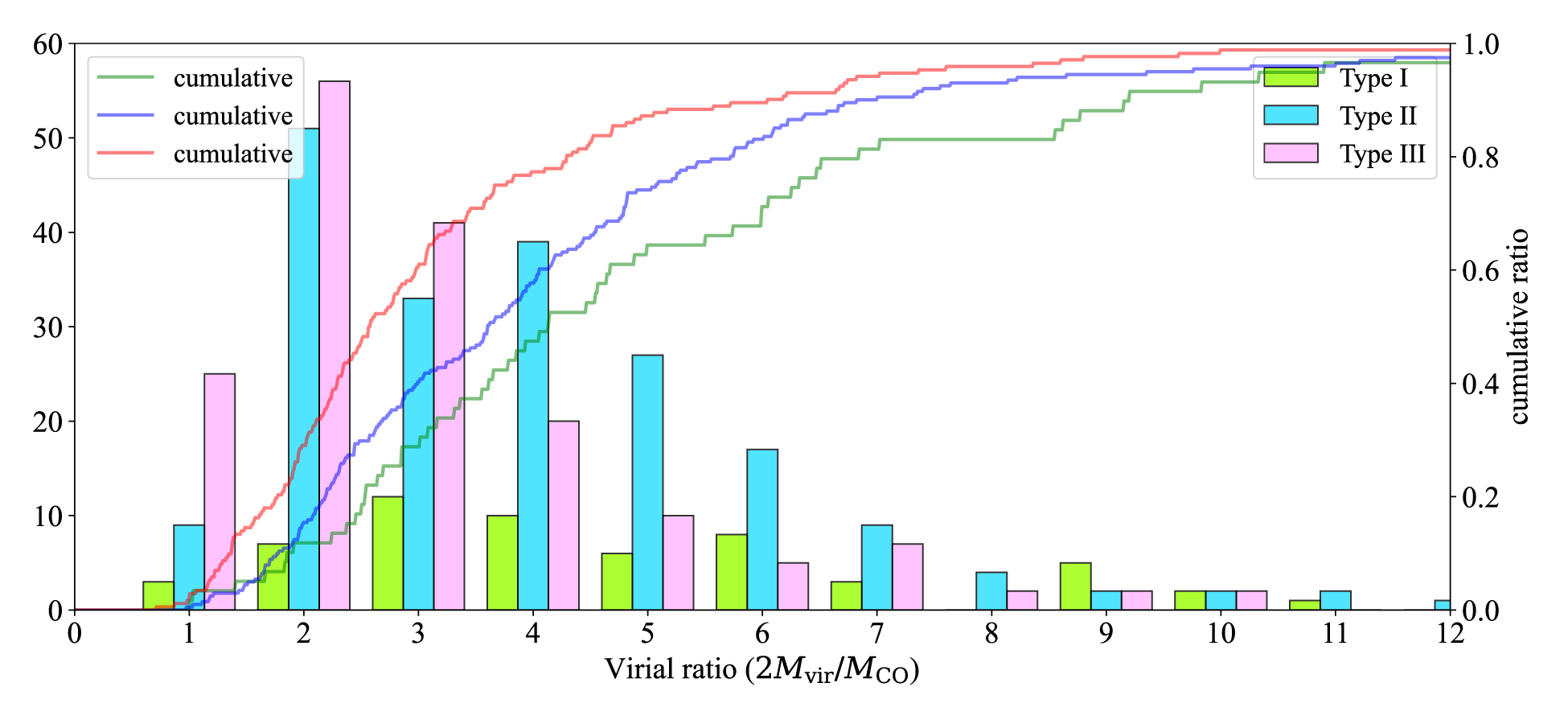}
    \end{center}
    \caption{Histogram of the virial ratio $\alpha_\mathrm{vir} = 2M_\mathrm{vir}/M_\mathrm{CO}$. The light green, light blue, and pink-filled areas show the virial ratios of Type I, II, and III, respectively. The green, blue, and red line plots show the cumulative distributions of Type I, II, and III, respectively.}
    \label{alpha_vir}
\end{figure}

Figures \ref{Mco_Mvir} shows the scatter plots of the virial mass $M_\mathrm{vir}$ vs. half of the luminous mass $M_\mathrm{CO}/2$ for each GMC Type.
\texttt{PYCPROPS} uses a tow-dimensional Gaussian cloud model and estimates radius and velocity dispersion.
This means $M_\mathrm{vir}$ estimated within the FWHM is compared with half $M_\mathrm{CO}$, and we recognize that $M_\mathrm{vir}$ decreases with Type for  a given $M_\mathrm{CO}$.
The trend is most significant for Type III at higher $M_\mathrm{CO}$ or larger $M_\mathrm{vir}$ (Figure \ref{Mco_Mvir}(c)), as is consistent with $M_\mathrm{vir}$ smallest at large $M_\mathrm{CO}/2$ above $10^5~M_\odot$.
Figure \ref{alpha_vir} shows the histograms of GMCs as functions of the virial ratio $\alpha_\mathrm{vir} = 2 M_\mathrm{vir}/M_\mathrm{CO}$, which indicates that the fraction of the GMCs with lower $\alpha_\mathrm{vir}$ increases from Type I to Type III.
Figures \ref{Mco_Mvir} and \ref{alpha_vir} indicate that the gravitational relaxation increases from Type I to Type III.
Therefore, Type III GMCs are in the most gravitationally relaxed state with the smallest virial ratio peaked at $\alpha_\mathrm{vir} \sim 1$, compared with Type I and Type II.

\subsection{Comparison with the JWST $21 \ \mathrm{\mu m}$ emission}
\begin{figure}[htbp]
    \begin{center}
        \includegraphics[width=0.85\linewidth,clip]{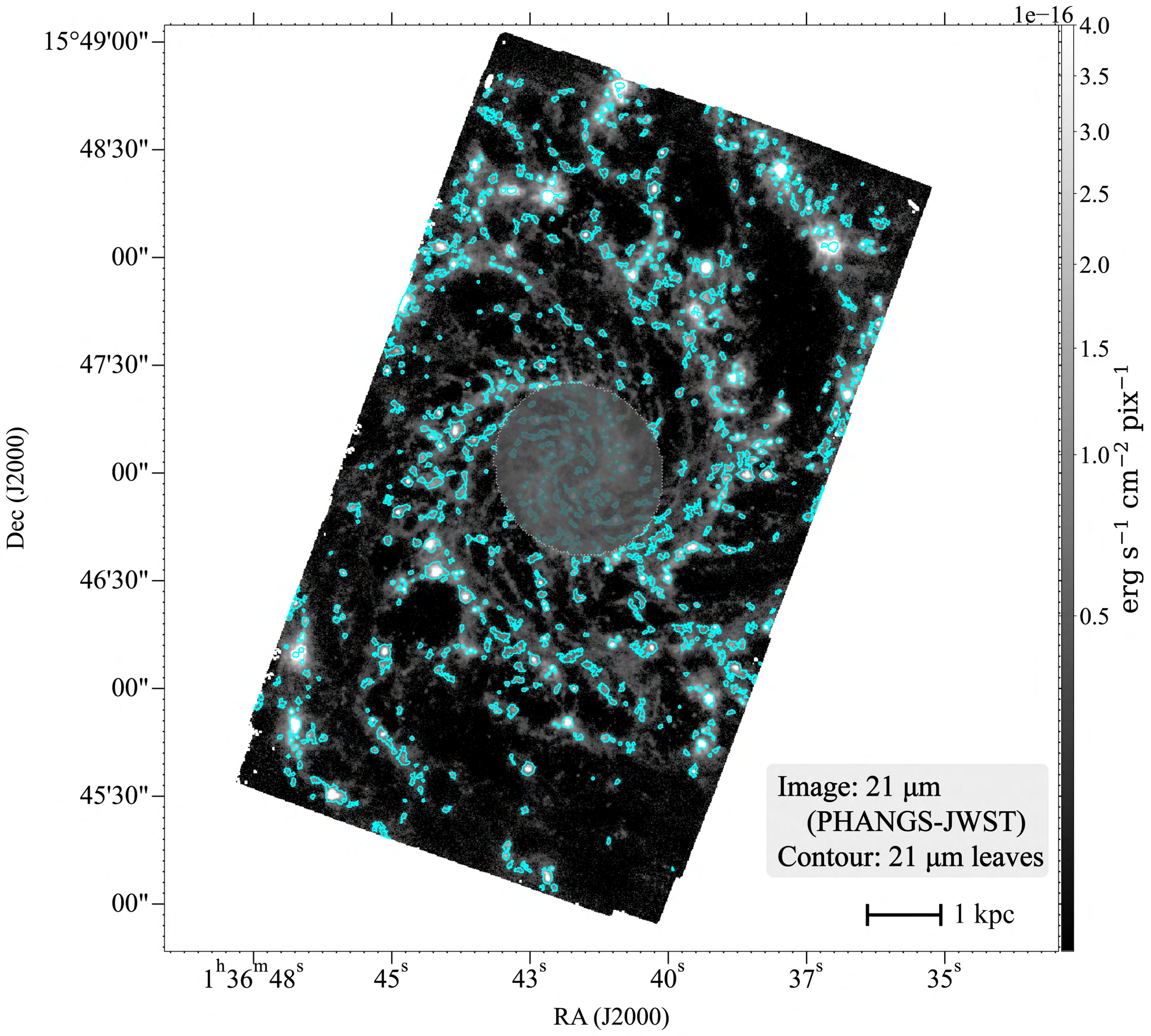}
    \end{center}
    \caption{692 leaves identified by \texttt{Astrodendro} are shown by cyan contours on the 21 $\mathrm{\mu m}$ image.}
    \label{21um_map}
\end{figure}

The JWST was used to observe M74 at the resolution of $0\farcs67$ at infrared wavelengths in the range of 18.48 to 23.16 $\mathrm{\mu m}$ using F2100W.
The 21 $\mathrm{\mu m}$ emission is generally considered as the dust emission heated by young stellar clusters \citep{Calzetti2007,Kennicutt2007}, and we expect that a comparison of the 21 $\mathrm{\mu m}$ data with the present results will provide a new perspective to the present study.
We applied the \texttt{Astrdendro} algorithm to the JWST data with the parameters as follows: \texttt{min\_value} $= 5\sigma$ ($\sigma = 0.25 \ \mathrm{MJy \ sr^{-1}}$ is the typical noise level), \texttt{min\_delta} $= 3\sigma$, \texttt{min\_npix} $= 10$ (the number of pixels comparable to the resolution), and obtained 692 leaves, similar to the other data including the H$\alpha$.
Figure \ref{21um_map} shows the results of the 21 $\mathrm{\mu m}$ leaves, where the observed area is relatively smaller than in H$\alpha$.
In addition, it can be seen in Figure \ref{21um_map} that the spiral arms are traced-well by the 21 $\mathrm{\mu m}$ leaves in contours.

\begin{figure*}[htbp]
    \begin{center}
        \includegraphics[width=0.65\linewidth,clip]{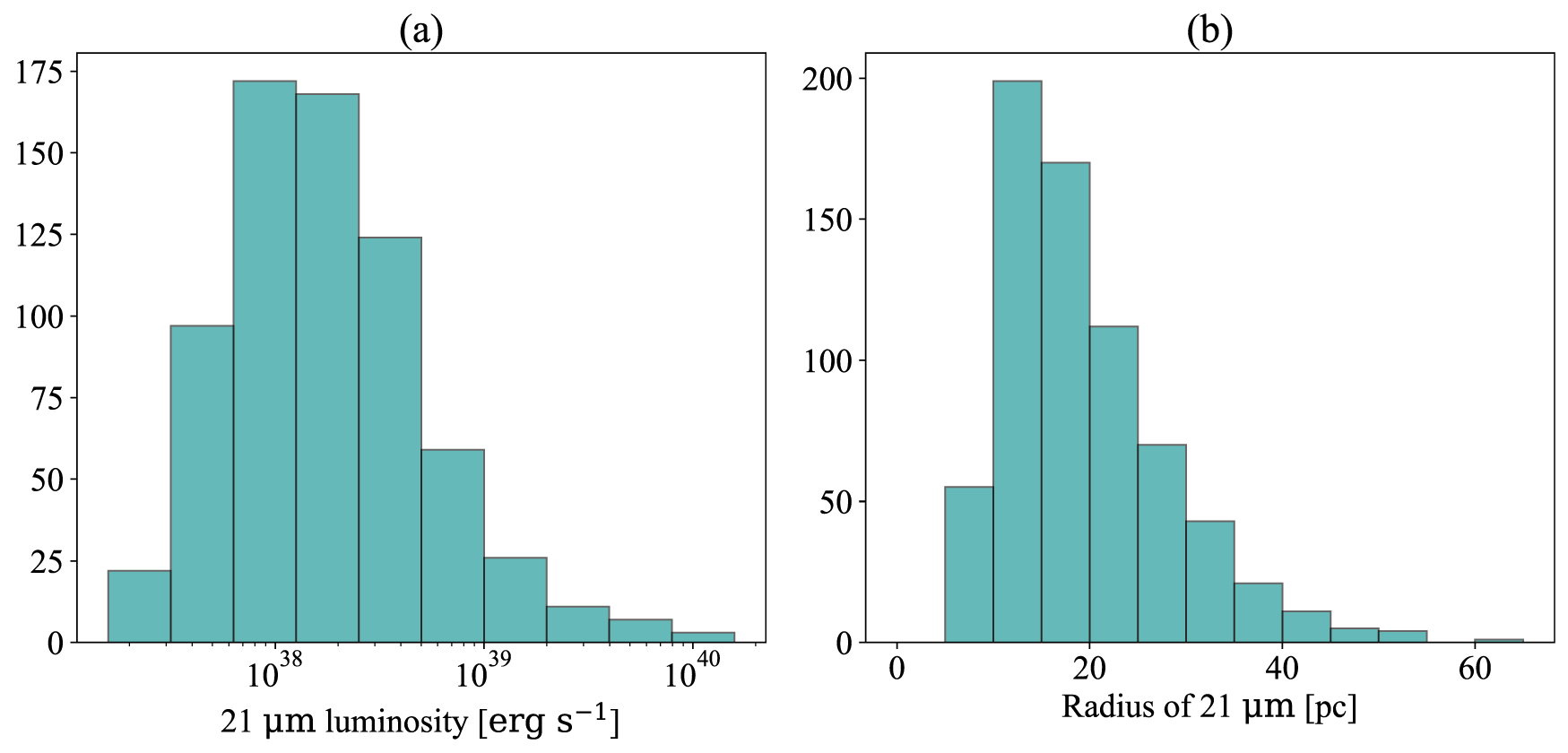}
    \end{center}
    \caption{Distributions of (a) the 21 $\mathrm{\mu m}$ luminosity and (b) radius of the 21 $\mathrm{\mu m}$ leaves.}
    \label{21um_prop}
\end{figure*}

\begin{figure}[htbp]
    \begin{center}
        \includegraphics[width=0.8\linewidth,clip]{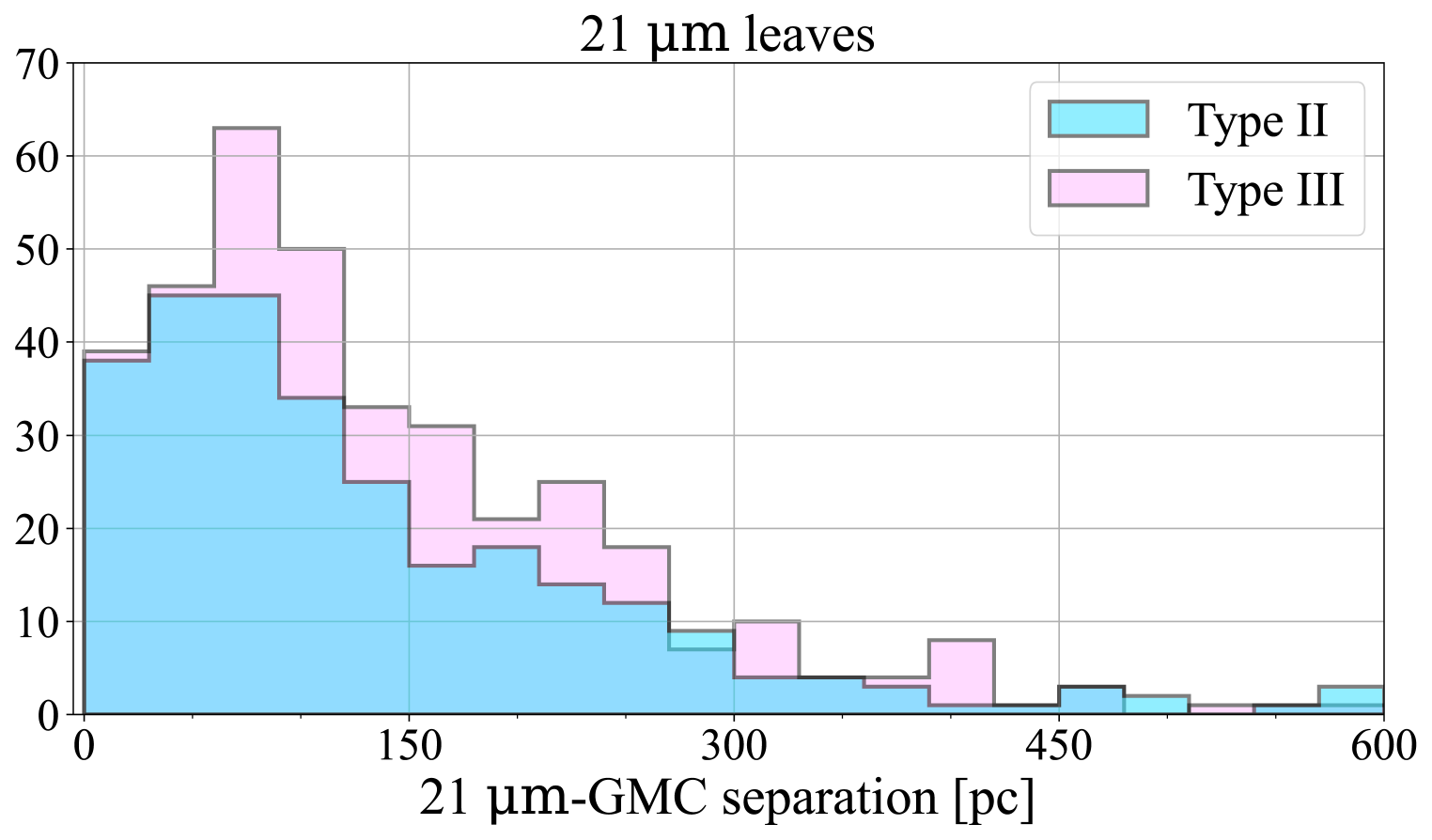}
    \end{center}
    \caption{Histogram of the separation between the nearest GMCs and 21 $\mathrm{\mu m}$ leaves.
        The separation is shown in blue or red when the nearest GMC is Type II or Type III, respectively.}
    \label{dist_21um_from_GMC}
\end{figure}

Figures \ref{21um_prop} (a) and (b) show histograms of the luminosity and the radius of the 21 $\mathrm{\mu m}$ leaves.
The histograms are similar to those of H$\alpha$ (Figure \ref{HII_prop}).
The 21 $\mathrm{\mu m}$ luminosity is larger than the H$\alpha$ luminosity by two orders of magnitude, whereas the radius is similar to that of H$\alpha$ leaves.
Figure \ref{dist_21um_from_GMC} shows the histograms of the separation between the 21 $\mathrm{\mu m}$ leaves and Type II/Type III GMCs and indicates that the 21 $\mathrm{\mu m}$ leaves are well associated with the star-forming GMCs.
The results show that 64\% of the 21 $\mathrm{\mu m}$ leaves are located within 150 pc from a GMC; as a result, the 21 $\mathrm{\mu m}$ leaves are correlated with star formation in a similar, but not identical, manner with H$\alpha$.
We discuss more details of the 21 $\mathrm{\mu m}$ leaves in section \ref{dis_21um}.

\section{Effect of extinction in the H$\alpha$ leaves}
\label{extinction}
The H$\alpha$ emission can be affected by extinction and may be regarded with caution as a tracer of the \htwo \ regions.
However, previous studies that investigated the star formation rate in galaxies at z=1--3 (e.g., \cite{Glazebrook1999,Dominguez2013}) have reported that the extinction in H$\alpha$ is moderate with $A _\mathrm{v}=$0.5--1 mag and has no significant effect on the H$\alpha$ measurements.
To independently confirm the effect of the extinction in the H$\alpha$ leaves of M74, we compared the \LHa \ emission between the pixels with and without CO emission.
This approach is possible because the resolution of 50 pc is sufficiently high to resolve the two regions in the present GMCs.
Assuming that half of the H$\alpha$ leaves lie in front of the GMCs, when the extinction is significant, we expect the pixels with CO to have lower \LHa \ than those without CO owing to extinction.

\begin{figure*}[htbp]
    \begin{center}
        \includegraphics[width=0.6\linewidth,clip]{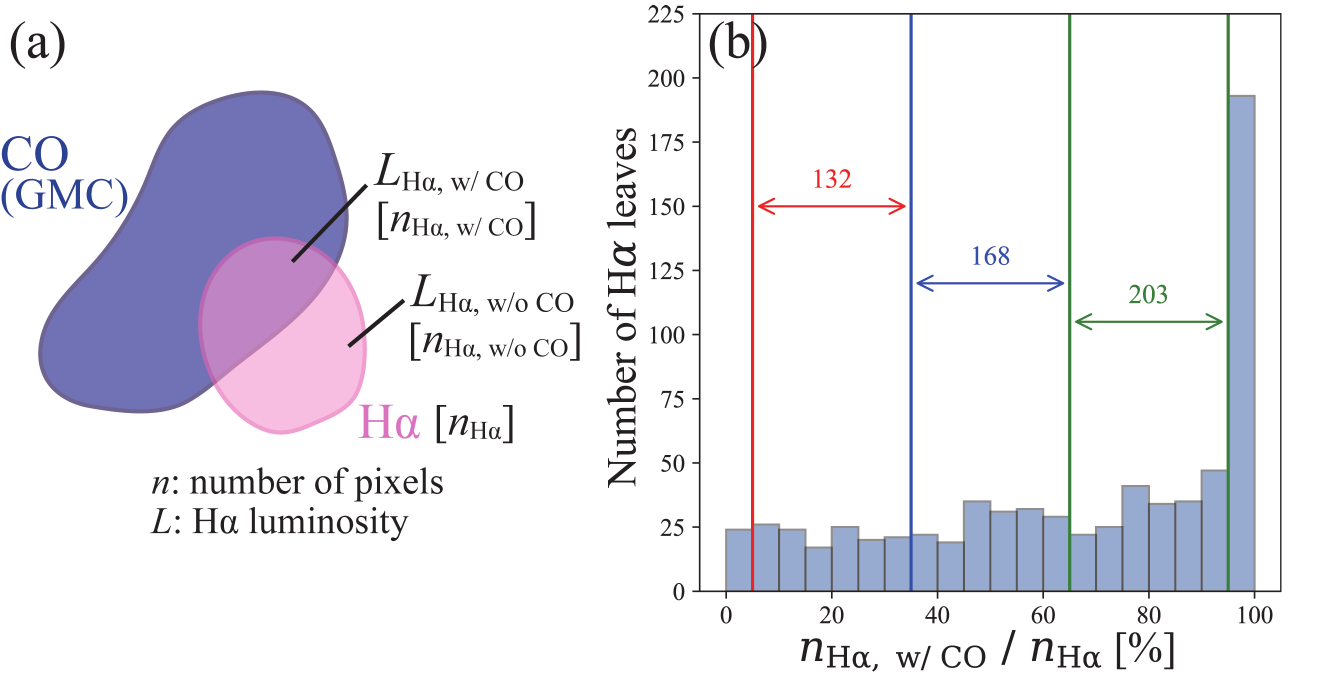}
    \end{center}
    \caption{(a) A schematic demonstrating the relationship between CO and H$\alpha$.
        The H$\alpha$ leaf toward no CO is shown in pink and has the number of pixels $n_\mathrm{H\alpha, \ w/o \ CO}$.
        The H$\alpha$ leaf overlapped with CO is shown in dark pink and has the number of pixels with CO emission $n_\mathrm{H\alpha, \ w/ \ CO}$.
        (b) A histogram of $n_\mathrm{H\alpha, \ w/ \ CO}/n_\mathrm{H\alpha}$.
        The H$\alpha$ leaves having a overlapping area fraction $n_\mathrm{H\alpha, \ w/ \ CO}/n_\mathrm{H\alpha}$ of 5--35\%, 35--65\%, and 65--95\% are 18.3\%, 23.3\%, and 28.1\% of the total number, respectively. }
    \label{picture_ovl}
\end{figure*}

\begin{figure*}[htbp]
    \begin{center}
        \includegraphics[width=0.95\linewidth,clip]{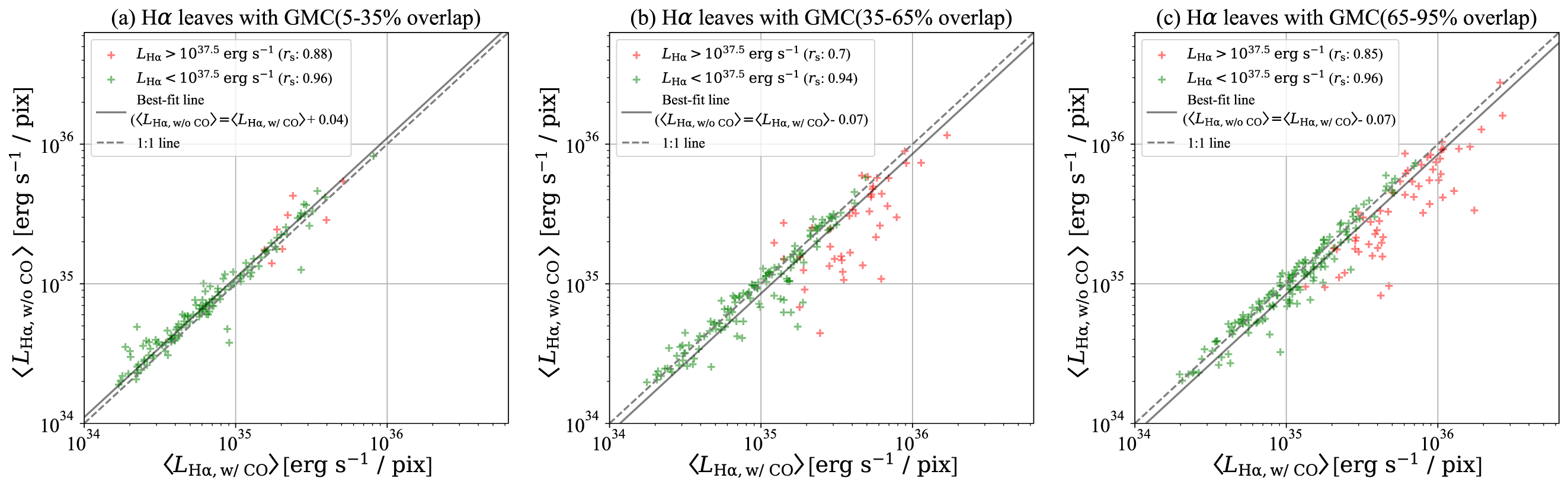}
    \end{center}
    \caption{Scatter plots for average \LHa \ between the H$\alpha$ leaves with and without CO emission, which are drived by $\langle L_\mathrm{H\alpha, w/ \ CO} \rangle = L_\mathrm{H\alpha, w/ \ CO} / n_\mathrm{H\alpha, w/ \ CO}$ and $\langle L_\mathrm{H\alpha, w/o \ CO} \rangle = L_\mathrm{H\alpha, w/o \ CO} / n_\mathrm{H\alpha, w/o \ CO}$. (a) is for the H$\alpha$ leaves, 5--35\% of which are with CO, (b) those, 35--65\% of which are with CO, and (c) those, 65--95\% of which are with CO. Green crosses show the leaves with $\LHa < 10^{37.5} \ \ergs$ and the red crosses show those with $>10^{37.5} \ \ergs$. Black lines indicate the results of the orthogonal distance regression fitting with a slope of 1 between $\langle L_\mathrm{H\alpha, \ w/ \ CO}\rangle$ and $\langle L_\mathrm{H\alpha, \ w/o \ CO}\rangle$. The dashed lines indicate that for the two variables are equal to each other.}
    \label{scatter_ovl}
\end{figure*}

Figure \ref{picture_ovl}(a) shows a schematic demonstrating the relationship between CO and H$\alpha$.
The H$\alpha$ leaf without CO (abbreviated as w/o CO subsequently) is pink and has a number of pixels $n_\mathrm{H\alpha, \ w/o \ CO}$.
The H$\alpha$ leaf with CO (abbreviated as w/ CO subsequently) is dark pink and has a number of pixels $n_\mathrm{H\alpha, \ w/ \ CO}$.
Then the whole H$\alpha$ leaf has a number of pixels $n_\mathrm{H\alpha} = n_\mathrm{H\alpha, \ w/o \ CO} + n_\mathrm{H\alpha, \ w/ \ CO}$.
Figure \ref{picture_ovl}(b) shows a histogram of $n_\mathrm{H\alpha, \ w/ \ CO}$ / $n_\mathrm{H\alpha}$, which is divided into three regions according to a fraction of the number of pixels of the H$\alpha$ leaves with CO as follows: $n_\mathrm{H\alpha, \ w/ \ CO}$ / $n_\mathrm{H\alpha}=$ 5--35\%, 35--65\%, and 65--95\%.
We except H$\alpha$ leaves that have a fraction with CO of $<5\%$ and $>95\%$ because it is difficult to calculate the averaged \LHa \ of these pixels with CO and without CO, respectively.
Figure \ref{scatter_ovl} shows the scatter plots between the average \LHa \ with CO and that without CO, which are derived by
$\langle L_\mathrm{H\alpha, w/ \ CO} \rangle = L_\mathrm{H\alpha, w/ \ CO} / n_\mathrm{H\alpha, w/ \ CO}$ and $\langle L_\mathrm{H\alpha, w/o \ CO} \rangle = L_\mathrm{H\alpha, w/o \ CO} / n_\mathrm{H\alpha, w/o \ CO}$.
Figures \ref{scatter_ovl}(a), (b), and (c) correspond to the H$\alpha$ leaves, 5--35\%, 35--65\%, and 65--95\% of which are with CO leaves, respectively.
Green crosses show the H$\alpha$ leaves with $\LHa < 10^{37.5} \ \ergs$ (Type II and III) and red crosses show those with $\LHa > 10^{37.5} \ \ergs$ (Type III).
Black straight lines indicate the results of the orthogonal distance regression fitting with a slope of 1.
The straight dashed lines indicate that the two values are equal to each other.
The observed \LHa \ with CO shows only a small deviation within $\sim 10$\% from the dashed line toward the high extinction regime, indicating that extinction is nonsignificant in the present GMCs.
On the contrary, in star-forming Type II and Type III GMCs shown in red (Figures \ref{scatter_ovl}(b) and (c)) the that \LHa \ shows large deviations in the opposite direction, indicating a strong enhancement in \LHa \ with CO, which is likely due to enhanced star formation in the GMCs.

\citet{Groves2023} studied 19 galaxies including M74, and calculated the median $A _\mathrm{v}$ to be $\sim 0.7$ mag by using the H$\alpha$ and H$\beta$ data obtained with MUSE, which is consistent with the other works showing $A _\mathrm{v}$ is typically $\sim 1$ mag for different galaxies.
In fact, the distribution of GMCs is compared with H$\alpha$ in Figure \ref{example_ovl} of Appendix \ref{ex_ovl}, showing that extinction does not affect much H$\alpha$ distribution.
Supposing that the extinction is small in LMC and significant only in the M74, we investigated the relationship between \htwo \ regions and H$\alpha$ leaves in \citet{Groves2023}, and applied the correction for extinction to the H$\alpha$ luminosity by using the intensity ratio of the Balmer decrement.
We found 62 Type II GMCs have the corrected H$\alpha$ luminosity ($L_\mathrm{H\alpha,corr}$) larger than $10^{37.5} \ \ergs$.
These GMCs can be classified to Type III using $L_\mathrm{H\alpha,corr}$, then the number ratio is changed from 59 Type I, 201 Type II, and 172 Type III to 59 Type I, 139 Type II, and 234 Type III,  but the trends of the physical parameters in Figures \ref{GMC_prop} and \ref{alpha_vir} remains essentially similar.
We therefore infer that the extinction correction does not alter the present conclusions.

\section{Discussion}\label{discussions}
\subsection{Type classification based on H$\alpha$ luminosity}
In the present work on M74, we analyzed the \twelvecoh \ data at 53 pc resolution, the H$\alpha$ image at 44 pc resolution, and the JWST 21 $\mathrm{\mu m}$ image at 32 pc resolution. 
We obtained the resolved physical properties of 432 GMCs, 1351 H$\alpha$ leaves, and 692 21 $\mathrm{\mu m}$ leaves. 
Based on these results, we classified the GMCs into three Types according to the level of high-mass star formation as originally adopted by F99. 
We did not use clusters in the classification because they are not always cataloged in galaxies, particularly for those more distant than 10 Mpc. 
Therefore, only H$\alpha$ (\htwo \ regions) were used in the present study for the measurements, which can be extended to other galaxies. 
To examine the H$\alpha$-based scheme, we compared the results with the clusters identified by HST images and the JWST mid-IR images. 
The results for the clusters were found to be similar to those of the LMC (F99; K09), owing to the fact that the developed clusters are associated with Type III but not with Type II. 
This supports the H$\alpha$-based scheme, which is applicable not only to the dwarfs, LMC and M33, but also to a grand design spiral. 
We further explored the physical properties and evolutionary trend of the GMCs and star/cluster formations. 
Tables \ref{GMC_prop_table} and \ref{association_table} summarize the GMC properties and star formation signatures associated with GMCs, along with the number of relevant objects, \htwo \ regions, 1--4 Myr clusters, and 21 $\mathrm{\mu m}$ leaves that show significant association with the GMCs. 
The physical and star-formation properties of each GMC Type are summarized as below.

\begin{description}
\item[Type I: ] $M_\mathrm{CO}$ and the radius of the GMCs is relatively small and the virial ratio is large. 
No \htwo \ regions or clusters are associated, indicating the absence of high-mass star formation.
\item[Type II: ] $M_\mathrm{CO}$, the radius and virial ratio of the GMCs are intermediate. 
High-mass star formation is observed, \LHa \ is small at approximately $10^{37} \ergs$ according to the Type definition, and 1--4 Myr clusters with low mass ($\sim 10^3 M_\odot$) including a few O stars are associated with $\sim 20\%$ of Type II GMCs.
\item[Type III: ] $M_\mathrm{CO}$ and the radius of the GMCs is large and the virial ratio is small. 
More active high-mass star formation is observed, \LHa \ is large at approximately $10^{38} \ergs$, and 1--4 Myr clusters with high mass ($\sim 10^4 M_\odot$) including a few tens of O stars are associated with $\sim 50\%$ of Type III GMCs.
\end{description}

\begin{figure}[htbp]
    \begin{center}
    \includegraphics[width=0.8\linewidth,clip]{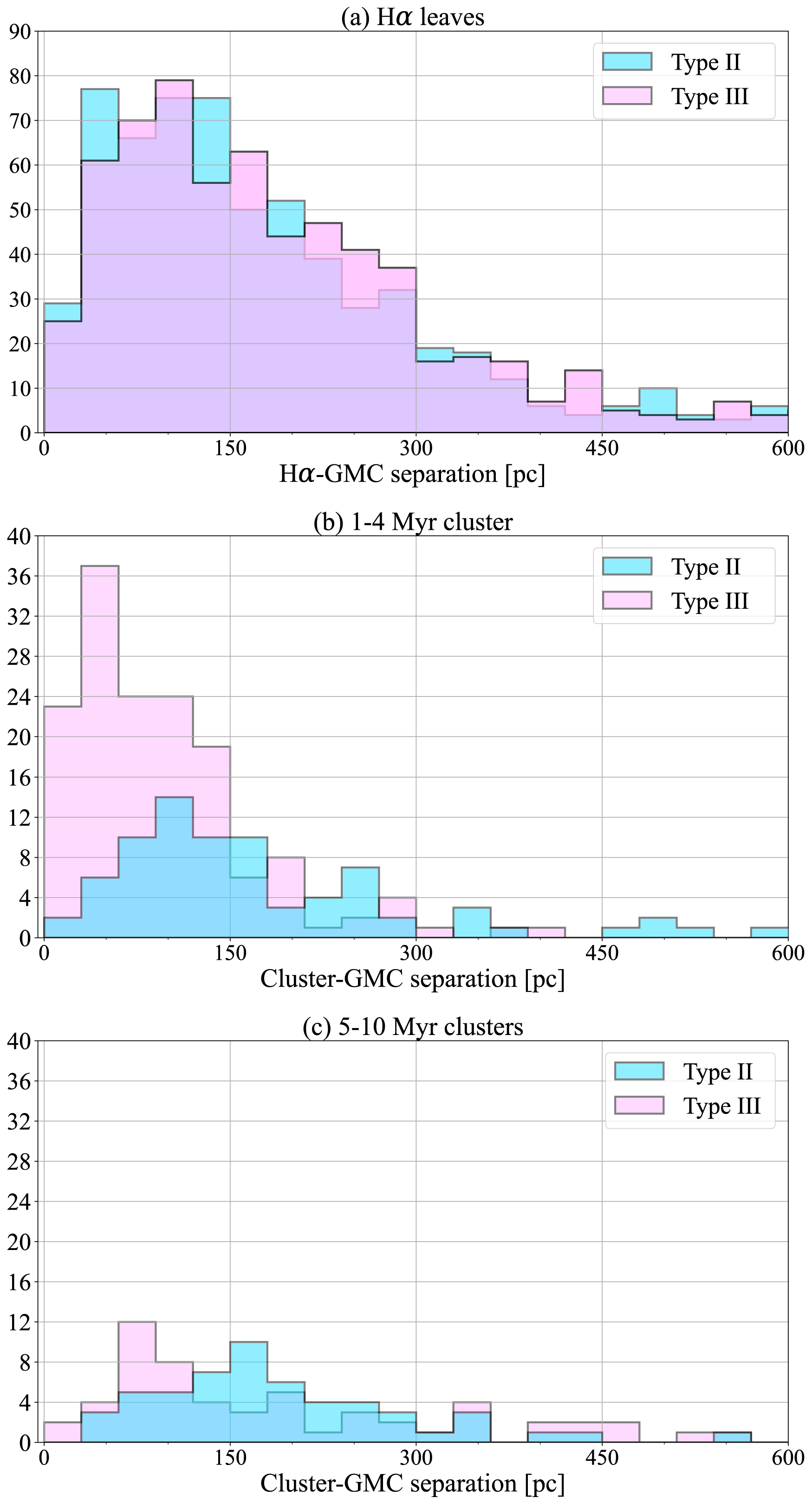}
    \end{center}
    \caption{Histogram of the separation between the nearest GMCs and (a)\htwo \ regions, (b)1--4 Myr clusters, and (c) 5--10 Myr clusters. If the nearest GMC is Type II and Type III, the separation is shown in light blue and pink, respectively.}
    \label{hist_separation_all}
\end{figure}

\begin{figure}[htbp]
    \begin{center}
    \includegraphics[width=0.8\linewidth,clip]{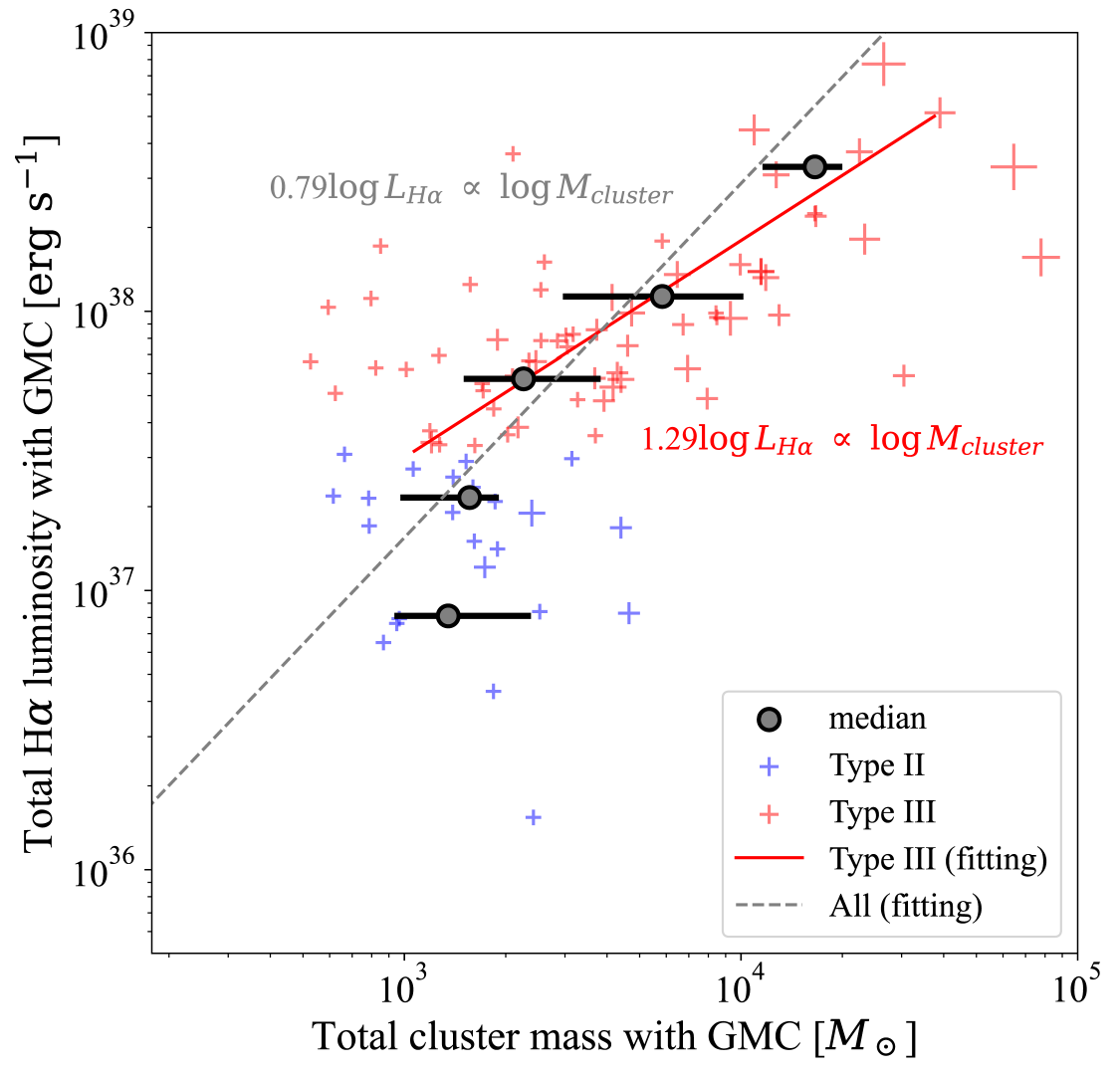}
    \end{center}
    \caption{A scatter plot between the total 1--4 Myr cluster mass ($M_\mathrm{cluster}$) and the total H$\alpha$ luminosity ($L_{\mathrm{H\alpha}}$) in each GMC. 
    Blue and red crosses indicate Type II and III GMCs, respectively, and the size of the marker is proportional to the number of clusters with each GMC, the max number is 9. 
    gray circles and black bars indicate the median value of $M_\mathrm{cluster}$ within 0.4-decade bins along the \LHa-axis, and the horizontal bars indicate their IQRs. 
    The red-solid and gray-dashed lines indicate the least squares fitting line with the median \LHa \ as the independent variable and $M_\mathrm{cluster}$ as the dependent variable weighted by the number of data points in each bin;
    Type III: $1.29\log L_{\mathrm{H \alpha}} \propto \log M_{\mathrm{cluster}}$,
    all: $0.79 \log L_{\mathrm{H \alpha}} \propto \log M_{\mathrm{cluster}}$.}
    \label{Ha_Mstar}
\end{figure}

These results are summarized in Figures \ref{hist_separation_all} and \ref{Ha_Mstar}.
Figure \ref{hist_separation_all} shows the histogram of the separation between the nearest GMCs and (a) H$\alpha$ leaves, (b) 1--4 Myr clusters, and (c) 5--10 Myr clusters. 
When the nearest GMCs are Type II and Type III, the separation is shown in light blue and pink, respectively. 
This Figure indicates that the H$\alpha$ leaves and the young clusters of 1--4 Myr age are strongly correlated with the Type II/III GMCs within 150 pc separation.

Figure \ref{Ha_Mstar} shows a scatter plot for the total mass of the 1--4 Myr cluster ($M_\mathrm{cluster}$) and the total H$\alpha$ luminosity (\LHa) in each GMC. 
The blue and red crosses indicate Type II and Type III GMCs, respectively, where the size of the cross corresponds to the number of clusters with each GMC, as indicated within the figure caption. 
The gray circles indicate the median values of $M_\mathrm{cluster}$ within 0.4-decade bins along the \LHa-axis, and the horizontal bars indicate their interquartile ranges (IQRs). 
The red-solid and gray-dashed lines indicate the lines determined by least squares fitting with the median \LHa \  as an independent variable and $M_\mathrm{cluster}$ as a dependent variable weighted by the number of data points in each bin, as expressed for Type III: $1.29\log L_{\mathrm{H \alpha}} \propto \log M_{\mathrm{cluster}}$, and for all: $0.79\log L_{\mathrm{H \alpha}} \propto \log M_{\mathrm{cluster}}$. 

\citet{Fukui1999} and \citet{Kawamura2009} made Type classification of GMCs in the LMC based on the star formation activity by using associated young clusters and the number of O stars where \htwo \ regions were used as a measure as summarized by \citet{FukuiKawamura2010}. 
\citet{Fukui1999} noted that the H$\alpha$ luminosity of the \htwo \ regions well corresponds to the association of clusters; i.e., GMCs with associated clusters have higher H$\alpha$ luminosity without exception and GMCs without clusters have lower H$\alpha$ luminosity, where the threshold H$\alpha$ luminosity was found to be $10^{37.0-37.5} \ \ergs$. 
The threshold H$\alpha$ luminosity was examined further by \citet{Yamaguchi2001} and the correspondence above was confirmed. 
For application of the classification to other galaxies, clusters are not uniformly cataloged generally, whereas the H$\alpha$ luminosity is uniformly observed over a galaxy. 
We adopted only \LHa \ as a measure of GMC Type in the present work. 
A remaining problem is to examine possible contamination in \LHa \ by older stars which may cause uncertainty in \LHa \ ionized by young O stars. 
In order to estimate such contamination, we calculated \LHa \ by older clusters, where we used the PARSEC code \citep{Bressan2012} in order to estimate the \LHa \ from clusters with an age up to 20 Myr. 
We used a stellar evolution model of  \citet{Kroupa2001,Kroupa2002} IMF, and calculated \LHa \ from the number of OB stars following the method of \citet{Panagia1973}, which is converted into \LHa \ by the relationship given in the \htwo \ regions of the LMC by \citet{Wilcots1994}. 
As a result, we obtained \LHa \ for clusters of two total masses, $500 \ M_\odot$ and $5000 \ M_\odot$, as follows. 
A $500 \ M_\odot$ cluster of an age of 1 Myr has $\LHa \sim 10^{38} \ \ergs$  and that of an age of 5 Myr has $\LHa \sim 10^{36.5} \ \ergs$ ionized mainly by one O star, and that of an age of 10 Myr has $\LHa \sim 10^{34} \ \ergs$ ionized by only B stars.
On the other hand, a $5000 \ M_\odot$ cluster of an age of 1 Myr has $\LHa \sim 10^{39} \ \ergs$ ionized by 15 O stars, that of an age of 5 Myr has $\LHa \sim 10^{37} \ \ergs$ ionized by 5 O stars, and that of an age of 10 Myr has $\LHa \sim 10^{35} \ \ergs$ ionized by only B stars.
Massive clusters of an age of 5 Myr may contaminate with \htwo \ regions of Type II GMCs. 
However, as shown in Figure \ref{dist_from_GMC_1e2} and \ref{dist_from_GMC_1e3} in Appendix \ref{appendix_separation}, a $10^3 \ M_\odot$ cluster that evolved by $>$5 Myr tends to be separated from a GMC by more than 150 pc.
A H$\alpha$ leaf associated with a GMC therefore tends to reflect the ionization mostly by young O stars (Figure \ref{Ha_Mstar}).

We also checked the contamination of planetary nebulae (PNe) to H$\alpha$ leaves using the PNe catalog by \citet{Scheuermann2022}. 
On the whole M74, 139 PNe are catalogued. 
Only 22 (16\%) of 139 are included within 150 pc of GMCs. 
On the other hand, 631 (47\%) H$\alpha$ leaves are distributed within 150 pc of a GMC. 
This indicates the fraction of PNe that can be found in H$\alpha$ leaves is only a few percent, so H$\alpha$ leaves with GMCs are mostly \htwo \ regions formed by massive young stars.

Evolved stars such as wolf rayet stars (WR stars) may also contaminate H$\alpha$ leaves.
In the LMC, $\sim 150$ WR stars \citep{Breysacher1999,Neugent2018} and $\sim 1800$ massive stars \citep{Bonanos2009} have been discovered to date. 
Considering that Bonanos’ catalog is a compilation of past publications and the observations are strongly biased, the number ratio of WR stars to massive stars is less than 10\%. 
In addition, the timescale of WR stars is less than 1 Myr, more than an order of magnitude shorter than that of massive stars. 
These imply that the contamination of WR stars is negligible.

In summary, we confirmed that the present Type classification provides a scheme consistent with the original Type definitions given by F99 and K09 based on the level of high-mass star formation activity. 
The present work attempted to apply the Type classification scheme originated by F99 to the grand design spiral for the first time. 
This attempt became feasible thanks to the high resolution provided by ALMA under the PHANGS project. 
The results seem to be quite encouraging because several new aspects of the extra-galactic GMCs have been revealed, indicating promising steps to follow in our efforts to better understand the GMC evolution by more numerous sample galaxies already mapped with ALMA.

In the present work, we used the H$\alpha$ luminosity for Type classification to test the scheme usable to more galaxies for which cluster catalogs are not available. 
It is not clear how the worse resolution of CO and H$\alpha$ affects the association of GMCs and \htwo \ regions, and how the H$\alpha$ sensitivity limit of Type II affects the Type I definition.
At the present stage, Type classification based on H$\alpha$ luminosity can be applied under the limitation that there are $<100$--200 pc resolution datasets and sensitivity of H$\alpha$ is higher enough $10^{37.5} \ \ergs$.
We need to keep watching how these limitations of H$\alpha$ may be cured by more sensitive probes like JWST etc. in future works.

\begin{table*}[htbp]{
    \centering
    \tbl{Summary of GMC association with \htwo\ regions, 21 $\mathrm{\mu m}$ leaves and 1--4 Myr clusters.}{
    \begin{tabularx}{0.9\linewidth}{cllcccc}
        \hline
         &&  & Type I & Type II & Type III & All\\
        \hline
        (1) & H$\alpha$ observed area & $N_\mathrm{GMC}^*$ & 59 & 202 & 171 & 432\\
        && $N_\mathrm{H\alpha}^\dagger$ & 0 & 343 & 379 & 722 \\
        && $N_\mathrm{H\alpha}/N_\mathrm{GMC}$ & 0 & 1.70 & 2.22 & 1.67 \\
        && $L_\mathrm{H\alpha}/N_\mathrm{GMC}$ ($\mathrm{erg~s^{-1}}$) & -- & $10^{37.10}$ & $10^{38.15}$ & $10^{37.87}$ \\
        \hline
        (2) & HST observed area & $N_\mathrm{GMC}^*$ & 47 & 170 & 126 & 343 \\
        && \begin{tabular}{l}$N_\mathrm{GMC, w/ cluster}^\ddagger$ \\ ($N_\mathrm{GMC, w/ cluster}/N_\mathrm{GMC}$)\end{tabular} 
        & -- & 25 (15\%) & 65 (52\%) \\
        && $N_\mathrm{cluster}^\S$ & -- & 30 & 131 & 161 \\
        && $N_\mathrm{cluster}/N_\mathrm{GMC, w/ cluster}$ & -- & 1.20 & 2.02 & 1.79 \\
        && $M_\mathrm{cluster}/N_\mathrm{GMC, w/ cluster}$ ($M_\odot$) & -- & 1779 & 8328 \\
        \hline
        (3) & 21 $\mathrm{\mu m}$ observed area & $N_\mathrm{GMC}^*$ & 45 & 154 & 127 & 326 \\
        && \begin{tabular}{l}$N_\mathrm{GMC, w/ 21\mu m}^\|$ \\ ($N_\mathrm{GMC, w/ 21\mu m}/N_\mathrm{GMC}$)\end{tabular} 
        & 14 (31\%) & 119 (77\%) & 121 (95\%) & 254 \\
        && $N_\mathrm{21\mu m}^\#$ & 18 & 197 & 276 & 491 \\
        && $N_\mathrm{21\mu m}/N_\mathrm{GMC, w/ 21\mu m}$ & 1.29 & 1.66 & 2.28 & 1.93 \\
        && $L_\mathrm{21\mu m}/N_\mathrm{GMC, w/ 21\mu m}$ ($\mathrm{erg~s^{-1}}$) & $10^{38.30}$ & $10^{38.37}$ & $10^{38.96}$ & $10^{39.07}$ \\
        \hline
    \end{tabularx}
    }
    \begin{tabnote}
        Lines: Association between (1) GMCs in H$\alpha$ observed area and H$\alpha$ leaves, 
        (2) GMCs in HST observed area and 1--4 Myr clusters, 
        (3) GMCs in 21 $\mathrm{\mu m}$ observed area and 21 $\mathrm{\mu m}$ leaves.
        $*$ Number of GMCs. 
        $\dagger$ Number of H$\alpha$ leaves with GMCs.
        $\ddagger$ Number of GMCs with 1--4 Myr clusters.
        $\S$ Number of 1--4 Myr clusters with GMCs.
        $\|$ Number of GMCs with 21 $\mathrm{\mu m}$ leaves. 
        $\#$ Number of 21 $\mathrm{\mu m}$ leaves with GMCs.
    \end{tabnote}
    \label{association_table}
    }
\end{table*}

\subsection{Comparison with the 21 $\mathrm{\mu m}$ leaves}
\label{dis_21um}
\begin{figure}[htbp]
    \begin{center}
    \includegraphics[width=0.8\linewidth,clip]{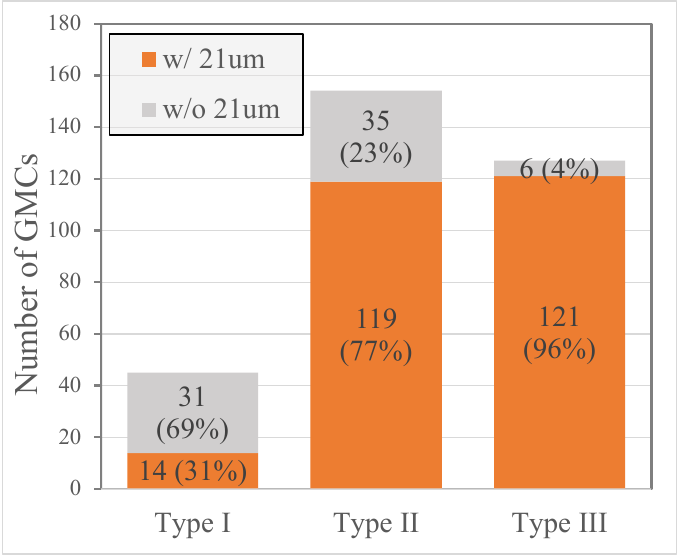}
    \end{center}
    \caption{Bar graph showing the relationship between Type I, II, III and 21 $\mathrm{\mu m}$ leaf association. Orange and gray represent the number of GMCs with and without 21 $\mathrm{\mu m}$ leaves, respectively. The labels on the graph indicate the number of GMCs, and the numbers in brackets indicate the percentage.}
    \label{bar_21um}
\end{figure}

To elucidate the young clusters in more detail, we compared the above results with those of the 21 $\mathrm{\mu m}$ leaves, and observed that association with the 21 $\mathrm{\mu m}$ leaves is not exactly the same with the H$\alpha$ leaves. 
Figure \ref{bar_21um} shows  the bar graphs for the fraction of the 21 $\mathrm{\mu m}$ leaves associated with each GMC Type. 
Only 6 among 127 Type III GMCs (4\%) are not associated with the 21 $\mathrm{\mu m}$ leaves, and 35 among 154 Type II GMCs (23\%) are not associated with the 21 $\mathrm{\mu m}$ leaves. 
For Type I GMCs, 14 among 45 GMCs (31\%) are associated with the 21 $\mathrm{\mu m}$ leaves. 
Therefore, we observe that small fractions of Type II and Type I GMCs show behavior different from association with H$\alpha$; i.e., Type II without 21 $\mathrm{\mu m}$ leaves and Type I with 21 $\mathrm{\mu m}$ leaves.

\begin{figure}[htbp]
    \begin{center}
    \includegraphics[width=0.8\linewidth,clip]{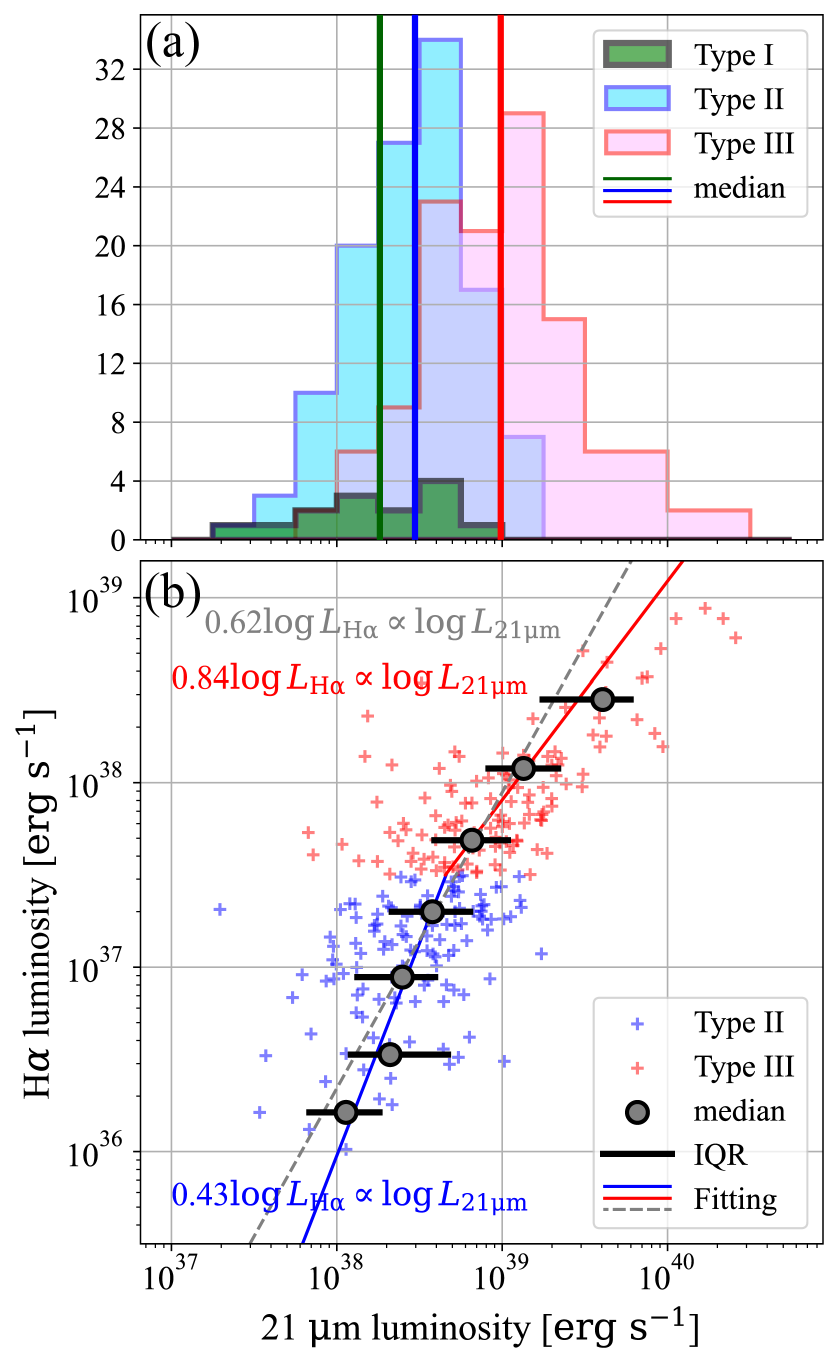}
    \end{center}
    \caption{(a) Histogram of the luminosities of the 21 $\mathrm{\mu}$m leaves associated with GMCs. Green and light blue and pink filled areas show the 21 $\mathrm{\mu}$m leaves associated with Type I, II, and III, respectively. The green, blue, and red vertical lines indicate the median value for Type I, II, and III, respectively. 
    (b) A scatter plot between the total 21 $\mathrm{\mu}$m luminosity ($L_\mathrm{21\mu m}$) and H$\alpha$ luminosity (\LHa) in each GMC. 
    The blue and red crosses indicate Type II and III GMCs, respectively. The gray circles indicate the median values of $L_\mathrm{21 \ \mu m}$ within 0.4-decade bins along the \LHa-axis, and the horizontal bars indicate their IQRs. The blue and red-solid and gray-dashed lines indicate the least squares fitting line with the median \LHa \ as the independent variable and $L_\mathrm{21 \ \mu m}$ as the dependent variable weighted by the number of data points in each bin;
    Type II: $0.43\log L_{\mathrm{H \alpha}} \propto \log L_{\mathrm{21\mu m}}$,
    Type III: $0.84\log L_{\mathrm{H \alpha}} \propto \log L_{\mathrm{21\mu m}}$,
    all: $0.62 \log L_{\mathrm{H \alpha}} \propto \log M_{\mathrm{21\mu m}}$.}
    \label{21um_Ha}
\end{figure}

To quantify the correlation of the 21 $\mathrm{\mu m}$ leaves, in Figure \ref{21um_Ha}(a) we show a histogram of the total luminosity of the 21 $\mathrm{\mu m}$ leaves, $L_\mathrm{21 \mu m}$, associated with three GMC Types. 
The green, light blue, and pink filled areas and the green, blue, and red vertical lines show $L_\mathrm{21 \mu m}$ and its median values for Type I, II, and III GMCs, respectively. 
Figure \ref{21um_Ha}(b) shows a scatter plot between $L_\mathrm{21 \mu m}$ and \LHa \ associated with a common GMC. 
The blue and red crosses indicate Type II and III GMCs, respectively. 
The gray circles indicate the median values of $L_\mathrm{21 \mu m}$ within 0.4-decade bins along the \LHa-axis, and the horizontal bars indicate their IQRs. 
The blue and red solid and gray-dashed lines indicate the least squares fitting line with the median \LHa \ as an independent variable and $L_\mathrm{21 \mu m}$ as a dependent variable weighted by the number of data points in each bin. 
Their slopes are given as follows; 
Type II: $0.43\log L_{\mathrm{H \alpha}} \propto \log L_{\mathrm{21\mu m}}$,
Type III: $0.84\log L_{\mathrm{H \alpha}} \propto \log L_{\mathrm{21\mu m}}$,
all: $0.62 \log L_{\mathrm{H \alpha}} \propto \log M_{\mathrm{21\mu m}}$.

\begin{figure}[htbp]
    \begin{center}
    \includegraphics[width=0.8\linewidth,clip]{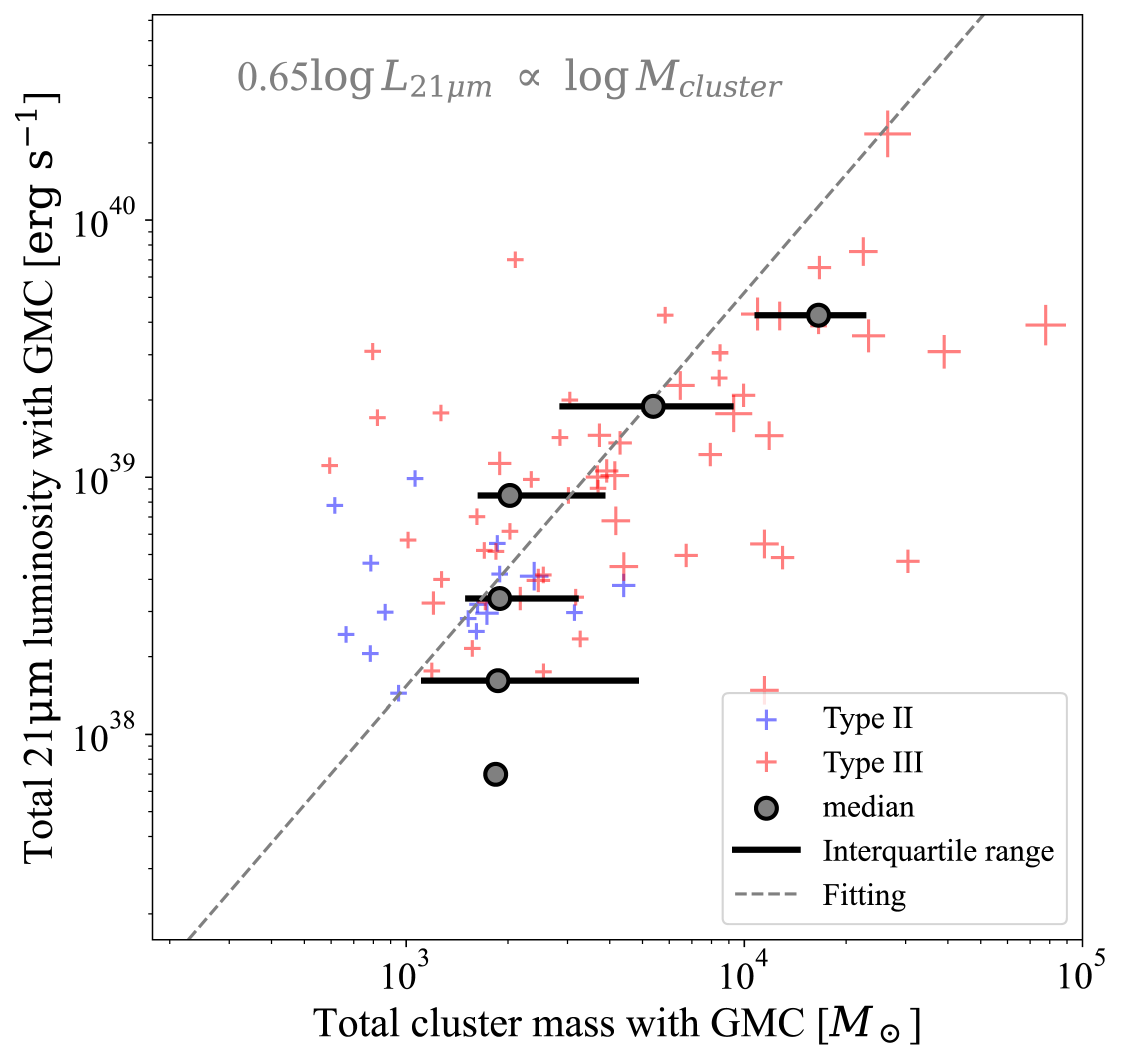}
    \end{center}
    \caption{A scatter plot between the total 1--4 Myr cluster mass ($M_\mathrm{cluster}$) and the total 21 $\mathrm{\mu}$m luminosity ($L_\mathrm{21 \mu m}$) in each GMC. Symbols and lines are the same as those used Figure \ref{Ha_Mstar}. The regression line: $0.65 \log L_{\mathrm{H \alpha}} \propto \log M_{\mathrm{cluster}}$.}
    \label{21_Mstar}
\end{figure}

\begin{figure}[htbp]
    \begin{center}
    \includegraphics[width=0.8\linewidth,clip]{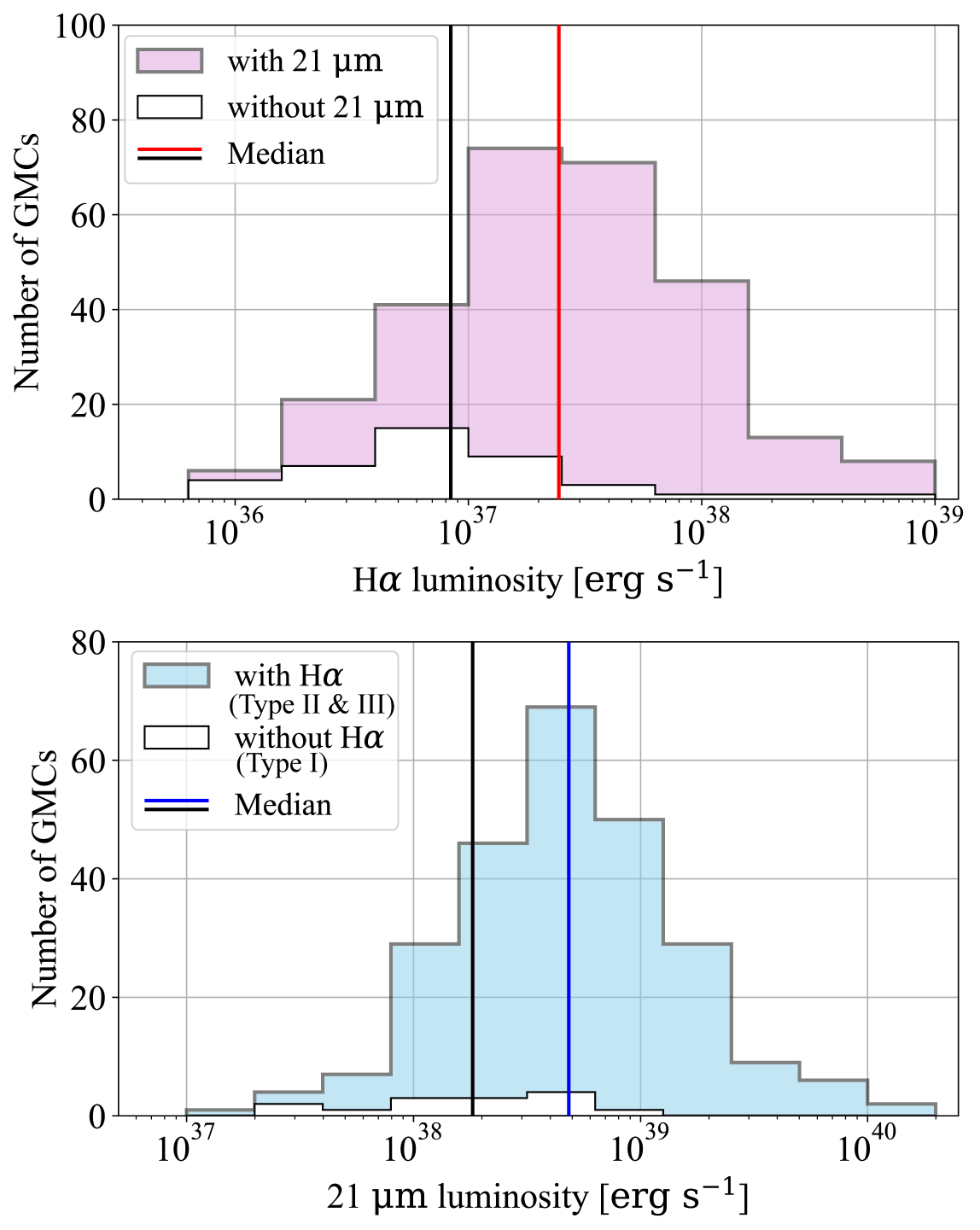}
    \end{center}
    \caption{(a) Histogram of H$\alpha$ luminosities associated with GMCs. The pink histogram shows all GMCs with 21 $\mathrm{\mu m}$ and the white shows GMCs not associated with 21 $\mathrm{\mu m}$ leaves. (b) Histogram of 21 $\mathrm{\mu m}$ luminosities associated with GMCs. The light blue histogram shows all GMCs with H$\alpha$ (Type II and III) and the white shows Type I not associated with H$\alpha$ leaves (Type I).}
    \label{hist_Ha21um_aboutTypeGMC}
\end{figure}

\begin{figure}[htbp]
    \begin{center}
    \includegraphics[width=0.8\linewidth,clip]{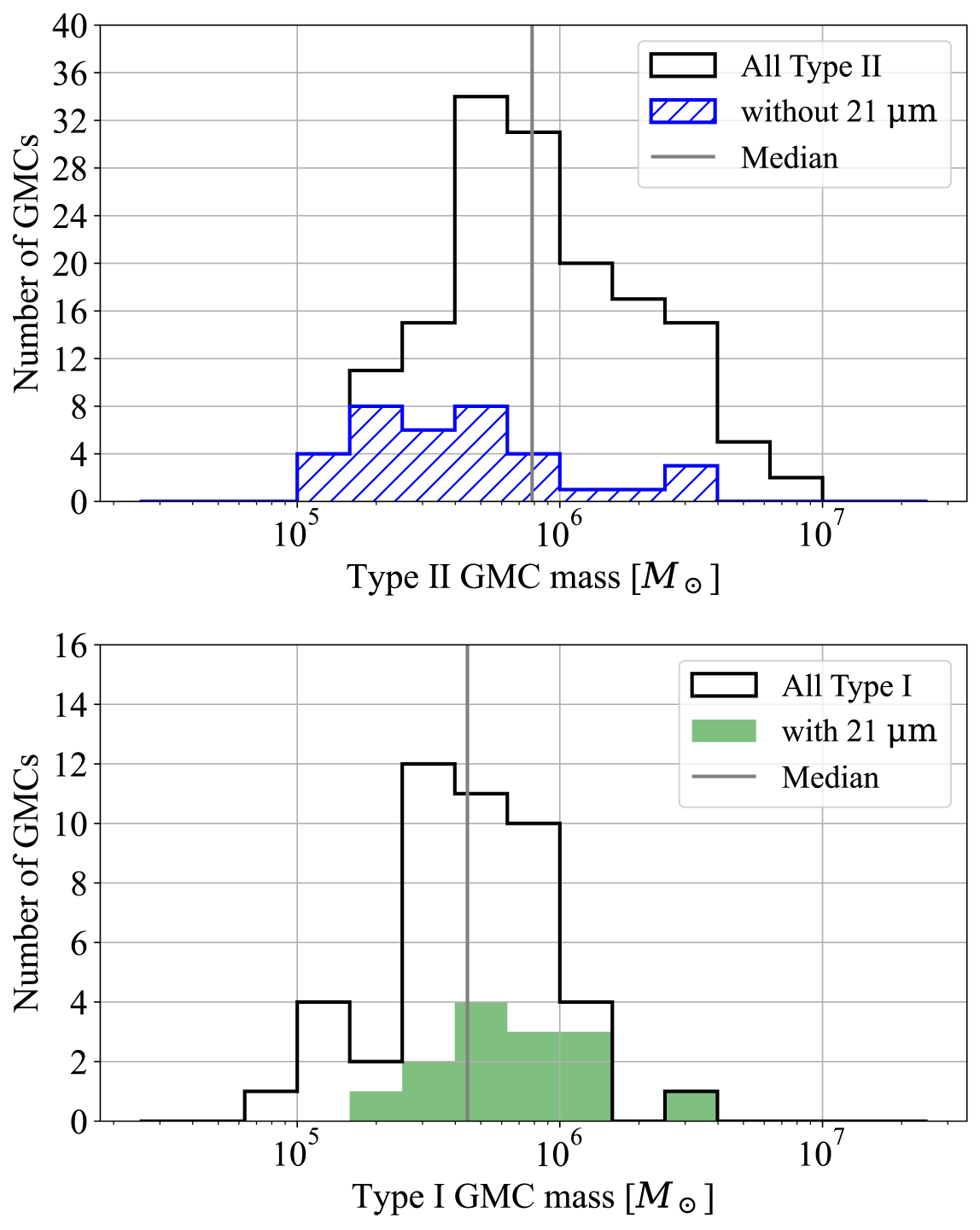}
    \end{center}
    \caption{(a) Histogram for mass of Type II. The black histogram shows all Type II and the blue shaded one shows Type II not associated with 21 $\mathrm{\mu m}$ leaves. (b) Histogram of mass of Type I. The black histogram shows all Type I and the green shaded histogram shows Type I associated with 21 $\mathrm{\mu m}$ leaves. }
    \label{hist_mass_comp21um}
\end{figure}

To obtain an approximate mass scale of the 21 $\mathrm{\mu m}$ leaves, we show in Figure \ref{21_Mstar} a scatter plot between the total 1--4 Myr cluster mass ($M_\mathrm{cluster}$) and $L_\mathrm{21 \mu m}$ in each GMC. 
The symbols and lines are the same as those in Figure \ref{Ha_Mstar}. 
Figure \ref{hist_Ha21um_aboutTypeGMC} (a) shows a histogram of \LHa \ associated with GMCs. 
The pink histogram shows all GMCs with 21 $\mathrm{\mu m}$ leaves and the white area shows GMCs not associated with 21 $\mathrm{\mu m}$ leaves. 
The GMCs without 21 $\mathrm{\mu m}$ leaves have lower \LHa, with a peak at $\sim 10^{36.8} \ \ergs$ in a range of $10^{35.7}$--$10^{37.8} \ \ergs$. 
Figure \ref{hist_Ha21um_aboutTypeGMC} (b) shows a histogram of $L_\mathrm{21 \mu m}$ associated with GMCs. 
The light blue histogram shows all the GMCs with H$\alpha$ (Type II and Type III), and the white histogram shows Type I not associated with the H$\alpha$ leaves (Type I). 
Figure \ref{hist_mass_comp21um} (a) shows a histogram of the Type II GMC mass. 
The black histogram shows all Type II and the blue area shows Type II not associated with 21 $\mathrm{\mu m}$ leaves. 
Figure \ref{hist_mass_comp21um} (b) shows a histogram of the Type I GMC mass. 
The black histogram shows all Type I and the green area shows Type I associated with the 21 $\mathrm{\mu m}$ leaves.

This trend can be interpreted to indicate an intermediary population of GMCs between Type I and II. 
The GMC population comprises two types; one is associated with a cluster of $\sim 10^3 \ M_\odot$ having no or only a few high-mass stars ($\LHa < 10^{36.5} \ \ergs$), and the other with a cluster of similar mass having several high-mass stars ($\LHa > 10^{36.5} \ \ergs$). 
The former GMC is classified as Type I by the low \LHa \ , whereas the cluster has a detectable 21 $\mathrm{\mu m}$ flux. 
The latter GMC is classified as Type II by the slightly higher \LHa, whereas the cluster mass is below the 21 $\mathrm{\mu m}$ detection limit.

Using this intermediate GMC population, we can find candidates in the solar neighborhood. 
They include M42, GM24, RCW120, and RCW79 and have molecular clouds with an \htwo \ region and high-mass star(s) in clusters. 
Most of them are likely formed by a cloud-cloud collision (CCC) trigger, as revealed by molecular studies \citep{Fukui2018_M42,Fukui2018_GM24,Torii2015,Ohama2018}. 
The number of high-mass stars is approximately proportional to the cluster mass, which has a wide range of orders of magnitude for a given number of high-mass stars, as reported by \citet{Fukui2021} and \citet{Enokiya2021}. 
According to recent studies on CCCs, the formation of high mass stars is a stochastic process with a typical time interval of several Myr.
Environmental conditions such as cloud number density determine the mean free time between CCCs (for a theoretical base see  \cite{HabeOhta1992,Inoue2013,Takahira2014,Dobbs2015,Kobayashi2018}). 
Therefore, in the present classification based on the number of high-mass stars, a wide range of cluster masses for a given number of high-mass stars is the natural outcome of stochastic high-mass star formation. 
This provides an explanation for the existence  of Type II GMCs without 21 $\mathrm{\mu m}$ detection and the presence of Type I GMCs with 21 $\mathrm{\mu m}$ detection. 
The confirmation of this in other galaxies is a future task that can be addressed by an extensive 21 $\mathrm{\mu m}$ study with JWST.

\subsection{Evolutionary time scales of GMCs}
Based on the above properties, we discuss the timescales of each GMC Type under the assumption of the steady-state evolution of GMCs. 
A simple assumption is that the Type III timescale is equal to a cluster age of 4 Myr. 
The timescales of Type I and II are proportional to the number of GMCs if we assume that all GMCs evolve steadily. 
The duration of Type I and Type II are estimated to be 1 Myr and 5 Myr, respectively, resulting in a total GMC lifetime of 10 Myr.
These timescales should be considered the lower limits, and the real timescale may be longer than these. 
This is because the sensitivity of cluster observations may not be sufficiently high to detect even older clusters with ages of several Myr \citep{Adamo2017}. 
Timescale of Type III may have a few Myr uncertainly because 20\% of 5--10 Myr clusters are associated with Type III.
Moreover, 76 1--4 Myr clusters, corresponding to 30\% fraction, are not associated with GMCs. 
These young clusters may be associated with low mass ($\sim 10^4  \ M_\odot$) GMCs, which are not detected in the observation because of lack of sensitivity.
There may be a few Myr uncertainties in the timescale as discussed above.
Alternatively, if we assume that the time scale of Type II is equal to that of Type II at 12 Myr, as in the LMC, the timescales of the Type I, Type II, and Type III GMCs are estimated to be 3 Myr, 12 Myr, and 10 Myr, respectively, giving a total GMC lifetime of 25 Myr, which is similar to that derived in the LMC (K09) and M33 \citep{Corbelli2017}. 
More sensitive CO observation and determination of the cluster age in the future is desirable to confirm this possibility.

\subsection{GMC evolution processes}
We further explored the physical processes of each GMC Type considering the aforementioned properties. 
The evolution from Type I to Type III is consistent with GMC growth (increase of mass and radius) and gravitational relaxation (decrease in the virial ratio). 
The viral ratio indicates that Type III GMCs have more gravitationally relaxed gas than Type II GMCs and that Type II GMCs have more relaxed gas than Type I GMCs. 
This evolution likely provides more gas with a high column density, which offers favorable conditions for the formation of higher--mass stars. 
The star formation efficiency is estimated by the ratio of the cluster mass to the GMC mass to be $10^3~M_\odot / 10^6~M_\odot \sim 10^{-3}$ in Type II or $10^4~M_\odot$/$10^6~M_\odot \sim 10^{-2}$ in Type III, if we assume the formation of a single cluster with mass of $10^3$ $M_\odot$ or $10^{4}$ $M_\odot$, respectively. 

Therefore, the GMC evolution is characterized by the mass increase as well as the gravitational relaxation over 10 Myr, which provides the conditions favorable to enhanced high-mass star formation with a higher star formation rate as observed. 
The evolution is well evidenced quantitatively by the relationship between \LHa, the number of O stars, and $L_\mathrm{21 \ \mu m}$, the cluster mass (Figure \ref{21um_Ha}(b)). 
The evolution is likely terminated at the end of Type III by violent GMC disruption by cluster feedback. 
Such disruptions are observed in GMCs like W3 in the Galaxy (Yamada et al. 2024), and in the LMC as those like N11 \citep{Israel2003} and N48 \citep{Fujii2014}. 
Based on the present results, we conclude that the Type classification of GMCs is a useful scheme to describe the evolutionary phases of GMCs, and provides a clear quantitative basis to pursue the GMC evolution.

\citet{Fukui2009} presented a mass growth picture of GMCs based on the accretion of \hone ~in the LMC, which was supported by theoretical studies (e.g., \cite{Inutsuka2015,Kobayashi2017}).
The accretion is consistent with the gravity of the GMCs and the \hone \ velocity dispersion \citep{Fukui2009}.
We found that the following parameters provide a possible picture of GMC growth over a time scale of approximately 10 Myr. 
In the LMC, for the GMC mass $10^5$ $M_\odot$ and radius 40 pc, the mass accretion rate $dM/dt$ required is $10^{-1}$ $M_\odot~\mathrm{yr^{-1}}$ per GMC. 
In M74, for GMC mass $10^6$ $M_\odot$ and radius 90 pc, the mass accretion rate required is 1 $M_\odot~\mathrm{yr^{-1}}$ per GMC. 
The higher mass accretion rate in M74 is consistent with the larger surface area of the GMC if the \hone ~density of M74 is higher than that of the LMC, whereas the low resolution of \hone ~mapping in M74 ($\sim 570$ pc, \cite{Walter2008}) does not allow direct testing. 
Because the star formation in Type II is limited in the surface area, the stellar feedback is not likely to limit the mass accretion. 
On the other hand, in Type III, in particular in its late phase, it is likely that the mass accretion is halted by the feedback. 
The gravitational relaxation over the GMC lifetime is driven by the energy loss of molecular cooling. 
A similar evolution of cloud cores was proposed for Ophiuchus cores by \citet{Tachihara2000,Tachihara2002}. 
Although the mass scale is much different between the GMCs and low mass cloud cores, the decay of turbulent motion in a core/cloud by molecular cooling followed by the gravitational relaxation occurs in essentially a similar fashion. 

A remaining issue is the early evolution prior to Type I.
Type I is probably preceded by a stage of small clouds of $M_{\rm CO} \sim 10^4$ $M_\odot$ similar to the Taurus complex. 
The stage may be even dominated by \hone \ gas.
These values were below the present detection limit, and we were not sensitive enough to probe this early stage. 
To reveal this stage, a detailed comprehensive study of nearby clouds, for example in the Milky Way, and higher resolution and sensitivity \hone\ observations in galaxies will be essential.
Such efforts are undertaken outside the solar circle aiming to grasp the phase earlier than Type I (e.g., Yamada et al. 2024, in prep).

\subsection{Effects of the spiral arm}
\begin{figure*}[htbp]
    \begin{center}
    \includegraphics[width=0.8\linewidth,clip]{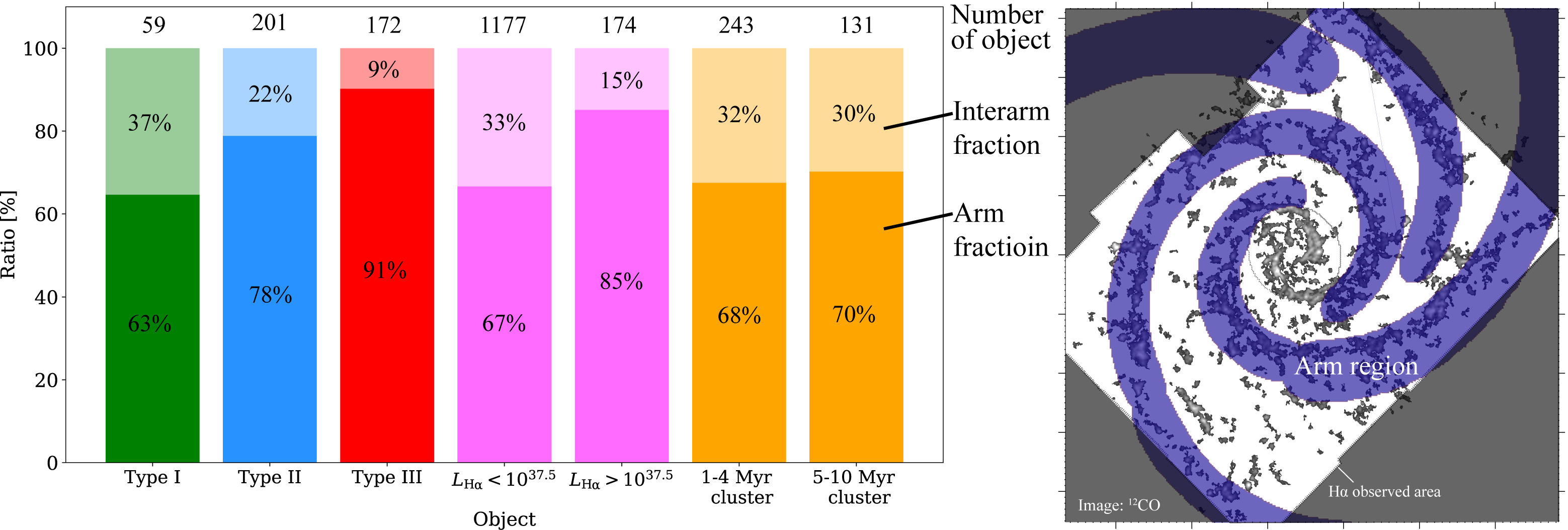}
    \end{center}
    \caption{The left panel shows a bar graph of the ratio of position of GMCs, \htwo\ regions, and clusters. The dark- and light-colored areas represent arms and interarms, respectively. The right panel shows the area of the arm region on the CO integrated intensity map.}
    \label{arm_ratio}
\end{figure*}

In the following, we focus on star formation mechanism including triggers. 
One of such triggers is driven by spiral arms \citep{Roberts1969} and the other by collisions between two clouds \citep{Fukui2021}.
In M74 the spiral arms may affect the formation of GMCs and star formation. 
\citet{Querejeta2021} identified arms and interarms by using the log-spiral pattern based on the Spitzer $3.6~\mathrm{\mu m}$ image, and we adopt the division of the galaxy into two parts. 
According to the present results, we find that the number of GMCs in the arm is 350 (81.0\%) and that in the interarm is 82 (19.0 \%), respectively, for a total GMCs of 432.
Figure \ref{arm_ratio} shows the fractions of various objects, the GMCs of each Type, and \htwo\ regions in the arm and interarm. 
It is evident that almost all objects are enhanced in arm than in the interarm. 
In particular, we find enhancements in Type II, and III GMCs and luminous \htwo\ regions in the arm.

\subsection{High mass star formation mechanism}
The transitions from Type I to Type II and from Type II to Type III occurred at 5--10 Myr by driving high-mass star formation. 
Recent investigations of high-mass star formation indicate that external triggers by cloud-cloud collisions (CCCs) or colliding flows play a role in driving high-mass star formation (see for a review \cite{Fukui2021} and references therein). 
Collisions between molecular clouds triggering high mass star/cluster formation are resolved with ALMA in more than ten regions in the LMC and M33 as well as the Antennae Galaxies, which efficiently compress gas into a small volume as an essential process for high-mass star formation (e.g., \cite{Fukui2015, Fukui2019, Saigo2017, Tokuda2019, Sano2021, Tokuda2020, Muraoka2020, Tsuge2021a,Tsuge2021b}). 
Some of these objects are resolved and physical parameters of the relevant objects are available. 
These processes have also been intensively investigated theoretically via numerical simulations of CCC hydrodynamics \citep{Inoue2013,Inoue2018,Takahira2014,Takahira2018,Haworth2015,Dobbs2015,Liow2020,Sakre2021,Fukui2021,Wu2017}.
The transition from Type I to Type II GMC probably accompanies the formation of 1--10 M42-type \htwo \ regions triggered by CCC. 
Such a CCC has a collision velocity of 10 \kms \ in the galaxy disk and can form 1--10 O stars for a compressed gas column density of $10^{23}~\mathrm{cm^{-2}}$ as found in the collisions of M42, NGC 6334, M20, and RCW120 etc. \citep{Enokiya2021}. 
In contrast, the transition from Type II to Type III requires the formation of ten times more O stars, as manifested by the extremely enhanced SFR in Type III GMCs. 
Such O star formation is achieved by a CCC having a compressed gas column density greater than $10^{23}~\mathrm{cm^{-2}}$ \citep{Enokiya2021}. 
It is probable that the advanced gravitational relaxation in Type III accompanies an increase of the gas column density beyond $10^{23}~\mathrm{cm^{-2}}$, leading to the formation of rich clusters with tens of O stars. 
The present ALMA resolution is not high enough to resolve such high column density in M74, while the situation may resemble the active high-mass star formation in W43, the mini-starburst (\cite{Kohno2021}; see also \cite{Motte2018}), or in the $\eta$ Car complex (e.g., \cite{Fujita2021,Yonekura2005}). 
The two regions of the Milky Way have $M_\mathrm{CO}$ of $10^6$--$10^7~M_\odot$ and high column density in excess of $10^{23}~\mathrm{cm^{-2}}$, consisting of several CO components that are colliding with each other over a few Myr intervals \citep{Kohno2021,Fujita2021}. 
These regions may be in a similar evolutionary stage with Type III having a high \LHa\ of $10^{38}~\ergs$ in M74. 
These two regions are characterized by multiple CCCs within the GMCs, which can facilitate CCCs more frequently than those with external clouds.

\subsection{Collision frequency}
In M74, we admit that the present resolution of 50 pc is not sufficiently high to resolve individual collision events, and we have yet to find direct evidence for CCCs. 
Instead, the collision frequency can be estimated using the volume density of the GMCs in the arms. 
In the areas of the arm and the interarm (see Figure \ref{arm_ratio}), as defined by \citet{Querejeta2021}, the surface densities of GMCs are 7.7$~\mathrm{kpc^{-2}}$ and 2.2$~\mathrm{kpc^{-2}}$, respectively. 
For a typical scale height of molecular clouds of 100 pc, the number densities of GMCs N are estimated to be 77 and 22$~\mathrm{kpc^{-3}}$, respectively. 
In addition, the median values of GMC radius R are 99 pc and 77 pc with velocity dispersion $\sigma_{v,l}$ of 14 and 14 \kms, respectively. 
If GMCs are assumed to be randomly moving in the disk at $\sigma_v=\sqrt{2}\sigma_{v,l}$, the mean free time, $1/(N\sigma_v \pi R^2)$, is estimated to be 20 Myr in the arm, and 122 Myr in the interarm. 
Table \ref{ccc_table} summarizes these parameters. 
As M74 is face-on with an inclination angle of $\ang{9;;}$, the velocity dispersion along the disk is likely to be larger than that perpendicular to the disk, suggesting that the mean free time can be even shorter. 
In addition, as shown by numerical simulations of spiral galaxies, the typical time scale of a major collision between GMCs is estimated to be $\sim 8$ Myr \citep{Dobbs2015}, which is similar to the timescale of each GMC Type. 
Further, the present sensitivity for GMCs is too low to detect smaller mass clouds less than 10$^5$ M$_{\odot}$, whereas such clouds are able to trigger CCCs (e.g., \cite{Kobayashi2018,Fukui2021}). 
CCCs involving less massive clouds of 10$^5$ M$_{\odot}$ form O star(s) if the column density is greater than 10$^{22}$ cm$^{-2}$ \citep{Enokiya2021}. 
The formation of high-mass stars triggered by CCCs occurs in mean free time and offers a reasonable explanation for the duration of each GMC Type. 
Such time-dependent high-mass star formation dramatically changed the SFR over 10 Myr for a given cloud mass and column density. 
Naturally, this breaks down the K--S law at the 100 pc scale.

\begin{table}[htbp]{
    \centering
    \tbl{The mean free time of GMCs in arm and interarm.}{
    \begin{tabularx}{0.95\linewidth}{ccccc}
        \hline
         &&& Arm & Interarm\\
        \hline
        (1) & Number density & $(\mathrm{kpc^{-3}})$ & 77 & 22 \\
        (2) & Radius & $(\mathrm{pc})$ & 99 & 77 \\
        (3) & Velocity dispersion & $(\mathrm{km~s^{-1}})$ & 14 & 14 \\
        \hline
        (4) & The mean free time & (Myr) & 20 & 122 \\
        \hline
    \end{tabularx}
    }
    \begin{tabnote}
        (1)The number density is estimated by measuring the surface number density of GMCs and assuming the scale height of the molecular clouds to be 100 pc. (2) (3)The radius and the velocity dispersion between GMCs are the median values in the arm or interarm. 
        (4) The mean free time $1/(N\sigma_v \pi R^2)$.
    \end{tabnote}
    \label{ccc_table}
    }
\end{table}

\subsection{Alternative interpretation}
\citet{Chevance2020} derived a time scale for the GMC evolution that is similar to that of F99 and K09, but there are some differences. 
First, the authors applied a method based on the breakdown of the K--S law below the kpc scale \citep{KruijssenLongmore2014,Kruijssen2018}. 
They measured the fraction of CO and H$\alpha$ within an aperture of varying a radius centered on the peak positions of CO and H$\alpha$, and obtained the fraction of pre-star formation, star-formation, and dispersed molecular gas period. 
Then, to convert these fractions to timescale, they assumed the duration of the H$\alpha$ emitting phase once the molecular gas has been dispersed is 4 Myr as timescale of dispersed molecular gas period based on a theoretical model of \htwo \ regions \citep{Haydon2020}. 
\citet{Chevance2020} showed that the pre-star formation period was as long as a few times 10 Myr, whereas the duration of star formation was short, as characterized by \htwo \ regions, resulting in rapid cloud destruction.  
The long pre-star formation duration reported by \citet{Chevance2020} is significantly different from the present results and those by F99, K09, \citet{Gratier2012,Corbelli2017}. 
Their formulation is rather complicated and it is not straight-forward to directly compare their results with those of this study. 
We infer that a possible cause of this difference is their method which does not define the GMC as a physical entity. 
If we suppose a GMC such Orion A, then the gas is dominated by no high-mass star-forming part and is slightly toward the \htwo \ region (see e.g., \cite{Nishimura2015}). 
If the GMC is defined as an entity, the age of the gas in the GMC is constrained by the age of the Orion Nebula Cluster. 
However, the age of the non-star-forming gas is not well constrained if the GMC boundary is not defined. 
This may have cased the difference in the timescales obtained using the two methods. 
We note that M74 and the other spiral galaxies in PHANGS--ALMA exhibit the same trend as LMC and M33 in terms of Type classification, which will be revealed by the resolved GMCs in future studies.  (Demachi et al. 2024 in prep.).

\subsection{Star formation vs GMC mass/column density ---Breakdown of the K--S law}
\begin{figure}[htbp]
    \begin{center}
    \includegraphics[width=0.9\linewidth,clip]{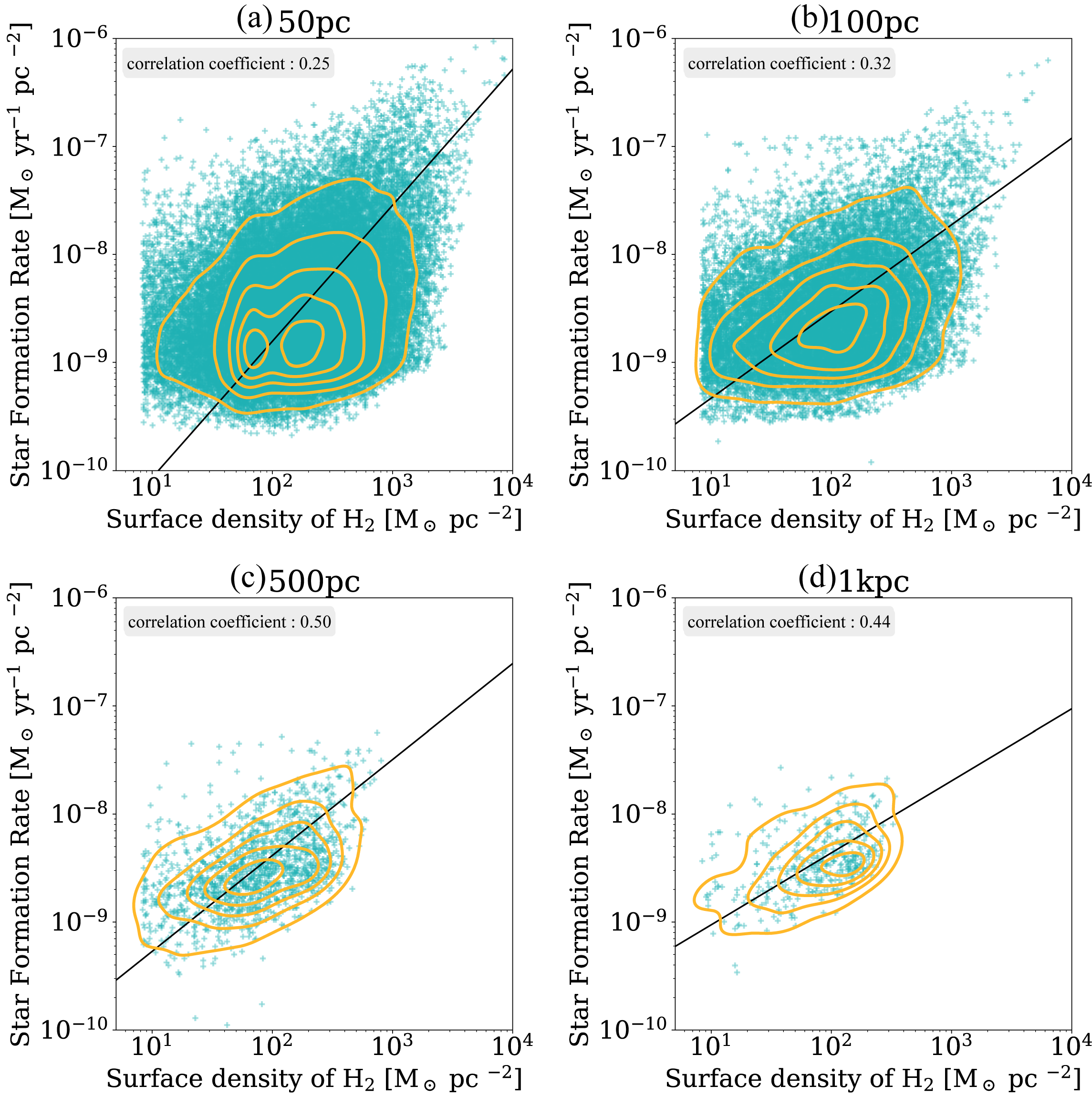}
    \end{center}
    \caption{Correlation between the surface density of $\mathrm{H_2}$ ($\Sigma_\mathrm{H_2}$) and SFR per unit area ($\Sigma_\mathrm{SFR}$) for four different resolutions: (a) 50 pc, (b) 100 pc, (c) 500 pc, and (d) 1 kpc. 
    Contours show the kernel density estimation of these plots and are plotted at every 20 \% from 10 \% of the data points. 
    Spearman’s rank correlation coefficients are (a)0.25, (b)0.32, (c)0.50, and (d)0.44. 
    The black lines indicate the regression line obtained by the least squares method for results with larger correlation coefficients; (c) $\log \Sigma_{\mathrm{SFR}} \propto 0.89 \log \Sigma_\mathrm{H_2}$ and
    (d) $\log \Sigma_{\mathrm{SFR}} \propto 0.67 \log \Sigma_\mathrm{H_2}$.}
    \label{KSlaw}
\end{figure}

In Figure \ref{KSlaw} we show SFR in M74, estimated by [$L_{\mathrm{H\alpha}}$] /$[\mathrm{gas\ column\ density}]$ for high mass stars capable of ionizing \htwo ~regions, as a function of gas column density for four different resolutions from 50 pc to 1 kpc. 
The SFR shows a large dispersion with column density at a 50--100 pc scale, whereas the dispersion is smeared out at a 1 kpc scale. 
The kpc-scale smoothed property results in the K--S law, which states that the SFR in galaxies is simply a function of the column density. 
The law breaks down into resolved GMCs at 50--100 pc, as first demonstrated by the divergent SFR depending on the GMC Type in the LMC (F99, \cite{Yamaguchi2001}, and K09), which was confirmed in M33 \citep{Gratier2012,Onodera2010,Schruba2010}, NGC 1614 \citep{Xu2015}, and M74 \citep{Kreckel2018}.
The present results support the breakdown based on GMC Types with significantly different SFRs in a grand design spiral.

\section{Conclusions}\label{conclusions}
To investigate the mechanism of star formation in GMCs, we investigated GMCs in the disk of a grand-design spiral NGC~628 (M74) among the PHANGS galaxies in comparison with \htwo\ regions and stellar clusters.
In the present study, we followed the Type classification of GMCs in the Large Magellanic Cloud (F99; K09) using \htwo \ regions and stellar clusters as measures of high-mass star formation, provided that the present approach used first the \htwo \ regions only in classification, followed by comparative detailed investigations of the associate clusters by HST as well as the recent JWST data.
The main conclusions derived from the results of this study are summarized below:

\begin{enumerate}
    \item We identified 432 GMCs in M74 in the observed field of the PHANGS--ALMA survey by using the PYCPROPS algorithm at resolution of 53 pc.
          Also, we have identified 1351 \htwo\ regions based on the H$\alpha$ image data by using Astrodendro algorithm.
          The identified GMCs and \htwo\ regions were cross-correlated with each other, and the GMCs were classified into the three GMC Types as follows:
          59 Type I, 201 Type II, and 172 Type III, where Type I has no \htwo\ region, Type II is associated with lower luminosity ($\LHa < 10^{37.5}~\ergs$) \htwo\ regions, and Type III with higher luminosity ($\LHa > 10^{37.5}~\ergs$) \htwo\ regions.
          The trends for each GMC Type are as follows:
          Type I GMCs have small mass, small radius, and large viral ratio.
          Type II GMCs have intermediate mass, radius, and virial ratio.
          Type III GMCs have large mass and radius and small virial ratio.
          The velocity dispersion shows only a small variation according to the GMC Type.

    \item The \htwo\ regions nearest to the GMC show strong concentrations within 150 pc of the GMCs, indicating that the \htwo\ regions are tightly associated with the GMCs.
          Clusters having aged 1--4 Myr also show strong concentrations within 150 pc of the nearest GMCs, whereas clusters older than 5 Myr show no significant association with GMCs.
          1--4 Myr clusters are associated with $\sim 20\%$ of Type II GMCs and with $\sim 50\%$ of Type III GMCs.
          This indicates that the present Type classification based on only \LHa\ is fairly consistent with the original Type definition in the LMC which used both \htwo\ regions and young clusters as the Type signatures.
          The cluster mass associated with the GMCs is roughly proportional to \LHa.
          Clusters associated with Type II is low mass ($\sim 10^3~M_\odot$) and low \LHa\ ($\sim 10^{37}~\ergs$), while those associated with Type III is high mass ($\sim 10^4~M_\odot$) and high \LHa\ ($\sim 10^{38}~\ergs$).

    \item A detailed comparison between the H$\alpha$ leaves toward CO emission and those without CO emission indicates that the CO luminosities show no significant decrease toward the CO emission.
          This indicates that the extinction by the CO gas is not significant in the present GMCs.
          This is likely caused by the filamentary and/or clumpy CO distributions along with the spatially extended Champaign flows of \htwo \ regions, which make the extinction of the H$\alpha$ emission unimportant at a 50 pc scale.

    \item The virial mass was compared to the luminous mass of GMCs.
          We found the following trends:
          Type I GMCs significantly deviated from virial equilibrium,
          Type III GMCs were in the most virialized state,
          and Type II GMCs were located in between.
          Accordingly, we suggest that the GMC Type classification corresponds to a sequence of gravitational relaxation from Type I to Type III.
          This is consistent with the evolutionary sequence for a typical relaxation time of approximately 10 Myr.
          The relaxation is driven by the mass increase of GMCs via \hone \ accretion by the cloud gravity followed by the internal energy release by molecular cooling.

    \item 30\% of Type I and more than 80\% of Type II and III GMCs are associated with 21 $\mathrm{\mu m}$ emissions, which are considered to trace hot dust heated by UV light from massive stars and reradiated in the mind-IR band in \htwo \ regions.
          H$\alpha$ luminosity and cluster mass, and 21 $\mathrm{\mu m}$ luminosity and cluster mass associated with a common GMC indicate a positive correlation between them.
          We interpreted that H$\alpha$ luminosity correspond the number of O stars and total mass of clusters formed in the GMC.
          Nearly 30\% of Type I GMCs are associated with 21 $\mathrm{\mu m}$, and they are only massive Type I GMCs.
          The remaining of Type I GMCs without 21 $\mathrm{\mu m}$ leaves are less massive.
          From these results, we suggest that transient GMCs near the boundary between Type I and Type II have the 21 $\mathrm{\mu m}$ leaves, which indicate clusters of $10^3 \ M_\odot$.
          Some of these clusters have \LHa \ detectable classified as Type II, while the rest have low \LHa \ under the detection limit classified as Type I: the difference is probably caused by the small number of O stars in the clusters, which fluctuates below or above the H$\alpha$ detection limit.

    \item The time scales of each GMC Type can be estimated based on the number of GMC Types, if a steady state in the GMC evolution is assumed.
          By adopting the timescale of the associated young clusters, $<4$ Myr, we estimated the timescales of Types I, II, and III to be 1, 5, and 4 Myr, respectively.
          This yields a total GMC lifetime of 10 Myr.
          On the other hand, the timescale of Type II GMCs in the LMC was estimated to be 10 Myr, which is probably not significantly affected by the GMC mass detection limit compared to Type I or Type III.
          An alternative time scale estimate is possible by assuming a timescale of 10 Myr for Type II in M74, the same with that in the LMC.
          The timescales may then become 3, 12, and 10 Myr, for each Type, resulting in a lifetime of approximately 25 Myr.
          We suggest that this may be possible if the detection limit of clusters is extended to lower mass clusters in older age.
          In summary, we suggest that the GMC lifetime in M74 is approximately 10--20 Myr, which is consistent with that in the LMC and M33.
          This is important because the present work is the first which derived the GMC lifetime in a grand design spiral to be consistent with the previous results on dwarfs. \label{conc_timescale}

    \item The GMC evolution described in conclusions \ref{conc_timescale} is able to provide the conditions favorable for high mass star formation, but may not be sufficient as an explanation of the GMC evolution which is characterized by the level of high-mass star formation activity.
          High mass star formation requires triggers of high mass star formation including cloud-cloud collisions according to the previous works on high mass star formation in the Galaxy as well as the Local Group galaxies and the Antennae Galaxies.
          We argue that the GMC--GMC collision frequency is high enough to drive the Type transitions in the Arm where the number density GMCs is enhanced.
          In addition, it is probable that the internal collisions within a Type III GMC, as well as the collisions by the clouds smaller than $10^5~M_\odot$, are important as the more efficient trigger of high mass star formation.
          These triggers are likely an important driver of the active high-mass star formation toward Type III.

    \item The K--S law states that the SFR is proportional to the [column density]$^{1.4}$ in galaxies at a kpc scale.
          The present work shows that the law breaks down at a GMC scale of $<100$ pc, where the star formation is driven by stochastic events triggered by cloud-cloud collisions.
          Star formation by the triggers appears according to the K--S law when averaged over kpc scales.

\end{enumerate}

\begin{ack}
    We would like to thank the anonymous referee for valuable comments that improved the manuscript.
    We are grateful to M\'{e}lanie Chevance for her helpful comments on the earlier version of the manuscript.
    This paper makes use of the following ALMA data:
    ADS/JAO.ALMA\#2012.1.00650.S.
    ALMA is a partnership of ESO (representing its member states), NSF (USA) and NINS
    (Japan), together with NRC (Canada), MOST and ASIAA (Taiwan), and KASI (Republic
    of Korea), in cooperation with the Republic of Chile. The Joint ALMA Observatory is
    operated by ESO, AUI/NRAO and NAOJ. The National Radio Astronomy Observatory is
    a facility of the National Science Foundation operated under cooperative agreement by
    Associated Universities, Inc.
    Based on observations taken as part of the PHANGS-MUSE large program (Emsellem et al. 2021).
    Based on data products created from observations collected at the European Organisation for
    Astronomical Research in the Southern Hemisphere under ESO programme(s) 1100.B-0651, 095.C-0473,
    and 094.C-0623 (PHANGS-MUSE; PI Schinnerer), as well as 094.B-0321 (MAGNUM; PI Marconi),
    099.B-0242, 0100.B-0116, 098.B-0551 (MAD; PI Carollo) and 097.B-0640 (TIMER; PI Gadotti).
    This research has made use of the services of the ESO Science Archive Facility.
    This paper makes use of the following JWST data: Project code: 02107, PI: J. Lee.
    This work was supported by Grants-in-Aid for Scientific Research (KAKENHI) of Japan Society for the Promotion of Science (JSPS; grant Nos. JP18H05440, JP20H01945, JP20H04739, JP21H00049, JP21H00040, JP21K13962, JP22H00152, JP22K14080, JP22J22931).
    K. Tokuda acknowledges support from the NAOJ ALMA Scientific Research (grant No. 2022-22B).
    F. Demachi was supported by an ALMA Japan Research Grant of NAOJ ALMA Project, NAOJ-ALMA310.
    This research made use of astrodendro, a Python package to compute dendrograms of Astronomical data (\url{http://www.dendrograms.org/)}
    This work made use of Astropy:\footnote{http://www.astropy.org} a community-developed core Python package and an ecosystem of tools and resources for astronomy \citep{astropy:2013, astropy:2018, astropy:2022}.
    We would like to thank Editage (\url{www.editage.jp}) for English language editing.

\end{ack}

\appendix
\section{GMC Identification}\label{cprops}
Subsequently, \texttt{PYCPROPS} identifies a local maximum and associates the surrounding emissions with the local peak.
\begin{enumerate}
    \item Define leaf by applying \texttt{Astrodendro} (see section \ref{analysisHII}) to PPV datacube, and use ``leaf'' (see for details section \ref{analysisHII}) as a candidate of the local maximum. Call the peak intensity of the leaf $T_\mathrm{max}$.
    \item Define the merge level ($T_\mathrm{merge}$) as the maximum level including one $T_\mathrm{max}$ and the adjoining leaves, and exclude it from candidates if $T_\mathrm{max} - T_\mathrm{merge} < \delta$ ($\delta$ is input marapeter)
    \item Estimate $n_\mathrm{pix}$, the number of pixels greater than $T_\mathrm{merge}$ in the spatial direction.
          Leave it if $n_\mathrm{pix} > min_\mathrm{pix}$ ($min_\mathrm{pix}$ is the input parameter) and exclude it from the candidates if $n_\mathrm{pix} < min_\mathrm{pix}$. Exclude leaf if the distribution smaller than the beam size by taking $n_\mathrm{pix}$ to be of the same size as the beam.
    \item Leaves were excluded if it is not separated by $d_\mathrm{min}$ in space or by $v_\mathrm{min}$ in velocity ($d_\mathrm{min}$ and $v_\mathrm{min}$ are input parameters).
    \item If two leaves in the candidate that satisfy the above criteria are combined and the flux and size are $s$ times different from those of the original reef, each leaf is a separate structure ($s$ is input the parameter).
    \item We adopt the leaf that satisfies the above five criteria as the local maximum. In the end, allocate the pixels around the local maximum to each peak by using the statistical method watershed based on ``compactness parameter’’ and determine the boundary of a GMC.
\end{enumerate}

The boundary of a GMC is affected by the parameters, in particular, by $\delta$ and the compactness parameter.
$\delta$ is a parameter that specifies the fineness of the division, and the compactness parameter is related to the smoothness of the GMC boundary.
In the present study, we tuned these two parameters to realize the natural definition of a GMC.
The other parameters are fixed as follows, $s=0$, $min_\mathrm{pix} = 5$, $d_\mathrm{min} = 10$, $v_\mathrm{min} = 1$.

\begin{figure}[htbp]
    \begin{center}
        \includegraphics[width=0.9\linewidth,clip]{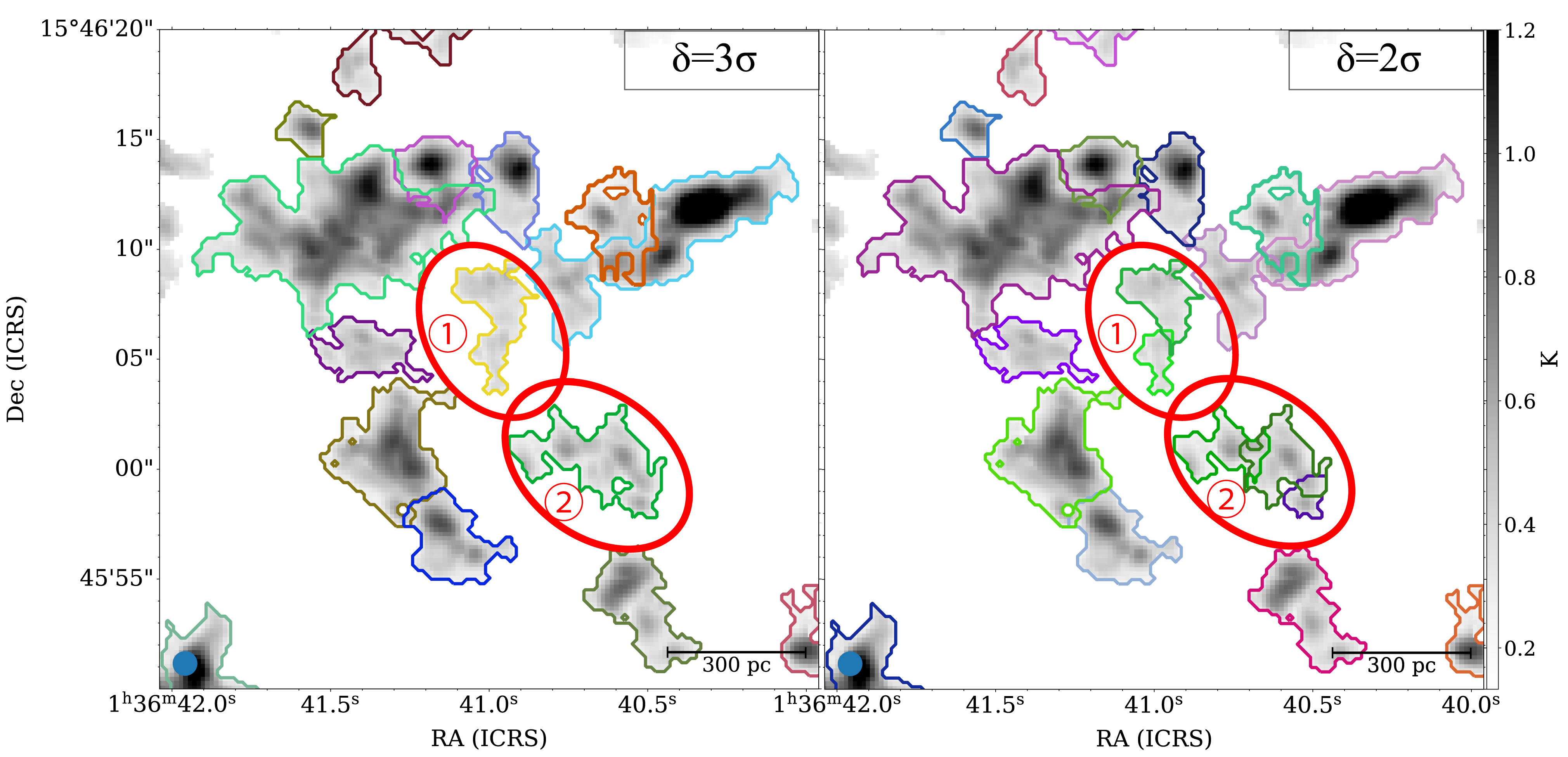}
    \end{center}
    \caption{Comparison of identified GMCs between $\delta = 3\sigma$ (left) and $\delta = 2\sigma$ (right). GMCs are shown by color contour on the close-up view of \twelvecoh \ peak intensity map. The red lines indicate the GMCs changed by $\delta$.}
    \label{zoom_delta}
\end{figure}

Figure \ref{zoom_delta} shows a GMC at $\delta = 2\sigma$ and $\delta = 3\sigma$.
The GMC at $\delta = 3\sigma$ is divided into too small pieces like 1 and 2 in the Figure.
In the Milky Way, W43 and $\eta$ Car complex are identified as a single GMC at 50 pc resolution, while they consist of several components less than a few 10 pc pieces.
It is therefore realistic not to divide a GMC into too small structures, and we adopted $\delta = 3\sigma$.

\begin{figure}[htbp]
    \begin{center}
        \includegraphics[width=0.9\linewidth,clip]{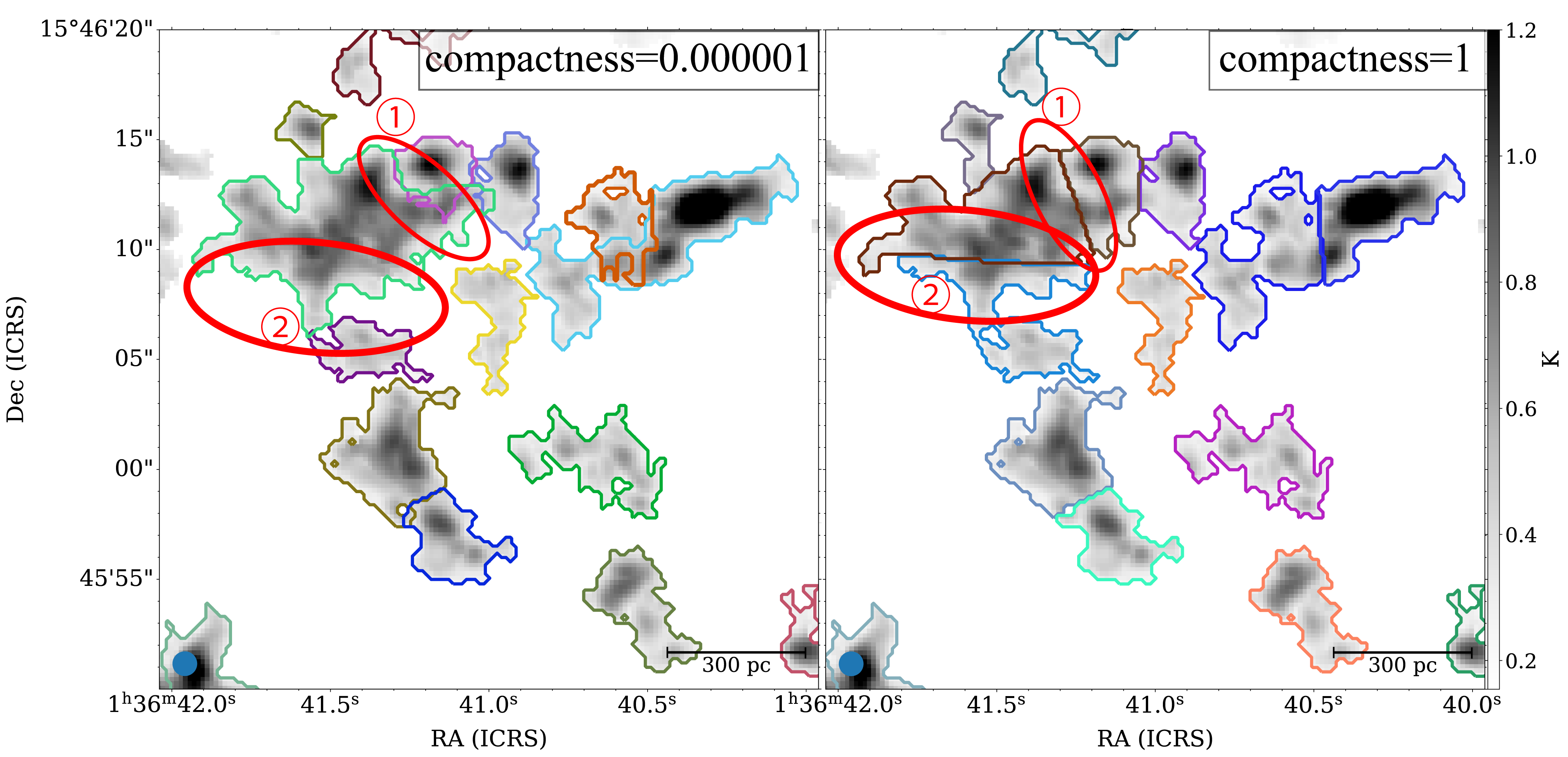}
    \end{center}
    \caption{Same as Figure \ref{zoom_delta}, but Comparison of identified GMCs between compactness $= 0.000001$ (left) and compactness $= 1$ (right).}
    \label{zoom_compactness}
\end{figure}

Compactness is defined as $4\pi \times$ area / $(\mathrm{size})^2$ and the GMC boundary is optimized to approach the specified compactness.
Figure \ref{zoom_compactness} shows GMCs with a compactness parameter between 0.00001 and 1.
This indicates that if the compactness parameter $= 1$, the GMC is divided into a straight line.
When the boundary line is complicated, compactness decreases.
Therefore, we adopt a compactness $= 0.00001$ to obtain the natural GMC boundary.

\section{Classification for Type GMC}\label{classification_apd}
As shown in section \ref{association}, we tested the association of a GMC with \htwo \ regions and found that the association is doubtful for 16 GMCs among the 432 GMCs.
Therefore, we corrected the association by eye inspection.

\begin{figure}[htbp]
    \begin{center}
        \includegraphics[width=0.7\linewidth,clip]{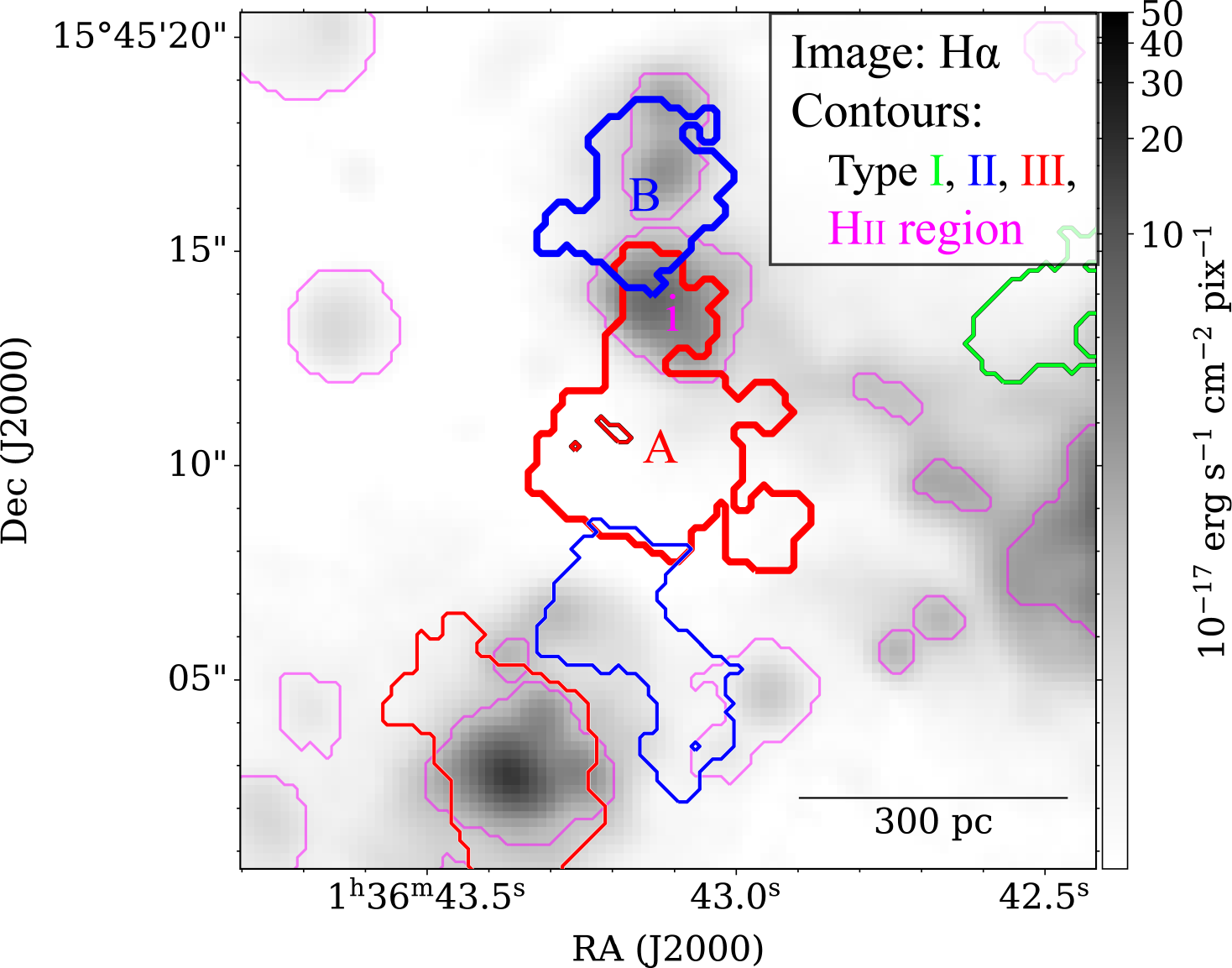}
    \end{center}
    \caption{The example of corrected association between GMCs and \htwo \ regions by eye inspection. Type I, Type II, Type III, and \htwo \ regions are shown by green, blue, red, and magenta, respectively on Ha image. The GMC A and B represented by bold lines are corrected in their association with \htwo \ region i.}
    \label{example_I_III}
\end{figure}

GMCs A and B in Figure \ref{example_I_III} are classified as Type I and Type III GMCs because the \htwo \ region i overlapped with both GMCs A and B, and GMC B is closer to the \htwo \ region.
However, by visual inspection, we recognized that the \htwo \ region i was clearly associated with GMC A, and we classified GMC A as Type III and GMC B as Type II.
There are three additional cases with similar reclassifications.

\begin{figure}[htbp]
    \begin{center}
        \includegraphics[width=0.7\linewidth,clip]{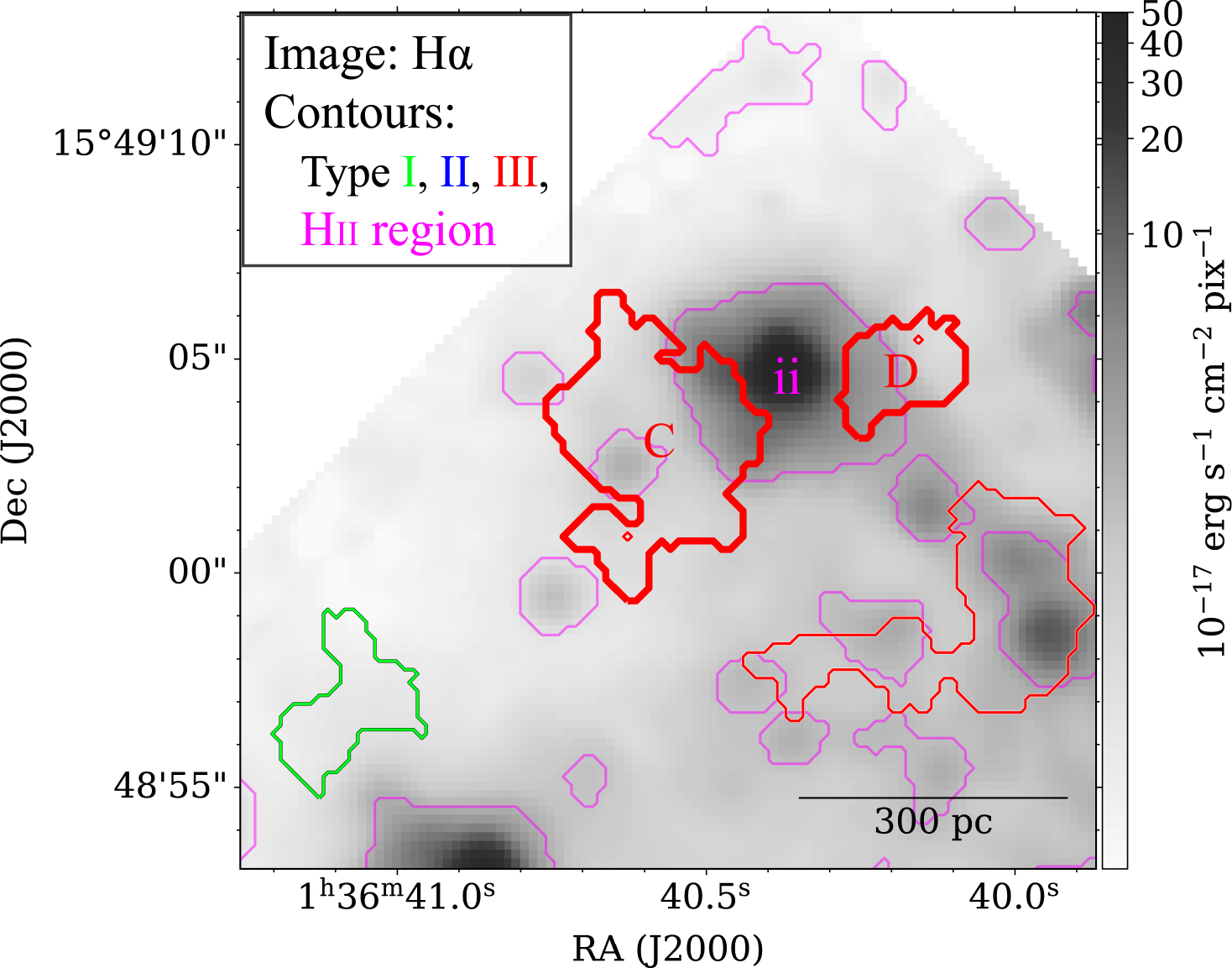}
    \end{center}
    \caption{Same as Figure \ref{example_I_III}, but in the case of GMC C and D and \htwo \ region ii.}
    \label{example_II_III}
\end{figure}

In Figure \ref{example_II_III}, GMCs C and D are classified as Type II and III, respectively.
This is because \htwo \ region ii overlaps with the two GMCs, but GMC D is closer to the \htwo \ region than GMC C, so \htwo \ region ii is determined to be associated with GMC D.
However, \htwo \ region ii is located at a similar distance from GMCs C and D, and we determined that the \htwo \ region was associated with both GMCs and classified both GMCs as Type III.
In this case, GMCs C and D were initially a single GMC, and the formed O stars ionized the GMC, resulting in two GMCs C and D.
Thirteen similar cases were identified.

\section{Examples of CO cloud overlapped with H$\alpha$}\label{ex_ovl}
\begin{figure}[htbp]
    \begin{center}
        \includegraphics[width=0.95\linewidth,clip]{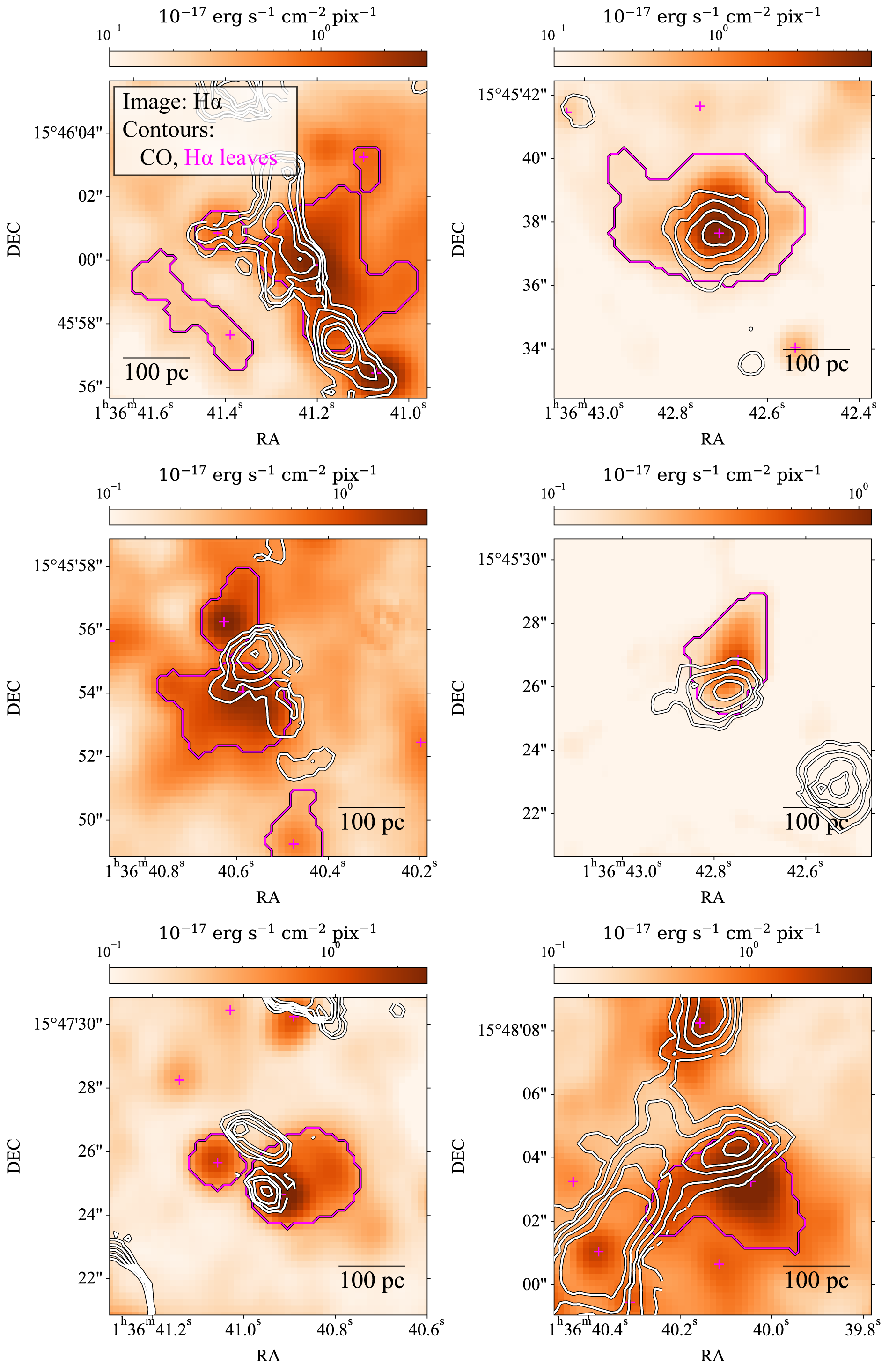}
    \end{center}
    \caption{The example of CO overlapped with H$\alpha$. CO integrated intensity map and H$\alpha$ leaves are shown by white and magenta contours, respectively on H$\alpha$ images.}
    \label{example_ovl}
\end{figure}

As shown in section \ref{association}, H$\alpha$ distribution is not heavily affected by extinction in the GMCs.
Some examples of CO clouds overlapped with H$\alpha$ are shown in Figure \ref{example_ovl}.
These Figures confirm no significant H$\alpha$ extinction toward CO.

\section{Spatial correlation between GMCs and star clusters by mass and age}\label{appendix_separation}
\begin{figure*}[htbp]
    \begin{center}
        \includegraphics[width=0.9\linewidth,clip]{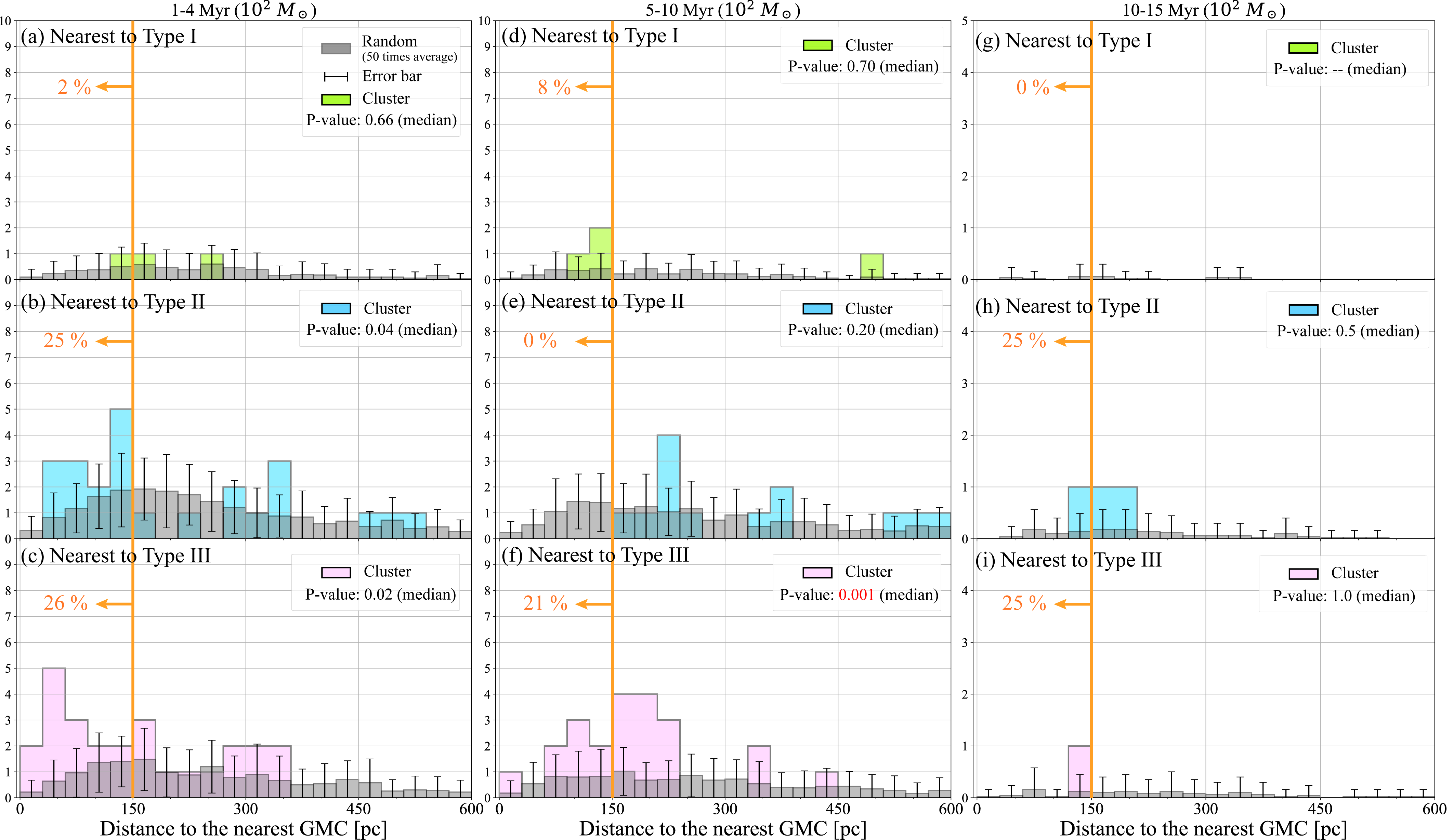}
    \end{center}
    \caption{Same as Figure \ref{dist_from_GMC}, but in the case of $10^{2-3} \ M_\odot$ clusters of (left) 1--4, (center) 5--10, and (right) 10--15 Myr.}
    \label{dist_from_GMC_1e2}
\end{figure*}

\begin{figure*}[htbp]
    \begin{center}
        \includegraphics[width=0.9\linewidth,clip]{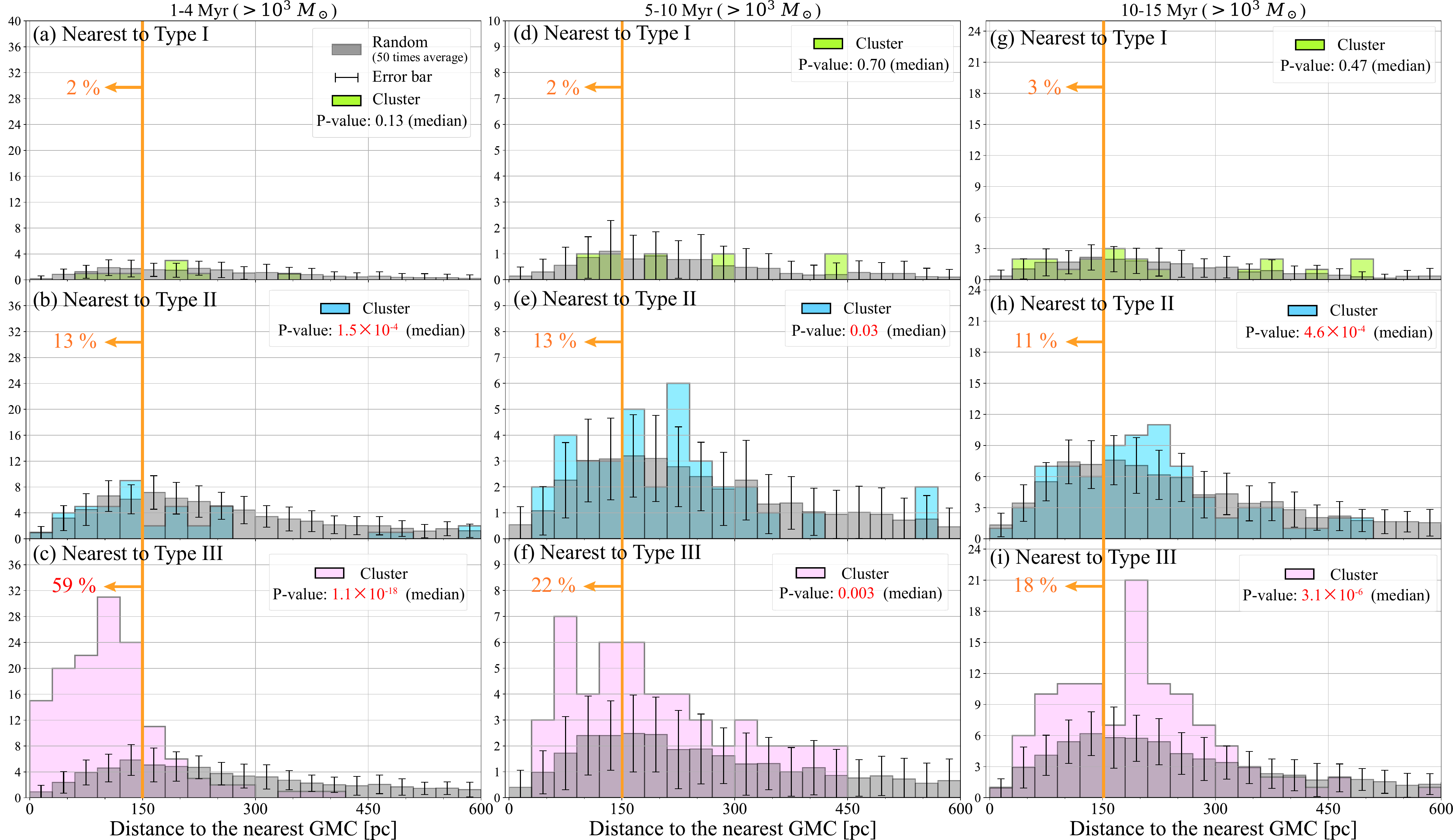}
    \end{center}
    \caption{Same as Figure \ref{dist_from_GMC}, but in the case of $10^{3-4} \ M_\odot$ clusters of (left) 1--4, (center) 5--10, and (right) 10--15 Myr.}
    \label{dist_from_GMC_1e3}
\end{figure*}

We performed the same analysis in Section \ref{sec_association} in more detail for clusters of $10^{2-3} \ M_\odot$ and $10^{3-4} \ M_\odot$.
Figures \ref{dist_from_GMC_1e2} and \ref{dist_from_GMC_1e3} show the results of 1--4 Myr, 5--10 Myr, 11--15 Myr.
Only 1--4 Myr clusters located within 150 pc from Type II and III GMCs, and we determine their distributions are not generated from the same probability distributions with random distributions because their P-value is smaller than 0.05.
Especially, 60\% of the clusters of $10^3 \ M_\odot$ are associated with Type III GMCs.
On the other hand, $>$5 Myr clusters have P-values smaller than 0.05, but less concentration on GMCs.
From these results, young and massive clusters are associated with GMCs, and older clusters are separated from GMCs whichever at any mass.

\begin{table*}[htbp]{
        \centering
        \tbl{Association of clusters of $10^{2-3} \ M_\odot$ and GMCs.}{
            \begin{tabularx}{0.85\linewidth}{cclccc}
                \hline
                    &                    &                                                                & Type I       & Type II      & Type III      \\
                \hline
                (1) & 1--4 Myr cluster   & $N_\mathrm{cluster}^*$                                         & 3            & 24           & 26            \\
                    &                    & $N_\mathrm{<150pc}^\dagger$ (fraction)                         & 1 (2\%)      & 13 (25\%)    & 14 (26\%)     \\
                    &                    & \begin{tabular}{l} Median of P-value $^\ddagger$ \end{tabular} & 0.66         & 0.04         & 0.02          \\
                    &                    & \begin{tabular}{l} (IQRs) \end{tabular}                        & (0.31--0.71) & (0.02--0.13) & (0.002--0.05) \\
                \hline
                (2) & 5--10 Myr cluster  & $N_\mathrm{cluster}^*$                                         & 4            & 13           & 22            \\
                    &                    & $N_\mathrm{<150pc}^\dagger$ (fraction)                         & 3 (8\%)      & 0 (0\%)      & 8 (21\%)      \\
                    &                    & \begin{tabular}{l} Median of P-value $^\ddagger$ \end{tabular} & 0.70         & 0.20         & 0.001         \\
                    &                    & \begin{tabular}{l} (IQRs) \end{tabular}                        & (0.24--0.82) & (0.12--0.36) & (0.002--0.05) \\
                \hline
                (3) & 10--15 Myr cluster & $N_\mathrm{cluster}^*$                                         & 0            & 3            & 1             \\
                    &                    & $N_\mathrm{<150pc}^\dagger$ (fraction)                         & 0 (0\%)      & 1 (25\%)     & 1 (25\%)      \\
                    &                    & \begin{tabular}{l} Median of P-value $^\ddagger$ \end{tabular} & --           & 0.5          & 1.0           \\
                    &                    & \begin{tabular}{l} (IQRs) \end{tabular}                        & --           & (0.30--0.90) & (0.67--1.0)   \\
                \hline
            \end{tabularx}}
        \begin{tabnote}
            Same as Table \ref{cluster_table}, but in the case of $10^{2-3} \ M_\odot$ clusters.
        \end{tabnote}
        \label{cluster_table_1e2Mo}}
\end{table*}

\begin{table*}[htbp]{
        \centering
        \tbl{Association of clusters of $10^{3-4} \ M_\odot$ and GMCs.}{
            \begin{tabularx}{0.85\linewidth}{cclccc}
                \hline
                    &                    &                                                                & Type I       & Type II                    & Type III                   \\
                \hline
                (1) & 1--4 Myr cluster   & $N_\mathrm{cluster}^*$                                         & 8            & 43                         & 139                        \\
                    &                    & $N_\mathrm{<150pc}^\dagger$ (fraction)                         & 3 (2\%)      & 24 (13\%)                  & 112 (59\%)                 \\
                    &                    & \begin{tabular}{l} Median of P-value $^\ddagger$ \end{tabular} & 0.13         & $10^{-3.8}$                & $10^{-18}$                 \\
                    &                    & \begin{tabular}{l} (IQRs) \end{tabular}                        & (0.05--0.22) & ($10^{-4.5}$--$10^{-3.3}$) & ($10^{-21}$--$10^{-16}$)   \\
                \hline
                (2) & 5--10 Myr cluster  & $N_\mathrm{cluster}^*$                                         & 5            & 37                         & 50                         \\
                    &                    & $N_\mathrm{<150pc}^\dagger$ (fraction)                         & 2 (2\%)      & 12 (13\%)                  & 20 (22\%)                  \\
                    &                    & \begin{tabular}{l} Median of P-value $^\ddagger$ \end{tabular} & 0.70         & 0.03                       & 0.003                      \\
                    &                    & \begin{tabular}{l} (IQRs) \end{tabular}                        & (0.56--0.91) & (0.004--0.07)              & ($10^{-3.3}$--0.01)        \\
                \hline
                (3) & 10--15 Myr cluster & $N_\mathrm{cluster}^*$                                         & 0            & 3                          & 1                          \\
                    &                    & $N_\mathrm{<150pc}^\dagger$ (fraction)                         & 7 (3\%)      & 24 (11\%)                  & 39 (18\%)                  \\
                    &                    & \begin{tabular}{l} Median of P-value $^\ddagger$ \end{tabular} & 0.47         & $10^{-3.3}$                & $10^{-5.5}$                \\
                    &                    & \begin{tabular}{l} (IQRs) \end{tabular}                        & (0.27--0.68) & ($10^{-4.0}$--0.002)       & ($10^{-7.1}$--$10^{-4.1}$) \\
                \hline
            \end{tabularx}}
        \begin{tabnote}
            Same as Table \ref{cluster_table}, but in the case of $10^{3-4} \ M_\odot$ clusters.
        \end{tabnote}
        \label{cluster_table_1e3Mo}}
\end{table*}

\bibliographystyle{apj}
\bibliography{M74_reference}

\begin{thebibliography}{}
\expandafter\ifx\csname natexlab\endcsname\relax\def\natexlab#1{#1}\fi

\bibitem[{{Adamo} {et~al.}(2017){Adamo}, {Ryon}, {Messa}, {Kim}, {Grasha}, {Cook}, {Calzetti}, {Lee}, {Whitmore}, {Elmegreen}, {Ubeda}, {Smith}, {Bright}, {Runnholm}, {Andrews}, {Fumagalli}, {Gouliermis}, {Kahre}, {Nair}, {Thilker}, {Walterbos}, {Wofford}, {Aloisi}, {Ashworth}, {Brown}, {Chandar}, {Christian}, {Cignoni}, {Clayton}, {Dale}, {de Mink}, {Dobbs}, {Elmegreen}, {Evans}, {Gallagher}, {Grebel}, {Herrero}, {Hunter}, {Johnson}, {Kennicutt}, {Krumholz}, {Lennon}, {Levay}, {Martin}, {Nota}, {{\"O}stlin}, {Pellerin}, {Prieto}, {Regan}, {Sabbi}, {Sacchi}, {Schaerer}, {Schiminovich}, {Shabani}, {Tosi}, {Van Dyk}, \& {Zackrisson}}]{Adamo2017}
{Adamo}, A., {Ryon}, J.~E., {Messa}, M., {et~al.} 2017, \apj, 841, 131

\bibitem[{{Anand} {et~al.}(2021){Anand}, {Lee}, {Van Dyk}, {Leroy}, {Rosolowsky}, {Schinnerer}, {Larson}, {Kourkchi}, {Kreckel}, {Scheuermann}, {Rizzi}, {Thilker}, {Tully}, {Bigiel}, {Blanc}, {Boquien}, {Chandar}, {Dale}, {Emsellem}, {Deger}, {Glover}, {Grasha}, {Groves}, {S. Klessen}, {Kruijssen}, {Querejeta}, {S{\'a}nchez-Bl{\'a}zquez}, {Schruba}, {Turner}, {Ubeda}, {Williams}, \& {Whitmore}}]{Anand2021}
{Anand}, G.~S., {Lee}, J.~C., {Van Dyk}, S.~D., {et~al.} 2021, \mnras, 501, 3621

\bibitem[{{Astropy Collaboration} {et~al.}(2013){Astropy Collaboration}, {Robitaille}, {Tollerud}, {Greenfield}, {Droettboom}, {Bray}, {Aldcroft}, {Davis}, {Ginsburg}, {Price-Whelan}, {Kerzendorf}, {Conley}, {Crighton}, {Barbary}, {Muna}, {Ferguson}, {Grollier}, {Parikh}, {Nair}, {Unther}, {Deil}, {Woillez}, {Conseil}, {Kramer}, {Turner}, {Singer}, {Fox}, {Weaver}, {Zabalza}, {Edwards}, {Azalee Bostroem}, {Burke}, {Casey}, {Crawford}, {Dencheva}, {Ely}, {Jenness}, {Labrie}, {Lim}, {Pierfederici}, {Pontzen}, {Ptak}, {Refsdal}, {Servillat}, \& {Streicher}}]{astropy:2013}
{Astropy Collaboration}, {Robitaille}, T.~P., {Tollerud}, E.~J., {et~al.} 2013, \aap, 558, A33

\bibitem[{{Astropy Collaboration} {et~al.}(2018){Astropy Collaboration}, {Price-Whelan}, {Sip{\H{o}}cz}, {G{\"u}nther}, {Lim}, {Crawford}, {Conseil}, {Shupe}, {Craig}, {Dencheva}, {Ginsburg}, {Vand erPlas}, {Bradley}, {P{\'e}rez-Su{\'a}rez}, {de Val-Borro}, {Aldcroft}, {Cruz}, {Robitaille}, {Tollerud}, {Ardelean}, {Babej}, {Bach}, {Bachetti}, {Bakanov}, {Bamford}, {Barentsen}, {Barmby}, {Baumbach}, {Berry}, {Biscani}, {Boquien}, {Bostroem}, {Bouma}, {Brammer}, {Bray}, {Breytenbach}, {Buddelmeijer}, {Burke}, {Calderone}, {Cano Rodr{\'\i}guez}, {Cara}, {Cardoso}, {Cheedella}, {Copin}, {Corrales}, {Crichton}, {D'Avella}, {Deil}, {Depagne}, {Dietrich}, {Donath}, {Droettboom}, {Earl}, {Erben}, {Fabbro}, {Ferreira}, {Finethy}, {Fox}, {Garrison}, {Gibbons}, {Goldstein}, {Gommers}, {Greco}, {Greenfield}, {Groener}, {Grollier}, {Hagen}, {Hirst}, {Homeier}, {Horton}, {Hosseinzadeh}, {Hu}, {Hunkeler}, {Ivezi{\'c}}, {Jain}, {Jenness}, {Kanarek}, {Kendrew}, {Kern}, {Kerzendorf}, {Khvalko}, {King}, {Kirkby}, {Kulkarni}, {Kumar}, {Lee}, {Lenz}, {Littlefair}, {Ma}, {Macleod}, {Mastropietro}, {McCully}, {Montagnac}, {Morris}, {Mueller}, {Mumford}, {Muna}, {Murphy}, {Nelson}, {Nguyen}, {Ninan}, {N{\"o}the}, {Ogaz}, {Oh}, {Parejko}, {Parley}, {Pascual}, {Patil}, {Patil}, {Plunkett}, {Prochaska}, {Rastogi}, {Reddy Janga}, {Sabater}, {Sakurikar}, {Seifert}, {Sherbert}, {Sherwood-Taylor}, {Shih}, {Sick}, {Silbiger}, {Singanamalla}, {Singer}, {Sladen}, {Sooley}, {Sornarajah}, {Streicher}, {Teuben}, {Thomas}, {Tremblay}, {Turner}, {Terr{\'o}n}, {van Kerkwijk}, {de la Vega}, {Watkins}, {Weaver}, {Whitmore}, {Woillez}, {Zabalza}, \& {Astropy Contributors}}]{astropy:2018}
{Astropy Collaboration}, {Price-Whelan}, A.~M., {Sip{\H{o}}cz}, B.~M., {et~al.} 2018, \aj, 156, 123

\bibitem[{{Astropy Collaboration} {et~al.}(2022){Astropy Collaboration}, {Price-Whelan}, {Lim}, {Earl}, {Starkman}, {Bradley}, {Shupe}, {Patil}, {Corrales}, {Brasseur}, {N{"o}the}, {Donath}, {Tollerud}, {Morris}, {Ginsburg}, {Vaher}, {Weaver}, {Tocknell}, {Jamieson}, {van Kerkwijk}, {Robitaille}, {Merry}, {Bachetti}, {G{"u}nther}, {Aldcroft}, {Alvarado-Montes}, {Archibald}, {B{'o}di}, {Bapat}, {Barentsen}, {Baz{'a}n}, {Biswas}, {Boquien}, {Burke}, {Cara}, {Cara}, {Conroy}, {Conseil}, {Craig}, {Cross}, {Cruz}, {D'Eugenio}, {Dencheva}, {Devillepoix}, {Dietrich}, {Eigenbrot}, {Erben}, {Ferreira}, {Foreman-Mackey}, {Fox}, {Freij}, {Garg}, {Geda}, {Glattly}, {Gondhalekar}, {Gordon}, {Grant}, {Greenfield}, {Groener}, {Guest}, {Gurovich}, {Handberg}, {Hart}, {Hatfield-Dodds}, {Homeier}, {Hosseinzadeh}, {Jenness}, {Jones}, {Joseph}, {Kalmbach}, {Karamehmetoglu}, {Ka{l}uszy{'n}ski}, {Kelley}, {Kern}, {Kerzendorf}, {Koch}, {Kulumani}, {Lee}, {Ly}, {Ma}, {MacBride}, {Maljaars}, {Muna}, {Murphy}, {Norman}, {O'Steen}, {Oman}, {Pacifici}, {Pascual}, {Pascual-Granado}, {Patil}, {Perren}, {Pickering}, {Rastogi}, {Roulston}, {Ryan}, {Rykoff}, {Sabater}, {Sakurikar}, {Salgado}, {Sanghi}, {Saunders}, {Savchenko}, {Schwardt}, {Seifert-Eckert}, {Shih}, {Jain}, {Shukla}, {Sick}, {Simpson}, {Singanamalla}, {Singer}, {Singhal}, {Sinha}, {Sip{H{o}}cz}, {Spitler}, {Stansby}, {Streicher}, {{ {S}}umak}, {Swinbank}, {Taranu}, {Tewary}, {Tremblay}, {Val-Borro}, {Van Kooten}, {Vasovi{'c}}, {Verma}, {de Miranda Cardoso}, {Williams}, {Wilson}, {Winkel}, {Wood-Vasey}, {Xue}, {Yoachim}, {Zhang}, {Zonca}, \& {Astropy Project Contributors}}]{astropy:2022}
{Astropy Collaboration}, {Price-Whelan}, A.~M., {Lim}, P.~L., {et~al.} 2022, apj, 935, 167

\bibitem[{{Blanc} {et~al.}(2013){Blanc}, {Weinzirl}, {Song}, {Heiderman}, {Gebhardt}, {Jogee}, {Evans}, {van den Bosch}, {Luo}, {Drory}, {Fabricius}, {Fisher}, {Hao}, {Kaplan}, {Marinova}, {Vutisalchavakul}, \& {Yoachim}}]{Blanc2013b}
{Blanc}, G.~A., {Weinzirl}, T., {Song}, M., {et~al.} 2013, \aj, 145, 138

\bibitem[{{Bonanos} {et~al.}(2009){Bonanos}, {Massa}, {Sewilo}, {Lennon}, {Panagia}, {Smith}, {Meixner}, {Babler}, {Bracker}, {Meade}, {Gordon}, {Hora}, {Indebetouw}, \& {Whitney}}]{Bonanos2009}
{Bonanos}, A.~Z., {Massa}, D.~L., {Sewilo}, M., {et~al.} 2009, \aj, 138, 1003

\bibitem[{{Bressan} {et~al.}(2012){Bressan}, {Marigo}, {Girardi}, {Salasnich}, {Dal Cero}, {Rubele}, \& {Nanni}}]{Bressan2012}
{Bressan}, A., {Marigo}, P., {Girardi}, L., {et~al.} 2012, \mnras, 427, 127

\bibitem[{{Breysacher} {et~al.}(1999){Breysacher}, {Azzopardi}, \& {Testor}}]{Breysacher1999}
{Breysacher}, J., {Azzopardi}, M., \& {Testor}, G. 1999, \aaps, 137, 117

\bibitem[{{Calzetti} {et~al.}(2007){Calzetti}, {Kennicutt}, {Engelbracht}, {Leitherer}, {Draine}, {Kewley}, {Moustakas}, {Sosey}, {Dale}, {Gordon}, {Helou}, {Hollenbach}, {Armus}, {Bendo}, {Bot}, {Buckalew}, {Jarrett}, {Li}, {Meyer}, {Murphy}, {Prescott}, {Regan}, {Rieke}, {Roussel}, {Sheth}, {Smith}, {Thornley}, \& {Walter}}]{Calzetti2007}
{Calzetti}, D., {Kennicutt}, R.~C., {Engelbracht}, C.~W., {et~al.} 2007, \apj, 666, 870

\bibitem[{{Calzetti} {et~al.}(2015){Calzetti}, {Lee}, {Sabbi}, {Adamo}, {Smith}, {Andrews}, {Ubeda}, {Bright}, {Thilker}, {Aloisi}, {Brown}, {Chandar}, {Christian}, {Cignoni}, {Clayton}, {da Silva}, {de Mink}, {Dobbs}, {Elmegreen}, {Elmegreen}, {Evans}, {Fumagalli}, {Gallagher}, {Gouliermis}, {Grebel}, {Herrero}, {Hunter}, {Johnson}, {Kennicutt}, {Kim}, {Krumholz}, {Lennon}, {Levay}, {Martin}, {Nair}, {Nota}, {{\"O}stlin}, {Pellerin}, {Prieto}, {Regan}, {Ryon}, {Schaerer}, {Schiminovich}, {Tosi}, {Van Dyk}, {Walterbos}, {Whitmore}, \& {Wofford}}]{Calzetti2015}
{Calzetti}, D., {Lee}, J.~C., {Sabbi}, E., {et~al.} 2015, \aj, 149, 51

\bibitem[{{Chevance} {et~al.}(2020){Chevance}, {Kruijssen}, {Hygate}, {Schruba}, {Longmore}, {Groves}, {Henshaw}, {Herrera}, {Hughes}, {Jeffreson}, {Lang}, {Leroy}, {Meidt}, {Pety}, {Razza}, {Rosolowsky}, {Schinnerer}, {Bigiel}, {Blanc}, {Emsellem}, {Faesi}, {Glover}, {Haydon}, {Ho}, {Kreckel}, {Lee}, {Liu}, {Querejeta}, {Saito}, {Sun}, {Usero}, \& {Utomo}}]{Chevance2020}
{Chevance}, M., {Kruijssen}, J.~M.~D., {Hygate}, A. P.~S., {et~al.} 2020, \mnras, 493, 2872

\bibitem[{{Colombo} {et~al.}(2014){Colombo}, {Meidt}, {Schinnerer}, {Garc{\'\i}a-Burillo}, {Hughes}, {Pety}, {Leroy}, {Dobbs}, {Dumas}, {Thompson}, {Schuster}, \& {Kramer}}]{Colombo2014}
{Colombo}, D., {Meidt}, S.~E., {Schinnerer}, E., {et~al.} 2014, \apj, 784, 4

\bibitem[{{Corbelli} {et~al.}(2017){Corbelli}, {Braine}, {Bandiera}, {Brouillet}, {Combes}, {Druard}, {Gratier}, {Mata}, {Schuster}, {Xilouris}, \& {Palla}}]{Corbelli2017}
{Corbelli}, E., {Braine}, J., {Bandiera}, R., {et~al.} 2017, \aap, 601, A146

\bibitem[{{den Brok} {et~al.}(2021){den Brok}, {Chatzigiannakis}, {Bigiel}, {Puschnig}, {Barnes}, {Leroy}, {Jim{\'e}nez-Donaire}, {Usero}, {Schinnerer}, {Rosolowsky}, {Faesi}, {Grasha}, {Hughes}, {Kruijssen}, {Liu}, {Neumann}, {Pety}, {Querejeta}, {Saito}, {Schruba}, \& {Stuber}}]{denBrok2021}
{den Brok}, J.~S., {Chatzigiannakis}, D., {Bigiel}, F., {et~al.} 2021, \mnras, 504, 3221

\bibitem[{{Dobbs} {et~al.}(2015){Dobbs}, {Pringle}, \& {Duarte-Cabral}}]{Dobbs2015}
{Dobbs}, C.~L., {Pringle}, J.~E., \& {Duarte-Cabral}, A. 2015, \mnras, 446, 3608

\bibitem[{{Dom{\'\i}nguez} {et~al.}(2013){Dom{\'\i}nguez}, {Siana}, {Henry}, {Scarlata}, {Bedregal}, {Malkan}, {Atek}, {Ross}, {Colbert}, {Teplitz}, {Rafelski}, {McCarthy}, {Bunker}, {Hathi}, {Dressler}, {Martin}, \& {Masters}}]{Dominguez2013}
{Dom{\'\i}nguez}, A., {Siana}, B., {Henry}, A.~L., {et~al.} 2013, \apj, 763, 145

\bibitem[{{Donovan Meyer} {et~al.}(2013){Donovan Meyer}, {Koda}, {Momose}, {Mooney}, {Egusa}, {Carty}, {Kennicutt}, {Kuno}, {Rebolledo}, {Sawada}, {Scoville}, \& {Wong}}]{DonovanMeyer2013}
{Donovan Meyer}, J., {Koda}, J., {Momose}, R., {et~al.} 2013, \apj, 772, 107

\bibitem[{{Emsellem} {et~al.}(2022){Emsellem}, {Schinnerer}, {Santoro}, {Belfiore}, {Pessa}, {McElroy}, {Blanc}, {Congiu}, {Groves}, {Ho}, {Kreckel}, {Razza}, {Sanchez-Blazquez}, {Egorov}, {Faesi}, {Klessen}, {Leroy}, {Meidt}, {Querejeta}, {Rosolowsky}, {Scheuermann}, {Anand}, {Barnes}, {Be{\v{s}}li{\'c}}, {Bigiel}, {Boquien}, {Cao}, {Chevance}, {Dale}, {Eibensteiner}, {Glover}, {Grasha}, {Henshaw}, {Hughes}, {Koch}, {Kruijssen}, {Lee}, {Liu}, {Pan}, {Pety}, {Saito}, {Sandstrom}, {Schruba}, {Sun}, {Thilker}, {Usero}, {Watkins}, \& {Williams}}]{Emsellem2022}
{Emsellem}, E., {Schinnerer}, E., {Santoro}, F., {et~al.} 2022, \aap, 659, A191

\bibitem[{{Enokiya} {et~al.}(2021){Enokiya}, {Torii}, \& {Fukui}}]{Enokiya2021}
{Enokiya}, R., {Torii}, K., \& {Fukui}, Y. 2021, \pasj, 73, S75

\bibitem[{{Faesi} {et~al.}(2018){Faesi}, {Lada}, \& {Forbrich}}]{Faesi2018}
{Faesi}, C.~M., {Lada}, C.~J., \& {Forbrich}, J. 2018, \apj, 857, 19

\bibitem[{{Freeman} {et~al.}(2017){Freeman}, {Rosolowsky}, {Kruijssen}, {Bastian}, \& {Adamo}}]{Freeman2017}
{Freeman}, P., {Rosolowsky}, E., {Kruijssen}, J.~M.~D., {Bastian}, N., \& {Adamo}, A. 2017, \mnras, 468, 1769

\bibitem[{{Fujii} {et~al.}(2014){Fujii}, {Minamidani}, {Mizuno}, {Onishi}, {Kawamura}, {Muller}, {Dawson}, {Tatematsu}, {Hasegawa}, {Tosaki}, {Miura}, {Muraoka}, {Sakai}, {Tsukagoshi}, {Tanaka}, {Ezawa}, \& {Fukui}}]{Fujii2014}
{Fujii}, K., {Minamidani}, T., {Mizuno}, N., {et~al.} 2014, \apj, 796, 123

\bibitem[{{Fujita} {et~al.}(2021){Fujita}, {Sano}, {Enokiya}, {Hayashi}, {Kohno}, {Tsuge}, {Tachihara}, {Nishimura}, {Ohama}, {Yamane}, {Ohno}, {Yamada}, \& {Fukui}}]{Fujita2021}
{Fujita}, S., {Sano}, H., {Enokiya}, R., {et~al.} 2021, \pasj, 73, S201

\bibitem[{{Fukui} {et~al.}(2021){Fukui}, {Habe}, {Inoue}, {Enokiya}, \& {Tachihara}}]{Fukui2021}
{Fukui}, Y., {Habe}, A., {Inoue}, T., {Enokiya}, R., \& {Tachihara}, K. 2021, \pasj, 73, S1

\bibitem[{{Fukui} \& {Kawamura}(2010)}]{FukuiKawamura2010}
{Fukui}, Y., \& {Kawamura}, A. 2010, \araa, 48, 547

\bibitem[{{Fukui} {et~al.}(1999){Fukui}, {Mizuno}, {Yamaguchi}, {Mizuno}, {Onishi}, {Ogawa}, {Yonekura}, {Kawamura}, {Tachihara}, {Xiao}, {Yamaguchi}, {Hara}, {Hayakawa}, {Kato}, {Abe}, {Saito}, {Mano}, {Matsunaga}, {Mine}, {Moriguchi}, {Aoyama}, {Asayama}, {Yoshikawa}, \& {Rubio}}]{Fukui1999}
{Fukui}, Y., {Mizuno}, N., {Yamaguchi}, R., {et~al.} 1999, \pasj, 51, 745

\bibitem[{{Fukui} {et~al.}(2008){Fukui}, {Kawamura}, {Minamidani}, {Mizuno}, {Kanai}, {Mizuno}, {Onishi}, {Yonekura}, {Mizuno}, {Ogawa}, \& {Rubio}}]{Fukui2008}
{Fukui}, Y., {Kawamura}, A., {Minamidani}, T., {et~al.} 2008, \apjs, 178, 56

\bibitem[{{Fukui} {et~al.}(2009){Fukui}, {Kawamura}, {Wong}, {Murai}, {Iritani}, {Mizuno}, {Mizuno}, {Onishi}, {Hughes}, {Ott}, {Muller}, {Staveley-Smith}, \& {Kim}}]{Fukui2009}
{Fukui}, Y., {Kawamura}, A., {Wong}, T., {et~al.} 2009, \apj, 705, 144

\bibitem[{{Fukui} {et~al.}(2015){Fukui}, {Harada}, {Tokuda}, {Morioka}, {Onishi}, {Torii}, {Ohama}, {Hattori}, {Nayak}, {Meixner}, {Sewi{\l}o}, {Indebetouw}, {Kawamura}, {Saigo}, {Yamamoto}, {Tachihara}, {Minamidani}, {Inoue}, {Madden}, {Galametz}, {Lebouteiller}, {Mizuno}, \& {Chen}}]{Fukui2015}
{Fukui}, Y., {Harada}, R., {Tokuda}, K., {et~al.} 2015, \apjl, 807, L4

\bibitem[{{Fukui} {et~al.}(2018{\natexlab{a}}){Fukui}, {Torii}, {Hattori}, {Nishimura}, {Ohama}, {Shimajiri}, {Shima}, {Habe}, {Sano}, {Kohno}, {Yamamoto}, {Tachihara}, \& {Onishi}}]{Fukui2018_M42}
{Fukui}, Y., {Torii}, K., {Hattori}, Y., {et~al.} 2018{\natexlab{a}}, \apj, 859, 166

\bibitem[{{Fukui} {et~al.}(2018{\natexlab{b}}){Fukui}, {Kohno}, {Yokoyama}, {Nishimura}, {Torii}, {Hattori}, {Sano}, {Ohama}, {Yamamoto}, \& {Tachihara}}]{Fukui2018_GM24}
{Fukui}, Y., {Kohno}, M., {Yokoyama}, K., {et~al.} 2018{\natexlab{b}}, \pasj, 70, S44

\bibitem[{{Fukui} {et~al.}(2019){Fukui}, {Tokuda}, {Saigo}, {Harada}, {Tachihara}, {Tsuge}, {Inoue}, {Torii}, {Nishimura}, {Zahorecz}, {Nayak}, {Meixner}, {Minamidani}, {Kawamura}, {Mizuno}, {Indebetouw}, {Sewi{\l}o}, {Madden}, {Galametz}, {Lebouteiller}, {Chen}, \& {Onishi}}]{Fukui2019}
{Fukui}, Y., {Tokuda}, K., {Saigo}, K., {et~al.} 2019, \apj, 886, 14

\bibitem[{{Gebel}(1968)}]{Gebel1968}
{Gebel}, W.~L. 1968, \apj, 153, 743

\bibitem[{{Glazebrook} {et~al.}(1999){Glazebrook}, {Blake}, {Economou}, {Lilly}, \& {Colless}}]{Glazebrook1999}
{Glazebrook}, K., {Blake}, C., {Economou}, F., {Lilly}, S., \& {Colless}, M. 1999, \mnras, 306, 843

\bibitem[{{Gratier} {et~al.}(2012){Gratier}, {Braine}, {Rodriguez-Fernandez}, {Schuster}, {Kramer}, {Corbelli}, {Combes}, {Brouillet}, {van der Werf}, \& {R{\"o}llig}}]{Gratier2012}
{Gratier}, P., {Braine}, J., {Rodriguez-Fernandez}, N.~J., {et~al.} 2012, \aap, 542, A108

\bibitem[{{Groves} {et~al.}(2023){Groves}, {Kreckel}, {Santoro}, {Belfiore}, {Zavodnik}, {Congiu}, {Egorov}, {Emsellem}, {Grasha}, {Leroy}, {Scheuermann}, {Schinnerer}, {Watkins}, {Barnes}, {Bigiel}, {Dale}, {Glover}, {Pessa}, {Sanchez-Blazquez}, \& {Williams}}]{Groves2023}
{Groves}, B., {Kreckel}, K., {Santoro}, F., {et~al.} 2023, \mnras, 520, 4902

\bibitem[{{Habe} \& {Ohta}(1992)}]{HabeOhta1992}
{Habe}, A., \& {Ohta}, K. 1992, \pasj, 44, 203

\bibitem[{{Haffner} {et~al.}(2009){Haffner}, {Dettmar}, {Beckman}, {Wood}, {Slavin}, {Giammanco}, {Madsen}, {Zurita}, \& {Reynolds}}]{Haffner2009}
{Haffner}, L.~M., {Dettmar}, R.~J., {Beckman}, J.~E., {et~al.} 2009, Reviews of Modern Physics, 81, 969

\bibitem[{{Haworth} {et~al.}(2015){Haworth}, {Tasker}, {Fukui}, {Torii}, {Dale}, {Shima}, {Takahira}, {Habe}, \& {Hasegawa}}]{Haworth2015}
{Haworth}, T.~J., {Tasker}, E.~J., {Fukui}, Y., {et~al.} 2015, \mnras, 450, 10

\bibitem[{{Haydon} {et~al.}(2020){Haydon}, {Kruijssen}, {Chevance}, {Hygate}, {Krumholz}, {Schruba}, \& {Longmore}}]{Haydon2020}
{Haydon}, D.~T., {Kruijssen}, J.~M.~D., {Chevance}, M., {et~al.} 2020, \mnras, 498, 235

\bibitem[{{Helfer} {et~al.}(2003){Helfer}, {Thornley}, {Regan}, {Wong}, {Sheth}, {Vogel}, {Blitz}, \& {Bock}}]{Helfer2003}
{Helfer}, T.~T., {Thornley}, M.~D., {Regan}, M.~W., {et~al.} 2003, \apjs, 145, 259

\bibitem[{{Inoue} \& {Fukui}(2013)}]{Inoue2013}
{Inoue}, T., \& {Fukui}, Y. 2013, \apjl, 774, L31

\bibitem[{{Inoue} {et~al.}(2018){Inoue}, {Hennebelle}, {Fukui}, {Matsumoto}, {Iwasaki}, \& {Inutsuka}}]{Inoue2018}
{Inoue}, T., {Hennebelle}, P., {Fukui}, Y., {et~al.} 2018, \pasj, 70, S53

\bibitem[{{Inutsuka} {et~al.}(2015){Inutsuka}, {Inoue}, {Iwasaki}, \& {Hosokawa}}]{Inutsuka2015}
{Inutsuka}, S.-i., {Inoue}, T., {Iwasaki}, K., \& {Hosokawa}, T. 2015, \aap, 580, A49

\bibitem[{{Israel} {et~al.}(2003){Israel}, {de Graauw}, {Johansson}, {Booth}, {Boulanger}, {Garay}, {Kutner}, {Lequeux}, {Nyman}, \& {Rubio}}]{Israel2003}
{Israel}, F.~P., {de Graauw}, T., {Johansson}, L.~E.~B., {et~al.} 2003, \aap, 401, 99

\bibitem[{{Jim{\'e}nez-Donaire} {et~al.}(2019){Jim{\'e}nez-Donaire}, {Bigiel}, {Leroy}, {Usero}, {Cormier}, {Puschnig}, {Gallagher}, {Kepley}, {Bolatto}, {Garc{\'\i}a-Burillo}, {Hughes}, {Kramer}, {Pety}, {Schinnerer}, {Schruba}, {Schuster}, \& {Walter}}]{Jimenez-Donaire2019}
{Jim{\'e}nez-Donaire}, M.~J., {Bigiel}, F., {Leroy}, A.~K., {et~al.} 2019, \apj, 880, 127

\bibitem[{{Kawamura} {et~al.}(2009){Kawamura}, {Mizuno}, {Minamidani}, {Filipovi{\'c}}, {Staveley-Smith}, {Kim}, {Mizuno}, {Onishi}, {Mizuno}, \& {Fukui}}]{Kawamura2009}
{Kawamura}, A., {Mizuno}, Y., {Minamidani}, T., {et~al.} 2009, \apjs, 184, 1

\bibitem[{{Kennicutt}(1998)}]{Kennicutt1998}
{Kennicutt}, Robert~C., J. 1998, \apj, 498, 541

\bibitem[{{Kennicutt} {et~al.}(2003){Kennicutt}, {Armus}, {Bendo}, {Calzetti}, {Dale}, {Draine}, {Engelbracht}, {Gordon}, {Grauer}, {Helou}, {Hollenbach}, {Jarrett}, {Kewley}, {Leitherer}, {Li}, {Malhotra}, {Regan}, {Rieke}, {Rieke}, {Roussel}, {Smith}, {Thornley}, \& {Walter}}]{Kennicutt2003}
{Kennicutt}, Robert~C., J., {Armus}, L., {Bendo}, G., {et~al.} 2003, \pasp, 115, 928

\bibitem[{{Kennicutt} {et~al.}(2007){Kennicutt}, {Calzetti}, {Walter}, {Helou}, {Hollenbach}, {Armus}, {Bendo}, {Dale}, {Draine}, {Engelbracht}, {Gordon}, {Prescott}, {Regan}, {Thornley}, {Bot}, {Brinks}, {de Blok}, {de Mello}, {Meyer}, {Moustakas}, {Murphy}, {Sheth}, \& {Smith}}]{Kennicutt2007}
{Kennicutt}, Robert~C., J., {Calzetti}, D., {Walter}, F., {et~al.} 2007, \apj, 671, 333

\bibitem[{{Kim} {et~al.}(2021){Kim}, {Chevance}, {Kruijssen}, {Schruba}, {Sandstrom}, {Barnes}, {Bigiel}, {Blanc}, {Cao}, {Dale}, {Faesi}, {Glover}, {Grasha}, {Groves}, {Herrera}, {Klessen}, {Kreckel}, {Lee}, {Leroy}, {Pety}, {Querejeta}, {Schinnerer}, {Sun}, {Usero}, {Ward}, \& {Williams}}]{Kim2021}
{Kim}, J., {Chevance}, M., {Kruijssen}, J.~M.~D., {et~al.} 2021, \mnras, 504, 487

\bibitem[{{Kim} {et~al.}(2022){Kim}, {Chevance}, {Kruijssen}, {Leroy}, {Schruba}, {Barnes}, {Bigiel}, {Blanc}, {Cao}, {Congiu}, {Dale}, {Faesi}, {Glover}, {Grasha}, {Groves}, {Hughes}, {Klessen}, {Kreckel}, {McElroy}, {Pan}, {Pety}, {Querejeta}, {Razza}, {Rosolowsky}, {Saito}, {Schinnerer}, {Sun}, {Tomi{\v{c}}i{\'c}}, {Usero}, \& {Williams}}]{Kim2022}
---. 2022, \mnras, 516, 3006

\bibitem[{{Kobayashi} {et~al.}(2017){Kobayashi}, {Inutsuka}, {Kobayashi}, \& {Hasegawa}}]{Kobayashi2017}
{Kobayashi}, M. I.~N., {Inutsuka}, S.-i., {Kobayashi}, H., \& {Hasegawa}, K. 2017, \apj, 836, 175

\bibitem[{{Kobayashi} {et~al.}(2018){Kobayashi}, {Kobayashi}, {Inutsuka}, \& {Fukui}}]{Kobayashi2018}
{Kobayashi}, M. I.~N., {Kobayashi}, H., {Inutsuka}, S.-i., \& {Fukui}, Y. 2018, \pasj, 70, S59

\bibitem[{{Kohno} {et~al.}(2021){Kohno}, {Tachihara}, {Torii}, {Fujita}, {Nishimura}, {Kuno}, {Umemoto}, {Minamidani}, {Matsuo}, {Kiridoshi}, {Tokuda}, {Hanaoka}, {Tsuda}, {Kuriki}, {Ohama}, {Sano}, {Hasegawa}, {Sofue}, {Habe}, {Onishi}, \& {Fukui}}]{Kohno2021}
{Kohno}, M., {Tachihara}, K., {Torii}, K., {et~al.} 2021, \pasj, 73, S129

\bibitem[{{Kreckel} {et~al.}(2018){Kreckel}, {Faesi}, {Kruijssen}, {Schruba}, {Groves}, {Leroy}, {Bigiel}, {Blanc}, {Chevance}, {Herrera}, {Hughes}, {McElroy}, {Pety}, {Querejeta}, {Rosolowsky}, {Schinnerer}, {Sun}, {Usero}, \& {Utomo}}]{Kreckel2018}
{Kreckel}, K., {Faesi}, C., {Kruijssen}, J.~M.~D., {et~al.} 2018, \apjl, 863, L21

\bibitem[{{Kroupa}(2001)}]{Kroupa2001}
{Kroupa}, P. 2001, \mnras, 322, 231

\bibitem[{{Kroupa}(2002)}]{Kroupa2002}
---. 2002, Science, 295, 82

\bibitem[{{Kruijssen} \& {Longmore}(2014)}]{KruijssenLongmore2014}
{Kruijssen}, J.~M.~D., \& {Longmore}, S.~N. 2014, \mnras, 439, 3239

\bibitem[{{Kruijssen} {et~al.}(2018){Kruijssen}, {Schruba}, {Hygate}, {Hu}, {Haydon}, \& {Longmore}}]{Kruijssen2018}
{Kruijssen}, J.~M.~D., {Schruba}, A., {Hygate}, A. P.~S., {et~al.} 2018, \mnras, 479, 1866

\bibitem[{{Kruijssen} {et~al.}(2019){Kruijssen}, {Schruba}, {Chevance}, {Longmore}, {Hygate}, {Haydon}, {McLeod}, {Dalcanton}, {Tacconi}, \& {van Dishoeck}}]{Kruijssen2019}
{Kruijssen}, J.~M.~D., {Schruba}, A., {Chevance}, M., {et~al.} 2019, \nat, 569, 519

\bibitem[{{Lang} {et~al.}(2020){Lang}, {Meidt}, {Rosolowsky}, {Nofech}, {Schinnerer}, {Leroy}, {Emsellem}, {Pessa}, {Glover}, {Groves}, {Hughes}, {Kruijssen}, {Querejeta}, {Schruba}, {Bigiel}, {Blanc}, {Chevance}, {Colombo}, {Faesi}, {Henshaw}, {Herrera}, {Liu}, {Pety}, {Puschnig}, {Saito}, {Sun}, \& {Usero}}]{Lang2020}
{Lang}, P., {Meidt}, S.~E., {Rosolowsky}, E., {et~al.} 2020, \apj, 897, 122

\bibitem[{{Lee} {et~al.}(2023){Lee}, {Sandstrom}, {Leroy}, {Thilker}, {Schinnerer}, {Rosolowsky}, {Larson}, {Egorov}, {Williams}, {Schmidt}, {Emsellem}, {Anand}, {Barnes}, {Belfiore}, {Be{\v{s}}li{\'c}}, {Bigiel}, {Blanc}, {Bolatto}, {Boquien}, {den Brok}, {Cao}, {Chandar}, {Chastenet}, {Chevance}, {Chiang}, {Congiu}, {Dale}, {Deger}, {Eibensteiner}, {Faesi}, {Glover}, {Grasha}, {Groves}, {Hassani}, {Henny}, {Henshaw}, {Hoyer}, {Hughes}, {Jeffreson}, {Jim{\'e}nez-Donaire}, {Kim}, {Kim}, {Klessen}, {Koch}, {Kreckel}, {Kruijssen}, {Li}, {Liu}, {Lopez}, {Maschmann}, {Chen}, {Meidt}, {Murphy}, {Neumann}, {Neumayer}, {Pan}, {Pessa}, {Pety}, {Querejeta}, {Pinna}, {Rodr{\'\i}guez}, {Saito}, {S{\'a}nchez-Bl{\'a}zquez}, {Santoro}, {Sardone}, {Smith}, {Sormani}, {Scheuermann}, {Stuber}, {Sutter}, {Sun}, {Teng}, {Tre{\ss}}, {Usero}, {Watkins}, {Whitmore}, \& {Razza}}]{Lee2023}
{Lee}, J.~C., {Sandstrom}, K.~M., {Leroy}, A.~K., {et~al.} 2023, \apjl, 944, L17

\bibitem[{{Leroy} {et~al.}(2009){Leroy}, {Walter}, {Bigiel}, {Usero}, {Weiss}, {Brinks}, {de Blok}, {Kennicutt}, {Schuster}, {Kramer}, {Wiesemeyer}, \& {Roussel}}]{Leroy2009}
{Leroy}, A.~K., {Walter}, F., {Bigiel}, F., {et~al.} 2009, \aj, 137, 4670

\bibitem[{{Leroy} {et~al.}(2016){Leroy}, {Hughes}, {Schruba}, {Rosolowsky}, {Blanc}, {Bolatto}, {Colombo}, {Escala}, {Kramer}, {Kruijssen}, {Meidt}, {Pety}, {Querejeta}, {Sandstrom}, {Schinnerer}, {Sliwa}, \& {Usero}}]{Leroy2016}
{Leroy}, A.~K., {Hughes}, A., {Schruba}, A., {et~al.} 2016, \apj, 831, 16

\bibitem[{{Leroy} {et~al.}(2021{\natexlab{a}}){Leroy}, {Schinnerer}, {Hughes}, {Rosolowsky}, {Pety}, {Schruba}, {Usero}, {Blanc}, {Chevance}, {Emsellem}, {Faesi}, {Herrera}, {Liu}, {Meidt}, {Querejeta}, {Saito}, {Sandstrom}, {Sun}, {Williams}, {Anand}, {Barnes}, {Behrens}, {Belfiore}, {Benincasa}, {Be{\v{s}}li{\'c}}, {Bigiel}, {Bolatto}, {den Brok}, {Cao}, {Chandar}, {Chastenet}, {Chiang}, {Congiu}, {Dale}, {Deger}, {Eibensteiner}, {Egorov}, {Garc{\'\i}a-Rodr{\'\i}guez}, {Glover}, {Grasha}, {Henshaw}, {Ho}, {Kepley}, {Kim}, {Klessen}, {Kreckel}, {Koch}, {Kruijssen}, {Larson}, {Lee}, {Lopez}, {Machado}, {Mayker}, {McElroy}, {Murphy}, {Ostriker}, {Pan}, {Pessa}, {Puschnig}, {Razza}, {S{\'a}nchez-Bl{\'a}zquez}, {Santoro}, {Sardone}, {Scheuermann}, {Sliwa}, {Sormani}, {Stuber}, {Thilker}, {Turner}, {Utomo}, {Watkins}, \& {Whitmore}}]{Leroy2021b}
{Leroy}, A.~K., {Schinnerer}, E., {Hughes}, A., {et~al.} 2021{\natexlab{a}}, \apjs, 257, 43

\bibitem[{{Leroy} {et~al.}(2021{\natexlab{b}}){Leroy}, {Hughes}, {Liu}, {Pety}, {Rosolowsky}, {Saito}, {Schinnerer}, {Schruba}, {Usero}, {Faesi}, {Herrera}, {Chevance}, {Hygate}, {Kepley}, {Koch}, {Querejeta}, {Sliwa}, {Will}, {Wilson}, {Anand}, {Barnes}, {Belfiore}, {Be{\v{s}}li{\'c}}, {Bigiel}, {Blanc}, {Bolatto}, {Boquien}, {Cao}, {Chandar}, {Chastenet}, {Chiang}, {Congiu}, {Dale}, {Deger}, {den Brok}, {Eibensteiner}, {Emsellem}, {Garc{\'\i}a-Rodr{\'\i}guez}, {Glover}, {Grasha}, {Groves}, {Henshaw}, {Jim{\'e}nez Donaire}, {Kim}, {Klessen}, {Kreckel}, {Kruijssen}, {Larson}, {Lee}, {Mayker}, {McElroy}, {Meidt}, {Mok}, {Pan}, {Puschnig}, {Razza}, {S{\'a}nchez-Bl'azquez}, {Sandstrom}, {Santoro}, {Sardone}, {Scheuermann}, {Sun}, {Thilker}, {Turner}, {Ubeda}, {Utomo}, {Watkins}, \& {Williams}}]{Leroy2021a}
{Leroy}, A.~K., {Hughes}, A., {Liu}, D., {et~al.} 2021{\natexlab{b}}, \apjs, 255, 19

\bibitem[{{Leroy} {et~al.}(2022){Leroy}, {Rosolowsky}, {Usero}, {Sandstrom}, {Schinnerer}, {Schruba}, {Bolatto}, {Sun}, {Barnes}, {Belfiore}, {Bigiel}, {den Brok}, {Cao}, {Chiang}, {Chevance}, {Dale}, {Eibensteiner}, {Faesi}, {Glover}, {Hughes}, {Jim{\'e}nez Donaire}, {Klessen}, {Koch}, {Kruijssen}, {Liu}, {Meidt}, {Pan}, {Pety}, {Puschnig}, {Querejeta}, {Saito}, {Sardone}, {Watkins}, {Weiss}, \& {Williams}}]{Leroy2022}
{Leroy}, A.~K., {Rosolowsky}, E., {Usero}, A., {et~al.} 2022, \apj, 927, 149

\bibitem[{{Liow} \& {Dobbs}(2020)}]{Liow2020}
{Liow}, K.~Y., \& {Dobbs}, C.~L. 2020, \mnras, 499, 1099

\bibitem[{{Motte} {et~al.}(2018){Motte}, {Bontemps}, \& {Louvet}}]{Motte2018}
{Motte}, F., {Bontemps}, S., \& {Louvet}, F. 2018, \araa, 56, 41

\bibitem[{{Muraoka} {et~al.}(2020){Muraoka}, {Kondo}, {Tokuda}, {Nishimura}, {Miura}, {Onodera}, {Kuno}, {Zahorecz}, {Tsuge}, {Sano}, {Fujita}, {Onishi}, {Saigo}, {Tachihara}, {Fukui}, \& {Kawamura}}]{Muraoka2020}
{Muraoka}, K., {Kondo}, H., {Tokuda}, K., {et~al.} 2020, \apj, 903, 94

\bibitem[{{Neugent} {et~al.}(2018){Neugent}, {Massey}, \& {Morrell}}]{Neugent2018}
{Neugent}, K.~F., {Massey}, P., \& {Morrell}, N. 2018, \apj, 863, 181

\bibitem[{{Nishimura} {et~al.}(2015){Nishimura}, {Tokuda}, {Kimura}, {Muraoka}, {Maezawa}, {Ogawa}, {Dobashi}, {Shimoikura}, {Mizuno}, {Fukui}, \& {Onishi}}]{Nishimura2015}
{Nishimura}, A., {Tokuda}, K., {Kimura}, K., {et~al.} 2015, \apjs, 216, 18

\bibitem[{{Ohama} {et~al.}(2018){Ohama}, {Kohno}, {Hasegawa}, {Torii}, {Nishimura}, {Hattori}, {Hayakawa}, {Inoue}, {Sano}, {Yamamoto}, {Tachihara}, \& {Fukui}}]{Ohama2018}
{Ohama}, A., {Kohno}, M., {Hasegawa}, K., {et~al.} 2018, \pasj, 70, S45

\bibitem[{{Onodera} {et~al.}(2010){Onodera}, {Kuno}, {Tosaki}, {Kohno}, {Nakanishi}, {Sawada}, {Muraoka}, {Komugi}, {Miura}, {Kaneko}, {Hirota}, \& {Kawabe}}]{Onodera2010}
{Onodera}, S., {Kuno}, N., {Tosaki}, T., {et~al.} 2010, \apjl, 722, L127

\bibitem[{{Pan} {et~al.}(2022){Pan}, {Schinnerer}, {Hughes}, {Leroy}, {Groves}, {Barnes}, {Belfiore}, {Bigiel}, {Blanc}, {Cao}, {Chevance}, {Congiu}, {Dale}, {Eibensteiner}, {Emsellem}, {Faesi}, {Glover}, {Grasha}, {Herrera}, {Ho}, {Klessen}, {Kruijssen}, {Lang}, {Liu}, {McElroy}, {Meidt}, {Murphy}, {Pety}, {Querejeta}, {Razza}, {Rosolowsky}, {Saito}, {Santoro}, {Schruba}, {Sun}, {Tomi{\v{c}}i{\'c}}, {Usero}, {Utomo}, \& {Williams}}]{Pan2022}
{Pan}, H.-A., {Schinnerer}, E., {Hughes}, A., {et~al.} 2022, \apj, 927, 9

\bibitem[{{Panagia}(1973)}]{Panagia1973}
{Panagia}, N. 1973, \aj, 78, 929

\bibitem[{{Querejeta} {et~al.}(2021){Querejeta}, {Schinnerer}, {Meidt}, {Sun}, {Leroy}, {Emsellem}, {Klessen}, {Mu{\~n}oz-Mateos}, {Salo}, {Laurikainen}, {Be{\v{s}}li{\'c}}, {Blanc}, {Chevance}, {Dale}, {Eibensteiner}, {Faesi}, {Garc{\'\i}a-Rodr{\'\i}guez}, {Glover}, {Grasha}, {Henshaw}, {Herrera}, {Hughes}, {Kreckel}, {Kruijssen}, {Liu}, {Murphy}, {Pan}, {Pety}, {Razza}, {Rosolowsky}, {Saito}, {Schruba}, {Usero}, {Watkins}, \& {Williams}}]{Querejeta2021}
{Querejeta}, M., {Schinnerer}, E., {Meidt}, S., {et~al.} 2021, \aap, 656, A133

\bibitem[{{Rebolledo} {et~al.}(2015){Rebolledo}, {Wong}, {Xue}, {Leroy}, {Koda}, \& {Donovan Meyer}}]{Rebolledo2015}
{Rebolledo}, D., {Wong}, T., {Xue}, R., {et~al.} 2015, \apj, 808, 99

\bibitem[{{Roberts}(1969)}]{Roberts1969}
{Roberts}, W.~W. 1969, \apj, 158, 123

\bibitem[{{Rosolowsky} \& {Leroy}(2006)}]{Rosolowsky2006}
{Rosolowsky}, E., \& {Leroy}, A. 2006, \pasp, 118, 590

\bibitem[{{Rosolowsky} {et~al.}(2021){Rosolowsky}, {Hughes}, {Leroy}, {Sun}, {Querejeta}, {Schruba}, {Usero}, {Herrera}, {Liu}, {Pety}, {Saito}, {Be{\v{s}}li{\'c}}, {Bigiel}, {Blanc}, {Chevance}, {Dale}, {Deger}, {Faesi}, {Glover}, {Henshaw}, {Klessen}, {Kruijssen}, {Larson}, {Lee}, {Meidt}, {Mok}, {Schinnerer}, {Thilker}, \& {Williams}}]{Rosolowsky2021}
{Rosolowsky}, E., {Hughes}, A., {Leroy}, A.~K., {et~al.} 2021, \mnras, 502, 1218

\bibitem[{{Rosolowsky} {et~al.}(2008){Rosolowsky}, {Pineda}, {Kauffmann}, \& {Goodman}}]{Rosolowsky2008}
{Rosolowsky}, E.~W., {Pineda}, J.~E., {Kauffmann}, J., \& {Goodman}, A.~A. 2008, \apj, 679, 1338

\bibitem[{{Rousseau-Nepton} {et~al.}(2018){Rousseau-Nepton}, {Robert}, {Martin}, {Drissen}, \& {Martin}}]{Rousseau-Nepton2018}
{Rousseau-Nepton}, L., {Robert}, C., {Martin}, R.~P., {Drissen}, L., \& {Martin}, T. 2018, \mnras, 477, 4152

\bibitem[{{Saigo} {et~al.}(2017){Saigo}, {Onishi}, {Nayak}, {Meixner}, {Tokuda}, {Harada}, {Morioka}, {Sewi{\l}o}, {Indebetouw}, {Torii}, {Kawamura}, {Ohama}, {Hattori}, {Yamamoto}, {Tachihara}, {Minamidani}, {Inoue}, {Madden}, {Galametz}, {Lebouteiller}, {Chen}, {Mizuno}, \& {Fukui}}]{Saigo2017}
{Saigo}, K., {Onishi}, T., {Nayak}, O., {et~al.} 2017, \apj, 835, 108

\bibitem[{{Sakre} {et~al.}(2021){Sakre}, {Habe}, {Pettitt}, \& {Okamoto}}]{Sakre2021}
{Sakre}, N., {Habe}, A., {Pettitt}, A.~R., \& {Okamoto}, T. 2021, \pasj, 73, S385

\bibitem[{{S{\'a}nchez} {et~al.}(2011){S{\'a}nchez}, {Rosales-Ortega}, {Kennicutt}, {Johnson}, {Diaz}, {Pasquali}, \& {Hao}}]{Sanchez2011}
{S{\'a}nchez}, S.~F., {Rosales-Ortega}, F.~F., {Kennicutt}, R.~C., {et~al.} 2011, \mnras, 410, 313

\bibitem[{{Sandstrom} {et~al.}(2013){Sandstrom}, {Leroy}, {Walter}, {Bolatto}, {Croxall}, {Draine}, {Wilson}, {Wolfire}, {Calzetti}, {Kennicutt}, {Aniano}, {Donovan Meyer}, {Usero}, {Bigiel}, {Brinks}, {de Blok}, {Crocker}, {Dale}, {Engelbracht}, {Galametz}, {Groves}, {Hunt}, {Koda}, {Kreckel}, {Linz}, {Meidt}, {Pellegrini}, {Rix}, {Roussel}, {Schinnerer}, {Schruba}, {Schuster}, {Skibba}, {van der Laan}, {Appleton}, {Armus}, {Brandl}, {Gordon}, {Hinz}, {Krause}, {Montiel}, {Sauvage}, {Schmiedeke}, {Smith}, \& {Vigroux}}]{Sandstrom2013}
{Sandstrom}, K.~M., {Leroy}, A.~K., {Walter}, F., {et~al.} 2013, \apj, 777, 5

\bibitem[{{Sano} {et~al.}(2021){Sano}, {Tsuge}, {Tokuda}, {Muraoka}, {Tachihara}, {Yamane}, {Kohno}, {Fujita}, {Enokiya}, {Rowell}, {Maxted}, {Filipovi{\'c}}, {Knies}, {Sasaki}, {Onishi}, {Plucinsky}, \& {Fukui}}]{Sano2021}
{Sano}, H., {Tsuge}, K., {Tokuda}, K., {et~al.} 2021, \pasj, 73, S62

\bibitem[{{Santoro} {et~al.}(2022){Santoro}, {Kreckel}, {Belfiore}, {Groves}, {Congiu}, {Thilker}, {Blanc}, {Schinnerer}, {Ho}, {Kruijssen}, {Meidt}, {Klessen}, {Schruba}, {Querejeta}, {Pessa}, {Chevance}, {Kim}, {Emsellem}, {McElroy}, {Barnes}, {Bigiel}, {Boquien}, {Dale}, {Glover}, {Grasha}, {Lee}, {Leroy}, {Pan}, {Rosolowsky}, {Saito}, {Sanchez-Blazquez}, {Watkins}, \& {Williams}}]{Santoro2022}
{Santoro}, F., {Kreckel}, K., {Belfiore}, F., {et~al.} 2022, \aap, 658, A188

\bibitem[{{Scheuermann} {et~al.}(2022){Scheuermann}, {Kreckel}, {Anand}, {Blanc}, {Congiu}, {Santoro}, {Van Dyk}, {Barnes}, {Bigiel}, {Glover}, {Groves}, {Klessen}, {Kruijssen}, {Rosolowsky}, {Schinnerer}, {Schruba}, {Watkins}, \& {Williams}}]{Scheuermann2022}
{Scheuermann}, F., {Kreckel}, K., {Anand}, G.~S., {et~al.} 2022, \mnras, 511, 6087

\bibitem[{{Schinnerer} {et~al.}(2013){Schinnerer}, {Meidt}, {Pety}, {Hughes}, {Colombo}, {Garc{\'\i}a-Burillo}, {Schuster}, {Dumas}, {Dobbs}, {Leroy}, {Kramer}, {Thompson}, \& {Regan}}]{Schinnerer2013}
{Schinnerer}, E., {Meidt}, S.~E., {Pety}, J., {et~al.} 2013, \apj, 779, 42

\bibitem[{{Schinnerer} {et~al.}(2019){Schinnerer}, {Hughes}, {Leroy}, {Groves}, {Blanc}, {Kreckel}, {Bigiel}, {Chevance}, {Dale}, {Emsellem}, {Faesi}, {Glover}, {Grasha}, {Henshaw}, {Hygate}, {Kruijssen}, {Meidt}, {Pety}, {Querejeta}, {Rosolowsky}, {Saito}, {Schruba}, {Sun}, \& {Utomo}}]{Schinnerer2019}
{Schinnerer}, E., {Hughes}, A., {Leroy}, A., {et~al.} 2019, \apj, 887, 49

\bibitem[{{Schmidt}(1959)}]{Schmidt1959}
{Schmidt}, M. 1959, \apj, 129, 243

\bibitem[{{Schruba} {et~al.}(2010){Schruba}, {Leroy}, {Walter}, {Sandstrom}, \& {Rosolowsky}}]{Schruba2010}
{Schruba}, A., {Leroy}, A.~K., {Walter}, F., {Sandstrom}, K., \& {Rosolowsky}, E. 2010, \apj, 722, 1699

\bibitem[{{Sun} {et~al.}(2018){Sun}, {Leroy}, {Schruba}, {Rosolowsky}, {Hughes}, {Kruijssen}, {Meidt}, {Schinnerer}, {Blanc}, {Bigiel}, {Bolatto}, {Chevance}, {Groves}, {Herrera}, {Hygate}, {Pety}, {Querejeta}, {Usero}, \& {Utomo}}]{Sun2018}
{Sun}, J., {Leroy}, A.~K., {Schruba}, A., {et~al.} 2018, \apj, 860, 172

\bibitem[{{Sun} {et~al.}(2020){Sun}, {Leroy}, {Schinnerer}, {Hughes}, {Rosolowsky}, {Querejeta}, {Schruba}, {Liu}, {Saito}, {Herrera}, {Faesi}, {Usero}, {Pety}, {Kruijssen}, {Ostriker}, {Bigiel}, {Blanc}, {Bolatto}, {Boquien}, {Chevance}, {Dale}, {Deger}, {Emsellem}, {Glover}, {Grasha}, {Groves}, {Henshaw}, {Jimenez-Donaire}, {Kim}, {Klessen}, {Kreckel}, {Lee}, {Meidt}, {Sandstrom}, {Sardone}, {Utomo}, \& {Williams}}]{Sun2020}
{Sun}, J., {Leroy}, A.~K., {Schinnerer}, E., {et~al.} 2020, \apjl, 901, L8

\bibitem[{{Sun} {et~al.}(2022){Sun}, {Leroy}, {Rosolowsky}, {Hughes}, {Schinnerer}, {Schruba}, {Koch}, {Blanc}, {Chiang}, {Groves}, {Liu}, {Meidt}, {Pan}, {Pety}, {Querejeta}, {Saito}, {Sandstrom}, {Sardone}, {Usero}, {Utomo}, {Williams}, {Barnes}, {Benincasa}, {Bigiel}, {Bolatto}, {Boquien}, {Chevance}, {Dale}, {Deger}, {Emsellem}, {Glover}, {Grasha}, {Henshaw}, {Klessen}, {Kreckel}, {Kruijssen}, {Ostriker}, \& {Thilker}}]{Sun2022}
{Sun}, J., {Leroy}, A.~K., {Rosolowsky}, E., {et~al.} 2022, \aj, 164, 43

\bibitem[{{Tachihara} {et~al.}(2000){Tachihara}, {Mizuno}, \& {Fukui}}]{Tachihara2000}
{Tachihara}, K., {Mizuno}, A., \& {Fukui}, Y. 2000, \apj, 528, 817

\bibitem[{{Tachihara} {et~al.}(2002){Tachihara}, {Onishi}, {Mizuno}, \& {Fukui}}]{Tachihara2002}
{Tachihara}, K., {Onishi}, T., {Mizuno}, A., \& {Fukui}, Y. 2002, \aap, 385, 909

\bibitem[{{Takahira} {et~al.}(2018){Takahira}, {Shima}, {Habe}, \& {Tasker}}]{Takahira2018}
{Takahira}, K., {Shima}, K., {Habe}, A., \& {Tasker}, E.~J. 2018, \pasj, 70, S58

\bibitem[{{Takahira} {et~al.}(2014){Takahira}, {Tasker}, \& {Habe}}]{Takahira2014}
{Takahira}, K., {Tasker}, E.~J., \& {Habe}, A. 2014, \apj, 792, 63

\bibitem[{{Tokuda} {et~al.}(2019){Tokuda}, {Fukui}, {Harada}, {Saigo}, {Tachihara}, {Tsuge}, {Inoue}, {Torii}, {Nishimura}, {Zahorecz}, {Nayak}, {Meixner}, {Minamidani}, {Kawamura}, {Mizuno}, {Indebetouw}, {Sewi{\l}o}, {Madden}, {Galametz}, {Lebouteiller}, {Chen}, \& {Onishi}}]{Tokuda2019}
{Tokuda}, K., {Fukui}, Y., {Harada}, R., {et~al.} 2019, \apj, 886, 15

\bibitem[{{Tokuda} {et~al.}(2020){Tokuda}, {Muraoka}, {Kondo}, {Nishimura}, {Tosaki}, {Zahorecz}, {Onodera}, {Miura}, {Torii}, {Kuno}, {Fujita}, {Sano}, {Onishi}, {Saigo}, {Fukui}, {Kawamura}, \& {Tachihara}}]{Tokuda2020}
{Tokuda}, K., {Muraoka}, K., {Kondo}, H., {et~al.} 2020, \apj, 896, 36

\bibitem[{{Torii} {et~al.}(2015){Torii}, {Hasegawa}, {Hattori}, {Sano}, {Ohama}, {Yamamoto}, {Tachihara}, {Soga}, {Shimizu}, {Okuda}, {Mizuno}, {Onishi}, {Mizuno}, \& {Fukui}}]{Torii2015}
{Torii}, K., {Hasegawa}, K., {Hattori}, Y., {et~al.} 2015, \apj, 806, 7

\bibitem[{{Tsuge} {et~al.}(2021{\natexlab{a}}){Tsuge}, {Fukui}, {Tachihara}, {Sano}, {Tokuda}, {Ueda}, {Iono}, \& {Finn}}]{Tsuge2021a}
{Tsuge}, K., {Fukui}, Y., {Tachihara}, K., {et~al.} 2021{\natexlab{a}}, \pasj, 73, S35

\bibitem[{{Tsuge} {et~al.}(2021{\natexlab{b}}){Tsuge}, {Tachihara}, {Fukui}, {Sano}, {Tokuda}, {Ueda}, \& {Iono}}]{Tsuge2021b}
{Tsuge}, K., {Tachihara}, K., {Fukui}, Y., {et~al.} 2021{\natexlab{b}}, \pasj, 73, 417

\bibitem[{{Turner} {et~al.}(2022){Turner}, {Dale}, {Lilly}, {Boquien}, {Deger}, {Lee}, {Whitmore}, {Anand}, {Benincasa}, {Bigiel}, {Blanc}, {Chevance}, {Emsellem}, {Faesi}, {Glover}, {Grasha}, {Hughes}, {Klessen}, {Kreckel}, {Kruijssen}, {Leroy}, {Pan}, {Rosolowsky}, {Schruba}, \& {Williams}}]{Turner2022}
{Turner}, J.~A., {Dale}, D.~A., {Lilly}, J., {et~al.} 2022, \mnras, 516, 4612

\bibitem[{{Utomo} {et~al.}(2018){Utomo}, {Sun}, {Leroy}, {Kruijssen}, {Schinnerer}, {Schruba}, {Bigiel}, {Blanc}, {Chevance}, {Emsellem}, {Herrera}, {Hygate}, {Kreckel}, {Ostriker}, {Pety}, {Querejeta}, {Rosolowsky}, {Sandstrom}, \& {Usero}}]{Utomo2018}
{Utomo}, D., {Sun}, J., {Leroy}, A.~K., {et~al.} 2018, \apjl, 861, L18

\bibitem[{Virtanen {et~al.}(2020)Virtanen, Gommers, Oliphant, Haberland, Reddy, Cournapeau, Burovski, Peterson, Weckesser, Bright, {van der Walt}, Brett, Wilson, Millman, Mayorov, Nelson, Jones, Kern, Larson, Carey, Polat, Feng, Moore, {VanderPlas}, Laxalde, Perktold, Cimrman, Henriksen, Quintero, Harris, Archibald, Ribeiro, Pedregosa, {van Mulbregt}, \& {SciPy 1.0 Contributors}}]{Virtanen2020}
Virtanen, P., Gommers, R., Oliphant, T.~E., {et~al.} 2020, Nature Methods, 17, 261

\bibitem[{{Wakker} \& {Adler}(1995)}]{WakkerAdler1995}
{Wakker}, B.~P., \& {Adler}, D.~S. 1995, \aj, 109, 134

\bibitem[{{Walter} {et~al.}(2008){Walter}, {Brinks}, {de Blok}, {Bigiel}, {Kennicutt}, {Thornley}, \& {Leroy}}]{Walter2008}
{Walter}, F., {Brinks}, E., {de Blok}, W.~J.~G., {et~al.} 2008, \aj, 136, 2563

\bibitem[{{Ward} {et~al.}(2022){Ward}, {Kruijssen}, {Chevance}, {Kim}, \& {Longmore}}]{Ward2022}
{Ward}, J.~L., {Kruijssen}, J.~M.~D., {Chevance}, M., {Kim}, J., \& {Longmore}, S.~N. 2022, \mnras, 516, 4025

\bibitem[{{Whitmore} {et~al.}(2014){Whitmore}, {Brogan}, {Chandar}, {Evans}, {Hibbard}, {Johnson}, {Leroy}, {Privon}, {Remijan}, \& {Sheth}}]{Whitmore2014}
{Whitmore}, B.~C., {Brogan}, C., {Chandar}, R., {et~al.} 2014, \apj, 795, 156

\bibitem[{{Wilcots}(1994)}]{Wilcots1994}
{Wilcots}, E.~M. 1994, \aj, 107, 1338

\bibitem[{{Wu} {et~al.}(2017){Wu}, {Tan}, {Nakamura}, {Van Loo}, {Christie}, \& {Collins}}]{Wu2017}
{Wu}, B., {Tan}, J.~C., {Nakamura}, F., {et~al.} 2017, \apj, 835, 137

\bibitem[{{Xu} {et~al.}(2015){Xu}, {Cao}, {Lu}, {Gao}, {Diaz-Santos}, {Herrero-Illana}, {Meijerink}, {Privon}, {Zhao}, {Evans}, {K{\"o}nig}, {Mazzarella}, {Aalto}, {Appleton}, {Armus}, {Charmandaris}, {Chu}, {Haan}, {Inami}, {Murphy}, {Sanders}, {Schulz}, \& {van der Werf}}]{Xu2015}
{Xu}, C.~K., {Cao}, C., {Lu}, N., {et~al.} 2015, \apj, 799, 11

\bibitem[{{Yamaguchi} {et~al.}(2001){Yamaguchi}, {Mizuno}, {Mizuno}, {Rubio}, {Abe}, {Saito}, {Moriguchi}, {Matsunaga}, {Onishi}, {Yonekura}, \& {Fukui}}]{Yamaguchi2001}
{Yamaguchi}, R., {Mizuno}, N., {Mizuno}, A., {et~al.} 2001, \pasj, 53, 985

\bibitem[{{Yasuda} {et~al.}(2023){Yasuda}, {Kuno}, {Sorai}, {Muraoka}, {Miyamoto}, {Kaneko}, {Yajima}, {Tanaka}, {Morokuma-Matsui}, {Takeuchi}, \& {Kobayashi}}]{Yasuda2023}
{Yasuda}, A., {Kuno}, N., {Sorai}, K., {et~al.} 2023, \pasj, 75, 743

\bibitem[{{Yonekura} {et~al.}(2005){Yonekura}, {Asayama}, {Kimura}, {Ogawa}, {Kanai}, {Yamaguchi}, {Barnes}, \& {Fukui}}]{Yonekura2005}
{Yonekura}, Y., {Asayama}, S., {Kimura}, K., {et~al.} 2005, \apj, 634, 476

\end{thebibliography}


\end{document}